\documentclass[iop]{emulateapj}
\usepackage{appendix}

\begin{document}

\newcommand{\mirlum}{L_{\rm 8}}
\newcommand{\ebmv}{E(B-V)}
\newcommand{\lha}{L(H\alpha)}
\newcommand{\lir}{L_{\rm IR}}
\newcommand{\lbol}{L_{\rm bol}}
\newcommand{\luv}{L_{\rm UV}}
\newcommand{\rs}{{\cal R}}
\newcommand{\ugr}{U_{\rm n}G\rs}
\newcommand{\ks}{K_{\rm s}}
\newcommand{\gmr}{G-\rs}

\title{THE CHARACTERISTIC STAR FORMATION HISTORIES OF GALAXIES AT
  REDSHIFTS $\lowercase{z}\sim 2-7$\,\altaffilmark{*}} \author{\sc
  Naveen A. Reddy\altaffilmark{1,2,9}, Max Pettini\altaffilmark{3,4},
Charles C. Steidel\altaffilmark{5}, Alice E. Shapley\altaffilmark{6,10},
Dawn K. Erb\altaffilmark{7}, David R. Law\altaffilmark{8}}

\altaffiltext{*}{Based, in part, on data obtained at the W.M. Keck
Observatory, which is operated as a scientific partnership among the
California Institute of Technology, the University of California, and
NASA, and was made possible by the generous financial support of the
W.M. Keck Foundation.}

\altaffiltext{1}{National Optical Astronomy Observatory, 950 N Cherry 
Avenue, Tucson, AZ 85719, USA}
\altaffiltext{2}{Department of Physics and Astronomy, 
University of California, Riverside, 900 University Avenue, Riverside, CA 92521, USA}
\altaffiltext{3}{Institute of Astronomy, Madingley Road, Cambridge
CB3 OHA, UK}
\altaffiltext{4}{International Centre for Radio Astronomy Research,
University of Western Australia, 7 Fairway, Crawley, WA 6009, AU}
\altaffiltext{5}{Department of Astronomy, California Institute of Technology, 
MS 105--24, Pasadena,CA 91125, USA}
\altaffiltext{6}{Department of Physics and Astronomy, University of California,
Los Angeles, 430 Portola Plaza, Los Angeles, CA 90024, USA}
\altaffiltext{7}{Department of Physics, University of Wisconsin, Milwaukee, 1900 E
Kenwood Blvd, Milwaukee, WI 53211, USA}
\altaffiltext{8}{Dunlap Institute for Astronomy \& Astrophysics, 50 St. George St.,
Toronto, ON M5S 3H4, CA}

\altaffiltext{9}{Hubble Fellow.}
\altaffiltext{10}{David and Lucile Packard Fellow.}

\slugcomment{DRAFT: \today}

\begin{abstract}

A large sample of spectroscopically confirmed star-forming galaxies at
redshifts $1.4\le z_{\rm spec}\le 3.7$, with complementary imaging in
the near- and mid-IR from the ground and from the {\em Hubble} and
{\em Spitzer Space Telescopes}, is used to infer the average star
formation histories (SFHs) of typical galaxies from $z\sim 7$ to 2.
For a subset of 302 galaxies at $1.5\le z_{\rm spec}<2.6$, we perform
a detailed comparison of star formation rates (SFRs) determined from
SED modeling (SFRs[SED]) and those calculated from deep Keck UV and
{\em Spitzer}/MIPS $24$\,$\mu$m imaging (SFRs[IR+UV]).  Exponentially
declining SFHs yield SFRs[SED] that are $5-10\times$ lower on average
than SFRs[IR+UV], indicating that declining SFHs may not be accurate
for typical galaxies at $z\ga 2$.  The SFRs of $z\sim 2-3$ galaxies
are directly proportional to their stellar masses ($M_{\ast}$), with
unity slope---a result that is confirmed with {\em Spitzer}/IRAC
stacks of 1179 UV-faint ($\rs>25.5$) galaxies---for $M_{\ast}\ga
5\times 10^{8}$\,M$_\odot$ and SFRs $\ga 2$\,M$_\odot$\,yr$^{-1}$.  We
interpret this result in the context of several systematic biases that
can affect determinations of the SFR-$M_{\ast}$ relation.  The average
specific SFRs at $z\sim 2-3$ are remarkably similar within a factor of
two to those measured at $z\ga 4$, implying that the average SFH is
one where SFRs increase with time.  A consequence of these rising SFHs
is that (a) a substantial fraction of UV-bright $z\sim 2-3$ galaxies
had faint sub-$L^{\ast}$ progenitors at $z\ga 4$; and (b) gas masses
must increase with time from $z=7$ to 2, over which time the net cold
gas accretion rate---as inferred from the specific SFR and the
Kennicutt-Schmidt relation---is $\sim 2-3\times$ larger than the SFR .
However, if we evolve to higher redshift the SFHs and masses of the
halos that are expected to host $L^{\ast}$ galaxies at $z\sim 2$, then
we find that $\la 10\%$ of the baryons accreted onto typical halos at
$z\ga 4$ actually contribute to star formation at those epochs.  These
results highlight the relative inefficiency of star formation even at
early cosmic times when galaxies were first assembling.

\end{abstract}


\section{INTRODUCTION}
\label{sec:intro}

In the last few years, it has become a standard practice to decipher
the physical characteristics of distant galaxies by fitting broadband
photometry with spectral synthesis models.  Stellar population
modeling, as it is called, has been aided by the availability of deep
imaging in extragalactic fields across a large baseline in wavelength.
Comparison of the broadband spectral energy distribution (SED) of
galaxies with that of a population of stars with a given initial mass
function, star formation history, age, dust reddening, and
metallicity, can therefore yield important insights into the physical
properties of high-redshift galaxies.  This modeling has become more
sophisticated, with some versions allowing for the presence of strong
emission lines (or simultaneously fitting for such lines) that may
affect the broadband photometry (e.g., \citealt{schaerer09,
  schaerer10}).  Other models incorporate the full stellar and dust
SEDs in order to derive self-consistently the reddening of starlight
based upon direct dust indicators (e.g., such as the mid- or
far-infrared dust continuum), thus accounting for the total energy
budget when fitting for the stellar populations \citep{gordon01,
  misselt01,noll09}.  The latter have somewhat limited use for
high-redshift galaxies since it is only for the most infrared luminous
and dusty galaxies at $z\ga 2$ that individual detections at mid and
far-infrared wavelengths are attainable, thus allowing the modeling of
the full IR SED.

While there has been much progress in developing ever-sophisticated
methods of fitting the stellar populations of distant galaxies, the
one fundamental obstacle that affects most of these methods is the
inherent degeneracy between the star formation history, age, and dust
reddening, even when the redshift of the galaxy is known beforehand
(e.g., from spectroscopy).  Lack of redshift information will of
course only further hinder one's ability to robustly determine these
quantities.  It is difficult, if not impossible, to reliably
disentangle these effects based on broadband photometry alone, even
with the deepest optical and near-IR data, as has been discussed in
the first investigations that modeled the stellar populations of
high-redshift galaxies \citep{sawicki98, papovich01, shapley01}.  Full
spectral energy distribution modeling of the stellar and dust
components can break some of this degeneracy, but can also add a new
layer of complication given the increase in number of free parameters
that describe the stellar population and the dust properties and the
spatial distribution of that dust with respect to the stars in a
galaxy.

Finally, there are some inherent uncertainties in SED modeling that
will likely never be fully resolved.  In particular, even in the best
case with deep UV through near-IR photometry, the data are still
insufficient to distinguish simple star formation histories (such as
those parameterized as monotonic exponentially declining, rising, or
constant functions) from more complicated ones that include multiple
generations of bursts.  
In contrast with fossil studies of nearby resolved stellar populations
(e.g., \citealt{williams10}), and in the absence of detailed
spectroscopic abundance measurements, it is difficult to work
``backwards'' from the integrated light of the stellar populations in
a galaxy to a unique set of star formation histories for that galaxy.
Nonetheless, simple star formation histories that vary monotonically
with time have been widely used to infer the stellar population
parameters for high-redshift galaxies.  The most commonly adopted
function is one in which the star formation rate (SFR) of a galaxy
declines exponentially with time, as would be predicted from a closed
box model of galaxy evolution \citep{schmidt59, tinsley80}.  Such
exponentially declining models have been popular as they reproduce the
optical/near-IR colors of local spiral galaxies \citep{bell00} and
appear to reproduce the overall evolution in the SFR density at
redshifts $z\la 2$ (e.g., \citealt{nagamine00}).  As surveys of
dropout selected samples push to increasing redshifts, however, it has
become clear that galaxies at $z\ga 2$ have SFRs and stellar masses
that are inconsistent with them having formed stars according to an
exponentially declining or constant star formation (CSF) history {\em
  prior} to the epoch during which they are observed.

Recently, there has been substantial interest in the possibility that
high-redshift galaxies in general may follow ``rising'' star formation
histories, where the SFR increases exponentially or linearly with
time.  Circumstantial evidence for such rising star formation
histories comes from predictions of cosmological hydrodynamic
simulations (e.g., \citealt{finlator11, weinmann11}), the presence of
SFR versus stellar mass correlations at redshifts $z\ga 1$ (e.g.,
\citealt{renzini09, stark09, gonzalez10, lee11a}), and the increase in
the SFR density per comoving volume at early times (e.g.,
\citealt{papovich11}).

While such simple monotonic functions are unlikely to capture the full
diversity and complexity in the star formation histories of galaxies,
we can still make progress by addressing the average statistical
properties of galaxies across a wide range in redshift, or lookback
times, to effectively look back into the history of star formation and
thus attempt to deduce the way in which average galaxies are evolving.
One method is to use clustering measurements and halo abundance
matching to infer a ``duty cycle'' for star formation on a statistical
basis (e.g., \citealt{adelberger98, lee09}).  Another method is to use
multi-wavelength indicators of reddening and total SFR to constrain
certain parts of parameter space spanned by the SED fitting
parameters.  This approach is simpler than that taken by studies that
treat the dust properties and distribution of dust with respect to the
stars as additional free parameters in modeling the full spectral and
dust SEDs.  Further, the advantage of performing direct comparisons
between independent indicators of SFR and those derived from SED
fitting is that the method can be applied to individual galaxies, so
long as they are detected (or have meaningful upper limits) at
mid-infrared wavelengths, in order to independently measure the
fraction of dust obscured light.  And, unlike stellar masses which are
typically exclusively measured from the rest-frame near-IR light
(which can also have a significant contribution from current star
formation), there are many independent methods of estimating SFRs from
continuum emission (UV, infrared, radio) or nebular line emission
(e.g., H$\alpha$, Pa$\alpha$), thus allowing one to investigate the
systematics and cross check results from different methods.

In this paper we investigate the typical star formation histories of
spectroscopically confirmed UV selected star-forming galaxies at
redshifts $1.4\la z\la 3.7$.  We incorporate in our analysis deep {\em
  Spitzer}/MIPS $24$\,$\mu$m data that exist for a subset of galaxies
in our sample, in the redshift range $1.5\le z\le 2.6$; these mid-IR
data are used to place independent constraints on the SFRs and dust
reddening of galaxies in our sample, quantities that are then compared
to those obtained from the SED fitting given various assumptions of
the star formation history.  We then proceed to discuss this
comparison in light of recent results at higher redshifts ($z\ga 3$)
to form a consistent picture for the typical star formation history of
galaxies during the first $\sim 3$\,billion years of cosmic time.  Our
sample and analysis lend themselves uniquely to addressing these broad
questions because of the large number ($N=1951$) of spectroscopic
redshifts in the range $1.4\le z_{\rm spec}<3.7$; the deep UV,
optical, near-IR, and IRAC data necessary to model the stellar
populations; the deep MIPS $24$\,$\mu$m data, used as an independent
probe of dust attenuation and bolometric SFR; and the careful
consideration of the typical assumptions in SED modeling, and biases
in determining the relationship between SFR and stellar mass.

The outline of this paper is as follows.  In Section~\ref{sec:sample}
we briefly describe the color criteria used to select the sample of
$1.4\la z\la 3.7$ galaxies, and summarize the fields targeted.  In
addition, we present details of the multi-wavelength data in our
fields, including ground-based near-IR and {\em Hubble}/WFC3 imaging,
{\em Spitzer}/IRAC imaging, and {\em Spitzer}/MIPS $24$\,$\mu$m
observations.  The rest-frame UV through near-IR photometry is used to
constrain the stellar populations of spectroscopically confirmed
galaxies in our sample, as discussed in Section~\ref{sec:sedmodeling}.
A detailed comparison between the bolometric measures of star
formation obtained by combining the UV and MIPS $24$\,$\mu$m data,
with those obtained from the SED modeling, is presented in
Section~\ref{sec:sfrcomparison}.  Section~\ref{sec:sfrsyoung} focuses
on modeling the ``younger'' galaxies in our sample by taking into
account dynamical time constraints on the ages and a systematic
steepening of the UV attenuation curve with younger stellar population
age.  The systematic variations and random uncertainties in the ages
and masses of galaxies in the spectroscopically confirmed sample are
discussed in Section~\ref{sec:agemass}.  In addition, we present our
determination of the SFR versus stellar mass correlation, and show how
Malmquist bias can affect inferences of the slope of this relation at
high-redshift.  In Section~\ref{sec:mlratio}, we examine the
mass-to-light ($M/L$) ratios of $z\sim 2-3$ galaxies at rest-frame UV
through near-IR wavelengths, and present a stacking analysis of the
IRAC data on UV faint galaxies lying below our spectroscopic limit.
The $M/L$ ratios and stacking results are then used to infer the
stellar masses of UV faint galaxies.  In Section~\ref{sec:discussion}
we discuss the implications of our results for the typical star
formation histories of high-redshift galaxies; the progenitors of
$z\sim 2-3$ galaxies; and the time evolution of cold gas mass and net
gas accretion rate with redshift.  For ease of comparison with the
literature, we assume a \citet{salpeter55} initial mass function (IMF)
and adopt a cosmology with $H_{0}=70$\,km\,s$^{-1}$\,Mpc$^{-1}$,
$\Omega_{\Lambda}=0.7$, and $\Omega_{\rm m}=0.3$.

\section{SAMPLE}
\label{sec:sample}

\subsection{Galaxy Selection and Optical Spectroscopy}

Galaxies at redshifts $1.4\la z\la 3.7$ were selected using the BM,
BX, and Lyman break galaxy (LBG) rest-UV color criteria
\citep{steidel03, adelberger04, steidel04}.  The imaging data were
obtained mostly from using the Palomar Large Format Camera (LFC) or
the Keck Low Resolution Imaging Spectrograph (LRIS; \citealt{oke95,
  steidel04}).  The photometry and spectroscopic followup for this
survey are described in \citet{steidel03, steidel04, adelberger04}.
Rest-UV spectroscopy with Keck/LRIS was obtained for about $25\%$ of
the sample with $\rs \le 25.5$.  Over all of the fields of our $z\sim
2-3$ survey, the total numbers of photometrically selected BX and LBG
candidates that are detected in $G$ and $\rs$ with $>5$\,$\sigma$
significance are 25,359 and 16,655, respectively, to a typical depth
of $\rs\sim 26.5$.  While most of the subsequent analysis is based on
the spectroscopic sample, we also use the faint $\rs>25.5$ galaxy data
to infer the stellar masses of UV faint galaxies
(Section~\ref{sec:mlratio}).

\subsection{Near-IR Data}

Constraining the stellar population of a galaxy relies critically on
data that bracket the rest-frame spectral region between $\simeq 3600$
and $4000$\,\AA.  It is at these wavelengths that metal absorption
lines from F, G, and K type stars dominate the spectrum resulting in a
break around $4000$\,\AA; an additional absorption feature at
$3646$\,\AA\, marks the edge of the Balmer series and is strongest in
more massive A stars.  Both features are sensitive to age (though the
$3646$\,\AA\, break reaches a maximum at intermediate ages of $\simeq
0.3-1$\,Gyr).  To probe the strength of these features in $z\sim 2-3$
galaxies, we obtained $J$ and/or $\ks$ imaging in 14 fields of the LBG
survey, using the Palomar/WIRC and Magellan/PANIC instruments, to
typical $2\arcsec$ aperture $3$\,$\sigma$ depths of $24.4$ ($\ks$) and
$25.0$\,mag ($J$).  The data were reduced using IDL scripts customized
for WIRC data, and photometry was performed using Source Extractor
\citep{bertin96}.  The near-IR data and the reduction procedures are
discussed in \citet{shapley05}.

In addition, we have obtained {\em Spitzer}/IRAC data for 14 fields of
the survey through General Observer (GO) programs in Cycles 1, 3, and
7, and through DDT time.  When we include the GTO, IOC, and Legacy
programs, there is a total of 27 fields in the LBG survey with IRAC
data (Table~\ref{tab:surveyfields}).  The IRAC coverage of our
galaxies typically included either channel 1 ($3.6$\,$\mu$m) and 3
($5.8$\,$\mu$m) or channel 2 ($4.5$\,$\mu$m) and 4 ($8.0$\,$\mu$m),
with a small fraction of galaxies having coverage in all four channels
(e.g., such as galaxies in the GOODS-N field, or those that are at the
edges of the optical images).  The data were reduced using custom IDL
scripts to correct for artifacts and flat field the data.  Individual
images were mosaiced using the MOPEX software \citep{makovoz05}.  To
take advantage of sub-pixel dithering between individual exposures, we
drizzled the final mosaics onto a grid with a pixel scale of
$0.6\arcsec$ (half the native IRAC pixel scale), enabling higher
resolution images and finer sampling of the point spread function.
Photometry was performed using point-spread function (PSF) fitting to
prior positions determined from the higher resolution optical and
near-IR data.  Errors and bias in the photometry were calculated by
adding artificial sources to the IRAC images and recovering them using
the same PSF fitting software used to measure photometry.  The details
of the PSF fitting and IRAC photometry are provided in
\citet{reddy06b}.

\begin{deluxetable*}{lcrcccl}
\tabletypesize{\footnotesize}
\tablewidth{0pc}
\tablecaption{LBG Survey Fields with Near-IR or IRAC Data}
\tablehead{
\colhead{} &
\colhead{$\alpha$\tablenotemark{a}} &
\colhead{$\delta$\tablenotemark{b}} &
\colhead{Optical Field Size} &
\colhead{} &
\colhead{} &
\colhead{} \\
\colhead{Field Name} &
\colhead{(J2000.0)} &
\colhead{(J2000.0)} &
\colhead{(arcmin$^{2}$)} &
\colhead{{\em HST}/WFC3\tablenotemark{c}} &
\colhead{Near-IR\tablenotemark{d}} &
\colhead{IRAC}}
\startdata
CDFa	&	00 53 23 & 12 33 46 &	\*78.4	& ... & ... & GTO (PI: Fazio) \\	
Q0100	&	01 03 11 & 13 16 18 &	\*42.9	& F160W & WIRC:$\ks$ & DDT (PI: Erb) \\	
Q0105   &       01 08 06 & 16 35 50 &   \*38.7  & ... & ...& GO7 (PI: Reddy) \\
Q0142	&	01 45 17 & -09 45 09 &	\*40.1	& F160W & WIRC:$\ks$ & DDT (PI: Erb) \\	
Q0207   &       02 09 51 & -00 04 58 &  \*37.5  & ... & ... & GO7 (PI: Reddy) \\
Q0302	&	03 04 23 & -00 14 32 &	244.9	& ... & ... & GTO (PI: Fazio) \\	
Q0449	& 	04 52 14 & -16 40 12 &	\*32.1	& F160W & PANIC:$J$,$\ks$ & GO7 (PI: Reddy) \\	
Q0821   &       08 21 05 & 31 08 11 &   \* 39.8 & ... & ... & GO7 (PI: Reddy) \\
B20902	&	09 05 31 & 34 08 02 &	\*41.8	& ... & ... & GTO (PI: Fazio) \\	
Q0933	&	09 33 36 & 28 45 35 &	\*82.9	& ... & WIRC:$\ks$ & ... \\	
Q1009	&	10 11 54 & 29 41 34 &	\*38.3	& F160W & WIRC:$J$,$\ks$ & GO7 (PI: Reddy) \\	
Q1217	&	12 19 31 & 49 40 50 &	\*35.3	& F160W & WIRC:$J$,$\ks$ & GO7 (PI: Reddy) \\	
GOODS-N	&	12 36 51 & 62 13 14 &	155.3	& ... & WIRC:$J$,$\ks$ & Legacy (PI: Dickinson) \\	
Q1307	&	13 07 45 & 29 12 51 &	258.7	& ... & ... & GTO (PI: Fazio) \\	
Westphal &	14 17 43 & 52 28 49 &	226.9	& ... & ... & GTO (PI: Fazio) \\	
Q1422	&	14 24 37 & 22 53 50 &	113.0	& ... & WIRC:$\ks$ & GTO (PI: Fazio) \\	
Q1442   &       14 44 54 & 29 19 06 &   \*36.9  & ... & ... & GO7 (PI: Reddy) \\
3C324	&	15 49 50 & 21 28 48 &	\*44.1	& ... & ... & GTO (PI: Fazio) \\	
Q1549	&	15 51 52 & 19 11 03 &	\*37.3	& F160W & WIRC:$J$,$\ks$ & GO3 (PI: Steidel) \\	
Q1603   &       16 04 56 & 38 12 09 &   \*38.8  & ... & ... & GO7 (PI: Reddy) \\
Q1623	&	16 25 45 & 26 47 23 &	290.0	& F160W & WIRC:$J$,$\ks$ & GO1 (PI: Steidel) \\	
Q1700	&	17 01 01 & 64 11 58 &	235.3	& F160W & WIRC:$J$,$\ks$ & IOC (PI: Fazio) \\	
Q2206 	&	22 08 53 & -19 44 10 &	\*40.5	& F160W & PANIC:$J$,$\ks$ & GO7 (PI: Reddy) \\	
SSA22a	&	22 17 34 & 00 15 04 &	\*77.7	& ... & ... & GTO (PI: Fazio) \\
SSA22b	&	22 17 34 & 00 06 22 &	\*77.6	& ... & ... & GTO (PI: Fazio) \\
Q2233	&	22 36 09 & 13 56 22 &	\*85.6	& ... & ... & GTO (PI: Fazio) \\
DSF2237b &	22 39 34 & 11 51 39 &	\*81.7	& ... & ... & GTO (PI: Fazio) \\
Q2343	&	23 46 05 & 12 49 12 &	212.8	& F160W & WIRC:$J$,$\ks$ & GO3 (PI: Steidel) \\
Q2346	&	23 48 23 & 00 27 15 &	280.3	& ... & WIRC:$\ks$ & ... 
\enddata
\tablenotetext{a}{Right ascension in hours, minutes, and seconds.}
\tablenotetext{b}{Declination in degrees, arcminutes, and arcseconds.}
\tablenotetext{c}{PI: Law.}
\tablenotetext{d}{PIs: Steidel, Erb.}
\label{tab:surveyfields}
\end{deluxetable*}

Finally, we have obtained 8100\,sec {\em HST}/WFC3-F160W ({\em
  H}-band) imaging in 10 fields (with 14 pointings total) of the
$z\sim 2-3$ survey, as part of the Cycle 17 GO-11694 program (PI:
Law).  Details of the data acquisition and reduction are given in
\citet{law12}.  Briefly, nine 900 second exposures in each pointing
were reduced and combined using MultiDrizzle \citep{koekemoer02}.  The
individual exposures are sampled onto a grid with a pixel scale of
$0\farcs08$ to take advantage of the subpixel dithering between
exposures.  The typical $5$\,$\sigma$ depth of the combined images is
$\simeq 27.9$\,AB, assuming a $0\farcs2$ radius aperture.  The {\em
  Hubble} data are particularly advantageous because the combined
depth of the F160W images allows us to constrain the SEDs for fainter
objects that may otherwise be undetected in the ground-based $J/\ks$
or {\em Spitzer}/IRAC imaging.  The ground-based near-IR, {\em
  Hubble}/WFC3 F160W, and {\em Spitzer}/IRAC data, in conjunction with
our $\ugr$ optical data, are used to constrain the stellar populations
and stellar masses of galaxies in our sample, as described in the next
section.

\subsection{Mid-IR Data}

A key aspect of our analysis incorporates independent measurements of
the SFRs of high-redshift galaxies, based on direct tracers of dust
emission.  Six of the fields in our $z\sim 2-3$ survey contain deep
($\approx 4$\,hr) {\em Spitzer}/MIPS imaging at $24$\,$\mu$m with a
typical 3\,$\sigma$ depth of $\approx 12$\,$\mu$Jy: GOODS-N (PI:
Dickinson) and Westphal (PI: Fazio) fields; and the Q1549, Q1623,
Q1700, and Q2343 fields from Cycle 1 and 3 {\em Spitzer} GO
programs.\footnote{There are several additional fields in our survey
  that contain Guaranteed Time Observer (GTO) MIPS imaging.  These
  data are generally much shallower (typically just a few hundred
  seconds), and are not used in this analysis.}  These observations
probe the dust sensitive features around rest-frame $8$\,$\mu$m.  The
MIPS data and reduction are discussed in detail in \citet{reddy06a,
  reddy10a}.  Briefly, the data are flatfielded using a custom IDL
program, and combined with the MOPEX software.  Photometry is
performed using PSF fitting to prior positions defined by detection in
the higher resolution IRAC data.  Photometric bias and errors are
estimated from simulations where we have added artificial sources to
the images and recovered them using the same PSF fitting method.

\subsection{Subsamples and Redshift Ranges}

\begin{figure}[tbp]
\plotone{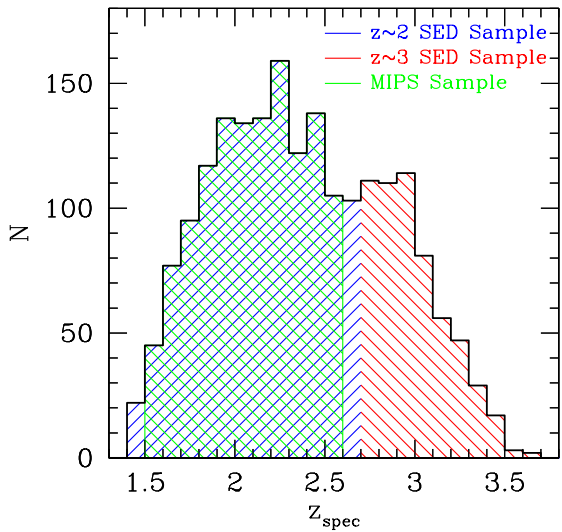}
\caption{Spectroscopic redshift distributions of 1959 galaxies at
  redshifts $1.4\le z_{\rm spec} \le 3.7$, color coded according to
  the subsamples considered (see Table~\ref{tab:subsamples}).}
\label{fig:specdist}
\end{figure}

Throughout this paper, we use different subsamples of the data in
different redshift ranges, with the following motivations.  In
general, the ``$z\sim 2$'' sample refers to those galaxies with
$1.5\le z_{\rm spec}\le 2.6$, or $1.4\le z_{\rm spec}<2.7$.  These two
different ranges are adopted depending on which sample (i.e., the MIPS
or SED sample) is being used.  The ``$z\sim 3$'' sample refers to
those galaxies with $2.7\le z_{\rm spec}<3.7$.  Our total sample with
available SED fits (i.e., have $\ugr$ data plus at least one
photometric point redward of the Balmer break) consists of 1959
galaxies with $1.4\le z_{\rm spec}\le 3.7$.  Of these, there are 302
galaxies with deep MIPS observations, 121 of which are detected
individually at $24$\,$\mu$m with $>3$\,$\sigma$ significance, that
allow for measurements of the rest-frame $8$\,$\mu$m emission,
specifically for those galaxies with spectroscopic redshifts in the
range $1.5\le z_{\rm spec}\le 2.6$.  This ``MIPS'' sample is used to
investigate the comparison between SED and multi-wavelength SFRs
(Section~\ref{sec:sfrcomparison}), and to investigate the relationship
between SFR and specific SFR and stellar mass for $z\sim 2$ galaxies
(Section~\ref{sec:sfrm}).  The comparison of the ages and stellar
masses derived assuming constant and rising star formation histories
is presented for the entire sample of 1959 galaxies in
Section~\ref{sec:compareagemass}.  In Section~\ref{sec:mlratio}, we
consider the mass-to-light ratios of galaxies in our sample at
different wavelengths.  To quantify the $M/L$ ratio at F160W and
$\ks$-band, we use only those galaxies with spectroscopic redshifts
such that F160W and $\ks$ lie longward of the $4000$\,\AA\, break (98
and 491 galaxies, respectively, over the redshifts ranges $1.4\le
z<2.5$ and $1.9\le z<3.7$).  We use similar subsets of the data that
have IRAC channel 1, 2, 3, or 4 data to quantify the $M/L$ ratio at
rest-frame $1.1-2.4$\,$\mu$m (643, 673, 180, and 190 galaxies,
respectively).  The mass-to-light ratios at UV wavelengths are
determined using 630 and 344 galaxies with IRAC channel 1 data over
the redshift ranges $1.4\le z<2.7$ and $2.7\le z <3.7$, respectively.
Finally, we also consider a UV-faint subsample with $\rs>25.5$,
consisting of 1179 candidates, as discussed in
Section~\ref{sec:massfaint}.  The subsamples, their redshift ranges,
and the number of objects, are summarized in
Table~\ref{tab:subsamples}.  The redshift distributions of the various
samples are shown in Figure~\ref{fig:specdist}.

\begin{deluxetable*}{lcr}
\tabletypesize{\footnotesize}
\tablewidth{0pc}
\tablecaption{Subsamples and Redshift Ranges}
\tablehead{
\colhead{Subsample Name} &
\colhead{Redshift Range} &
\colhead{$N_{\rm obj}$}}
\startdata
{\bf MIPS Sample}\tablenotemark{a} & $1.5\le z_{\rm spec} \le 2.6$ & 302 \\
{\bf SED Sample at $z\sim 2$} & $1.4\le z_{\rm spec} < 2.7$ & 1389 \\
{\bf SED Sample at $z\sim 3$} & $2.7\le z_{\rm spec} < 3.7$ & 570 \\
{\bf M/L Ratio at F160W}\tablenotemark{b} & $1.4\le z_{\rm spec} < 2.5$ & 98 \\
{\bf M/L Ratio at $\ks$-band}\tablenotemark{b} & $1.9\le z_{\rm spec} < 3.7$ & 491 \\
{\bf M/L Ratio at $3.6$\,$\mu$m}\tablenotemark{b} & $1.9 \le z_{\rm spec} < 3.7$ & 643 \\
{\bf M/L Ratio at $4.5$\,$\mu$m}\tablenotemark{b} & $1.9 \le z_{\rm spec} < 3.7$ & 673 \\
{\bf M/L Ratio at $5.8$\,$\mu$m}\tablenotemark{b} & $1.9 \le z_{\rm spec} < 3.7$ & 180 \\
{\bf M/L Ratio at $8.0$\,$\mu$m}\tablenotemark{b} & $1.9 \le z_{\rm spec} < 3.7$ & 190 \\
{\bf M/L Ratio at $1700$\,\AA\, with $3.6$\,$\mu$m Coverage} & $1.4\le z_{\rm spec}<3.7$ & 974 \\
{\bf Faint Sample with $\rs>25.5$} & BX/LBG Color Selection & 1179
\enddata
\tablenotetext{a}{The MIPS sample includes all galaxies that are
covered by {\em Spitzer}/MIPS $24$\,$\mu$m imaging, irrespective
of whether they were detected at $24$\,$\mu$m.}
\tablenotetext{b}{Includes only those galaxies in the SED sample that
are detected at F160W, $\ks$, or IRAC channels, and where the band
lies completely redward of the $4000$\,\AA\, break.}
\label{tab:subsamples}
\end{deluxetable*}

\section{STELLAR POPULATION MODELING: GENERAL PROCEDURE}
\label{sec:sedmodeling}

In this section, we discuss the general procedure used to model the
stellar populations of galaxies in our sample.  There are a number of
assumptions that enter into such modeling, such as the adopted star
formation history (e.g., constant, exponentially declining, or
rising), the imposition of a lower limit to the age of a galaxy, and
the choice of attenuation curve. In the subsequent sections, we
discuss and motivate our assumptions by utilizing the multi-wavelength
data that exist in a subset of the fields of our survey.

Stellar masses are inferred by modeling the broadband photometry of
galaxies, using the full rest-frame UV through near-IR photometry to
fit for their stellar populations.  For the fitting, we considered
only those galaxies that are directly detected at wavelengths longward
of rest-frame $4000$\,\AA\, which, for the majority of the sample
considered here, includes the F160W, $\ks$, and IRAC bands.  Further,
we excluded from the fitting any AGN that were identified with strong
UV emission lines (e.g., Ly$\alpha$, CIV) or had a power law SED
through the IRAC bands.  Previous efforts to model the stellar
populations of galaxies in our sample are described in
\citet{shapley05}, \citet{erb06b}, \citet{reddy06b}, and
\citet{reddy10a}.  The latest solar metallicity models of S. Charlot
\& G. Bruzual (in preparation, hereafter CB11) that include the
\citet{marigo07} prescription for the thermally-pulsating Asymptotic
Giant Branch (TP-AGB) evolution of low- and intermediate-mass stars
are used in the fitting.  The stellar masses obtained with these newer
models are generally lower than those based on the \citet{bruzual03}
models for galaxies with ages $\ga 200$\,Myr \citep{reddy10a}.  The
relative contribution of the TP-AGB phase is still debated (e.g.,
\citealt{muzzin09, kriek10, melbourne12}), and we note that adopting
the \citet{bruzual03} models does not significantly alter our results.

If such measurements were available, we corrected the broadband
photometry (optical and/or $\ks$-band) for the effect of Ly$\alpha$
emission/absorption and/or H$\alpha$ emission.  We did not explicitly
correct for [OIII] emission, which lies in the $\ks$-band at $z\sim 3$
and the F160W band at $z\sim 2$, as [OIII] line measurements were not
available.  However, neglecting the correction for Ly$\alpha$,
H$\alpha$, and/or [OIII] emission for most of the galaxies in our
sample results in differences in stellar masses and ages that are
substantially smaller than the marginalized errors on ages and stellar
masses \citep{reddy10a}.  This is due in part to the inclusion of the
IRAC data where line contamination is not an issue, and where such
data provide an additional lever arm to measure the strength of the
Balmer and $4000$\,\AA\, breaks.

For each galaxy, we considered a CSF model and exponentially declining
star formation histories with characteristic timescales $\tau_{\rm d} =$ 10,
20, 50, 100, 200, 500, 1000, 2000, and 5000 Myr.  For comparison, we
also investigated the effect on the stellar population parameters if
we adopt exponentially rising star formation histories, where the
SFR, $\Psi$, is expressed as:
\begin{equation}
\Psi(t)\equiv \Psi_{\rm o}{\rm exp}(t/\tau_{\rm r}), 
\label{eq:sfrequation}
\end{equation}
where $\Psi_{\rm o}$ is the normalization factor, $t$ is time (or
age), and $\tau_{\rm r}$ is the exponential timescale for the rising
history.  We have considered exponential timescales $\tau_{\rm r} =
100$, 200, 500, 1000, 2000, and 5000\,Myr.  These star formation
histories mimic linearly increasing ones if $t_{\rm age}\ll\tau_{\rm
  r}$.  We further considered a range of ages spaced roughly
logarithmically between 50 and 5000 Myr, excluding ages older than the
age of the universe at the redshift of each galaxy.  The lower age
limit of $50$\,Myr is adopted to reflect the dynamical timescale as
inferred from velocity dispersion and size measurements of $z\sim 2$
LBGs \citep{erb06c, law09, law12}; the imposition of this age limit
precludes galaxies from having unrealistic ages that are substantially
younger than the dynamical timescale.  As discussed below, we also
investigate the effect of relaxing this age constraint and show how
adopting a lower age limit (combined with a different attenuation
curve) can resolve the discrepant SED-inferred SFRs of young galaxies
relative to those obtained from direct measurements of the SFRs
derived from combining UV and {\em Spitzer}/MIPS $24$\,$\mu$m data.
Finally, reddening is taken into account by employing the
\citet{calzetti00} attenuation curve (but see below) and allowing
$\ebmv$ to range between 0.0 and 0.6.  The choice of the Calzetti
model is motivated by the good agreement between the Calzetti dust
corrections and those determined from {\em Spitzer}/MIPS $24$\,$\mu$m
and {\em Herschel}/PACS $100$ and $160$\,$\mu$m inferences of the
infrared luminosities (\citealt{reddy06a, reddy10a, reddy12}).

The model SED at each $\tau$ and age ($t_{\rm age}$) combination is
reddened, redshifted, and attenuated blueward of rest-frame
$1216$\,\AA\, for the opacity of the IGM using the \citet{madau95}
prescription.  The best-fit normalization of this model is determined
by minimizing its $\chi^2$ with respect to the observed
$\ugr+J\ks$+F160W+IRAC (3.6$-$8.0\,$\mu$m) photometry.  This
normalization then determines the SFR and stellar mass.  The model
(and normalization) that gives the lowest $\chi^2$ is taken to be the
best-fit SED.  Typically there are several best-fit models that may
adequately describe the observed photometry, even when the redshift is
fixed to the spectroscopic value, though there is generally less
variation in stellar mass than in the other parameters ($\tau$,
$t_{\rm age}$, $\ebmv$) among these best-fit models \citep{sawicki98,
  papovich01, shapley01, shapley05}.  Below, we consider a variety of
star formation history models with different assumptions for the age
limit and attenuation curve (Table~\ref{tab:sfrrelations}).

\begin{deluxetable}{llc}
\tabletypesize{\footnotesize}
\tablewidth{0pc}
\tablecaption{Description of SED Models}
\tablehead{
\colhead{Model} &
\colhead{Assumptions} &
\colhead{RMS\tablenotemark{a}}}
\startdata
{\bf Model A} & Constant Star Formation & 0.44\\
              & Calzetti Attenuation Curve \\
              & No Age Limit \\
\\
{\bf Model B} & Declining Star Formation & 0.48\\
              & Calzetti Attenuation Curve \\
              & No Age Limit \\
\\
{\bf Model C} & Rising Star Formation & 0.44\\
              & Calzetti Attenuation Curve \\
              & No Age Limit \\
\\
{\bf Model D} & Constant Star Formation & 0.44\\
              & Calzetti Attenuation Curve \\
              & $t_{\rm age}>50$\,Myr \\
\\
{\bf Model E} & Declining Star Formation & 0.48\\
              & Calzetti Attenuation Curve \\
              & $t_{\rm age}>50$\,Myr \\
\\
{\bf Model F} & Rising Star Formation & 0.46 \\
              & Calzetti Attenuation Curve \\
              & $t_{\rm age}>50$\,Myr \\
\\
{\bf Model G} & Constant Star Formation & 0.46\\
              & Calzetti/SMC Attenuation Curves\tablenotemark{b} \\
              & $t_{\rm age}>50$\,Myr \\
\\
{\bf Model H} & Declining Star Formation & 0.50\\
              & Calzetti/SMC Attenuation Curves\tablenotemark{b} \\
              & $t_{\rm age}>50$\,Myr \\
\\
{\bf Model I} & Rising Star Formation & 0.48\\
              & Calzetti/SMC Attenuation Curves\tablenotemark{b} \\
              & $t_{\rm age}>50$\,Myr 
\enddata
\tablenotetext{a}{RMS dispersion (in dex) about the best fit relation between
 MIPS+UV SFR and SED inferred SFRs, as determined from the expectation maximization (EM) parametric
estimator, for typical galaxies with $\lbol<10^{12}$\,L$_{\odot}$ and $t_{\rm age}>100$\,Myr.}
\tablenotetext{b}{In this model, we have assumed the \citet{calzetti00} attenuation
curve for galaxies with Calzetti-derived ages of $t_{\rm age}^{\rm Calz}>100$\,Myr.  For those
galaxies with $t_{\rm age}^{\rm Calz}<100$\,Myr, we remodeled their photometry assuming
an SMC attenuation curve.}
\label{tab:sfrrelations}
\end{deluxetable}

\section{MULTI-WAVELENGTH CONSTRAINTS ON THE SFRs AND
STELLAR POPULATIONS OF HIGH-REDSHIFT GALAXIES}
\label{sec:sfrcomparison}

In this section, we compare the SFRs derived from SED fitting
(SFR[SED]) with those calculated from combining the UV and MIPS data
(SFR[IR+UV]).  As discussed in Section~\ref{sec:sample}, there are 302
galaxies in our sample with MIPS $24$\,$\mu$m observations and
spectroscopic redshifts $1.5\le z\le 2.6$ (121 of these galaxies are
detected individually at $24$\,$\mu$m); it is at these redshifts where
the $24$\,$\mu$m fluxes are sensitive to the rest-frame 8\,$\mu$m
emission, which in turn can be converted to $\lir$.  Using the
procedure described in \citet{reddy10a}, we {\em k}-corrected the
$24$\,$\mu$m fluxes to estimate rest-frame 8\,$\mu$m luminosities
($\mirlum$).  We then assumed a ratio $\lir/\mirlum = 8.9\pm 1.3$, as
determined from a stacking analysis of the {\em Herschel}/PACS 100 and
160\,$\mu$m data for a subset of the same galaxies considered here
(i.e., those in the GOODS-North field; \citealt{reddy12}).  The
\citet{kennicutt98} relations are then used to convert $\lir$ and
$\luv$ (uncorrected for extinction) to SFRs, the sum of which gives
the bolometric SFR.

\begin{figure*}[tbp]
\plottwo{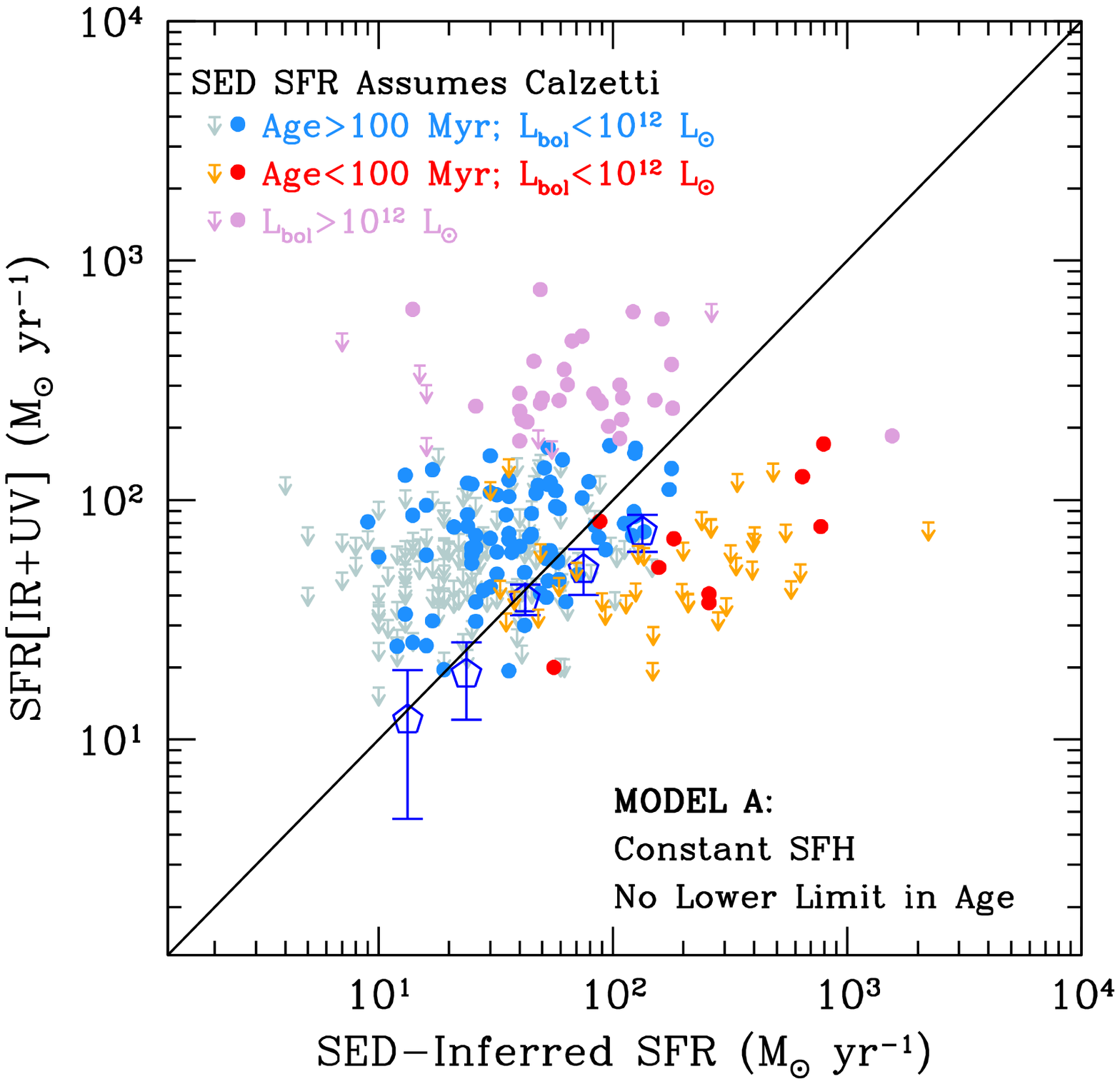}{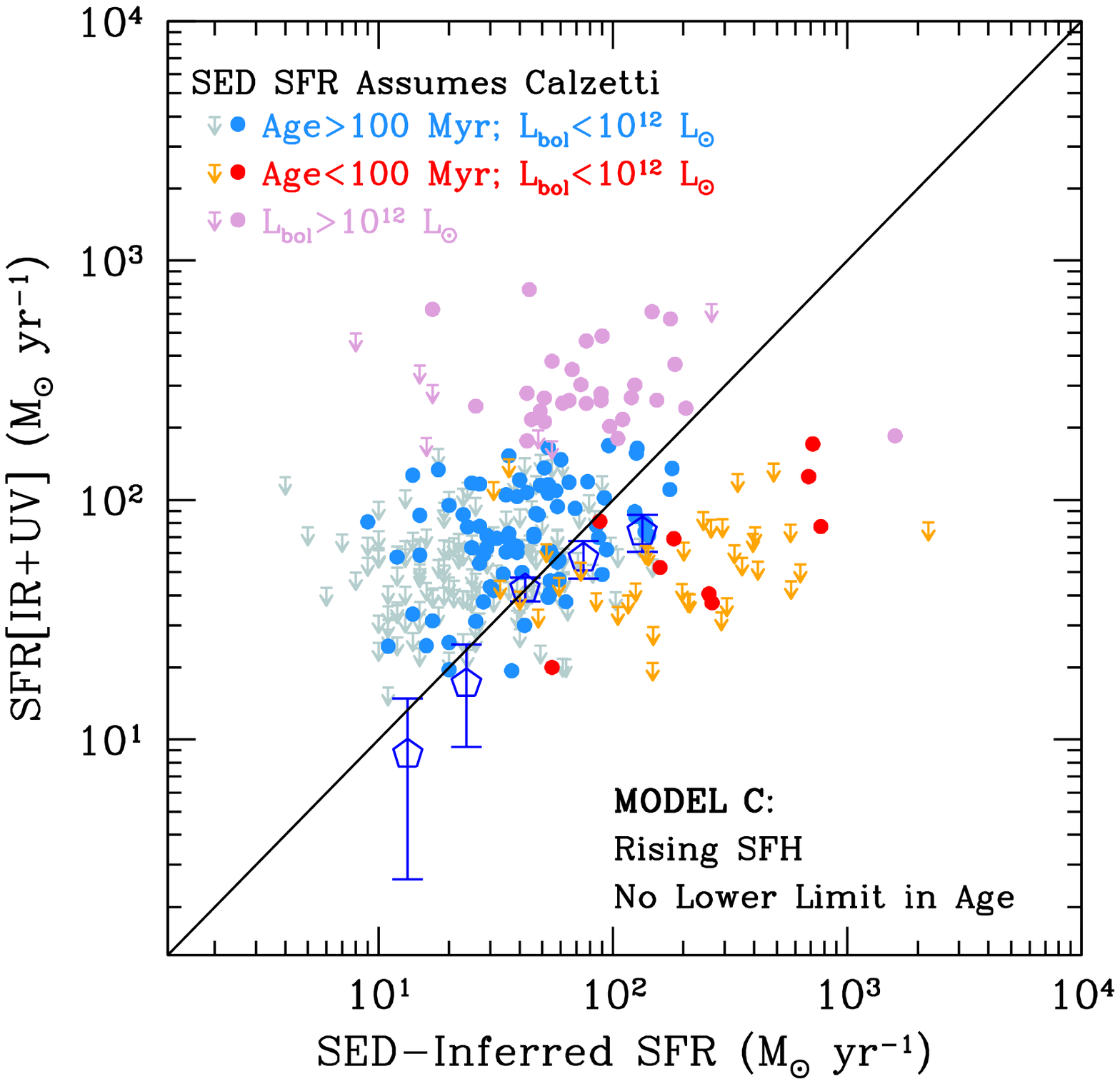}
\caption{Comparison between SFRs derived from SED fitting and those
  computed from combining MIPS $24$\,$\mu$m and UV data.  Points are
  color coded to differentiate galaxies with large bolometric
  luminosities ($\lbol > 10^{12}$\,L$_\odot$), typical galaxies that
  form the bulk of the sample ($\lbol \le 10^{12}$\,L$_\odot$ and
  $t_{\rm age}>100$\,Myr), and young galaxies with $t_{\rm age}\le
  100$\,Myr.  Upper limits are points denote galaxies with
  $24$\,$\mu$m nondetections and detections, respectively.  The large
  pentagons show the average $24$\,$\mu$m plus UV determined SFR in
  bins of SED-inferred SFR for typical galaxies, as determined from a
  $24$\,$\mu$m stacking analysis (see the text).  The results for a
  constant and rising star formation history, with no lower limit
  imposed on the age, are shown in the left and right panels,
  respectively.}
\label{fig:sfrallage}
\end{figure*}

Note that bolometric SFR computed in this way is not entirely
independent of the stellar population because the conversion from
$\luv$ to SFR will depend on the star formation history and age.  This
dependence is discussed in detail in Appendix~\ref{sec:uvconv}.  For
exponentially declining star formation histories where $t\gg\tau_{\rm
  d}$, the factor to convert $\luv$ to SFR will be substantially
smaller than the \citet{kennicutt98} value.  More generally, for all
of the models considered here (declining, constant, and rising), the
conversion factor is at least a factor of two larger than the
canonical value for ages $\la 10$\,Myr.  For convenience, in
Appendix~\ref{sec:uvconv} we provide formulae to compute the
conversion factor between $\luv$ and SFR for different star formation
histories and ages, assuming the CB11 stellar population synthesis
models, a \citet{salpeter55} IMF between 0.1-100\,M$_{\odot}$, and
solar metallicity.  Finally, we note that the factor to convert $\lir$
to SFR is also somewhat dependent on galaxy age, with the
\citet{kennicutt98} conversion valid for starbursts with ages $\la
100$\,Myr.  At older ages, there are the competing effects of lower
dust opacity and heating from older stellar populations that can
modulate the conversion factor between $\lir$ and SFR
\citep{kennicutt98}.  These effects at older ages are likely to be
negligible for most of the high-redshift galaxies studied here due to
their larger SFRs, and lower stellar masses, compared to the local
galaxies used to calibrate the relationship between $\lir$ and SFR.

For simplicity, we assume the canonical \citet{kennicutt98} relations
to convert $\lir$ and $\luv$ to SFR, and discuss below how changing
the $\luv$-SFR conversion factor affects our results.  The MIPS+UV
derived SFRs are compared to the SED-inferred SFRs in
Figure~\ref{fig:sfrallage}.  Below, we discuss in turn the three sets of
objects that differentiate themselves in the plane of SFR[IR+UV]
versus SFR[SED].

\subsection{SFRs of Typical Star-forming Galaxies at $z\sim 2$}

This subsection highlights the results found for typical star-forming
galaxies, defined as those galaxies with best-fit ages $>100$\,Myr and
bolometric luminosities $\lbol<10^{12}$\,L$_{\odot}$.  Here, $\lbol$
is taken as the sum of the UV and IR luminosities, where $\lir$ is
computed from the $24$\,$\mu$m data using the $\lir/\mirlum$ ratio
discussed above.  Because $60\%$ of the galaxies are not detected at
$24$\,$\mu$m, we have stacked the $24$\,$\mu$m data for typical
galaxies in bins of SFR[SED] to determine the average bolometric SFR.
The stacking procedure is discussed in \citet{reddy06a}.
Uncertainties in the stacked bolometric SFR are computed by combining
in quadrature the measurement uncertainty in the mean $24$\,$\mu$m
flux and the uncertainty in the mean UV luminosity of galaxies
contributing to the stack.

SFR[IR+UV] derived in this manner agrees well on average with SFR[SED]
assuming constant star formation and no age limit ({\em large
  pentagons} in the left panel of Figure~\ref{fig:sfrallage}).  Taking
into account upper limits using the ASURV statistical package
\citep{isobe86}, which includes the expectation maximization (EM)
parametric survival estimator for censored data, we compute an rms
dispersion about the best-fit linear relation between SFR[IR+UV] and
SFR[SED] of 0.44\,dex (Table~\ref{tab:sfrrelations}\footnote{The RMS
  values listed in Table~\ref{tab:sfrrelations} merely indicate the
  rms about the best-fit relation between SFR[SED] and SFR[IR+UV], and
  are not meant to indicate the ``goodness of fit'' between the two
  quantities.}).  Based on the stacking analysis and the survival
analysis, we conclude that there is a good agreement between the
MIPS+UV and SED derived SFRs for typical star-forming galaxies at
$z\sim 2$.  These results are not surprising because previous
investigations have shown that on average the \citet{calzetti00}
corrections for dust obscuration (which are assumed in the SED fitting
procedure) reproduce the values estimated from mid and far-infrared,
radio, and X-ray data for galaxies at $z\sim 2-3$ (e.g.,
\citealt{nandra02, seibert02, reddy04, reddy06a, daddi07a, pannella09,
  reddy10a, magdis10a, reddy12}).

\subsection{SFRs of Ultraluminous Infrared Galaxies at $z\sim 2$}

Turning our attention to ultraluminous infrared galaxies (ULIRGs), we
find that those galaxies with $\lbol>10^{12}$\,L$_{\odot}$ have
bolometric SFRs that exceed by up to a factor of ten the SFRs[SED].
The $24$\,$\mu$m fluxes of the IR-luminous sources in our sample tend
to over-predict their $\lir$ by a factor of $\approx 2$, relative to
the IR estimates obtained by including far-IR data (e.g., from {\em
  Herschel}; \citealt{reddy12}).  Further, as shown in several other
investigations, the \citet{calzetti00} dust corrections for such
objects are typically too low due to the fact that much of the star
formation is completely obscured by dust, and hence the UV slope
decouples from extinction for such highly obscured galaxies (e.g.,
\citealt{goldader02, reddy06a, reddy10a}).

\subsection{SFRs of Young Galaxies at $z\sim 2$}
\label{sec:youngsfrs}

A noted problem in stellar population modeling is the distribution of
unrealistically young ages derived for non-negligible fractions of
galaxies in high-redshift samples, particularly those selected in the
rest-UV or rest-optical \citep{shapley01, maraston10}.  This is
commonly referred to as the ``outshining'' problem, where the SED of a
galaxy may be dominated by the youngest stellar population even at
near-IR wavelengths, in which case the best-fit models (irrespective
of the star formation history) may have very young ($\la 100$\,Myr)
ages with very large values of SFR and reddening (e.g.,
\citealt{maraston10, wuyts11}).  We have investigated the validity of
these young models by cross-checking the SFRs derived using them with
those measured directly from MIPS $24$\,$\mu$m and UV data.

For such young ($t_{\rm age}<100$\,Myr) objects (the third set of
objects denoted in Figure~\ref{fig:sfrallage}), we find that SFRs[SED]
are systematically larger than SFRs[IR+UV].  This offset cannot be
completely accounted for by a change in the conversion between UV
luminosity and SFR.  In particular, the bulk of these young galaxies
have ages (derived with the CSF assumption) as short as $7$\,Myr.  For
these ages, the conversion between UV luminosity and SFR is about a
factor of two larger than in the CSF case where $t_{\rm age}>100$\,Myr
(Appendix~\ref{sec:uvconv}).  Given the typical dust attenuation of
these ``young'' galaxies of $\lir/\luv\approx 1$ (e.g.,
\citealt{reddy06a, reddy12}), modifying the UV-SFR conversion upward
by a factor of two will increase SFR[IR+UV] by $50\%$, or $0.18$\,dex.
This difference is not sufficient to account for the offset between
SFR[SED] and SFR[IR+UV] for the young subsample.

\begin{figure}[tbp]
\plotone{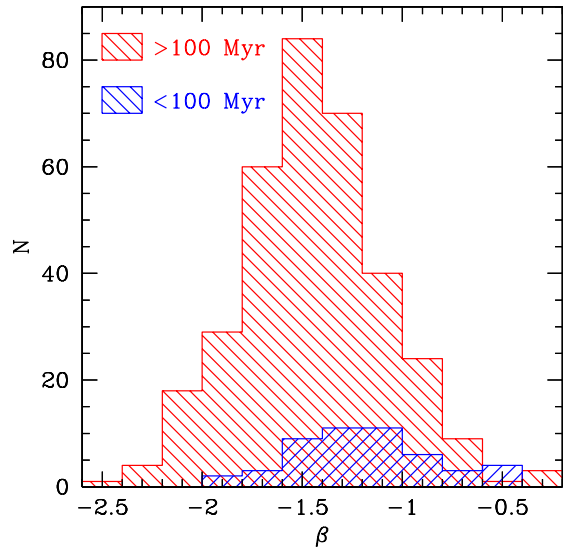}
\caption{$\beta$ distribution for galaxies with Calzetti-derived ages
  $>100$\,Myr (red histogram) compared to that of galaxies with
  Calzetti-derived ages $<100$\,Myr (blue histogram), in the sample of
  392 UV selected galaxies of \citet{reddy10a}.  The ages assume a CSF
  model.  A two-sided K-S test indicates a probability $P\la 0.01$
  that the two distributions are drawn from the same parent sample.
  On average, the younger galaxies have UV slopes that are $\langle
  \delta\beta\rangle \simeq 0.3$ redder than those of older galaxies.}
\label{fig:betadist}
\end{figure}

Galaxies at $z\sim 2$ that are identified as being young based on
their CSF model fits have UV slopes, $\beta$, which are on average
redder than the slopes of older galaxies at the same redshifts, with a
difference in $\beta$ of $\langle\delta \beta\rangle \simeq 0.3$
(Figure~\ref{fig:betadist}).  If we assume that a Calzetti attenuation
curve applies for such galaxies, then we would find them to be more
dust obscured, given their redder $\beta$, and hence to have larger
bolometric SFRs, than the ``typical'' galaxies discussed above.
Clearly, the very large dust obscuration and bolometric SFRs inferred
for such galaxies cannot be correct, based on the comparison of these
SFRs with those derived directly from combining the MIPS and UV data
(Figure~\ref{fig:sfrallage}).

\subsection{Results for Other Star Formation Histories}

\begin{figure*}[tbp]
\plottwo{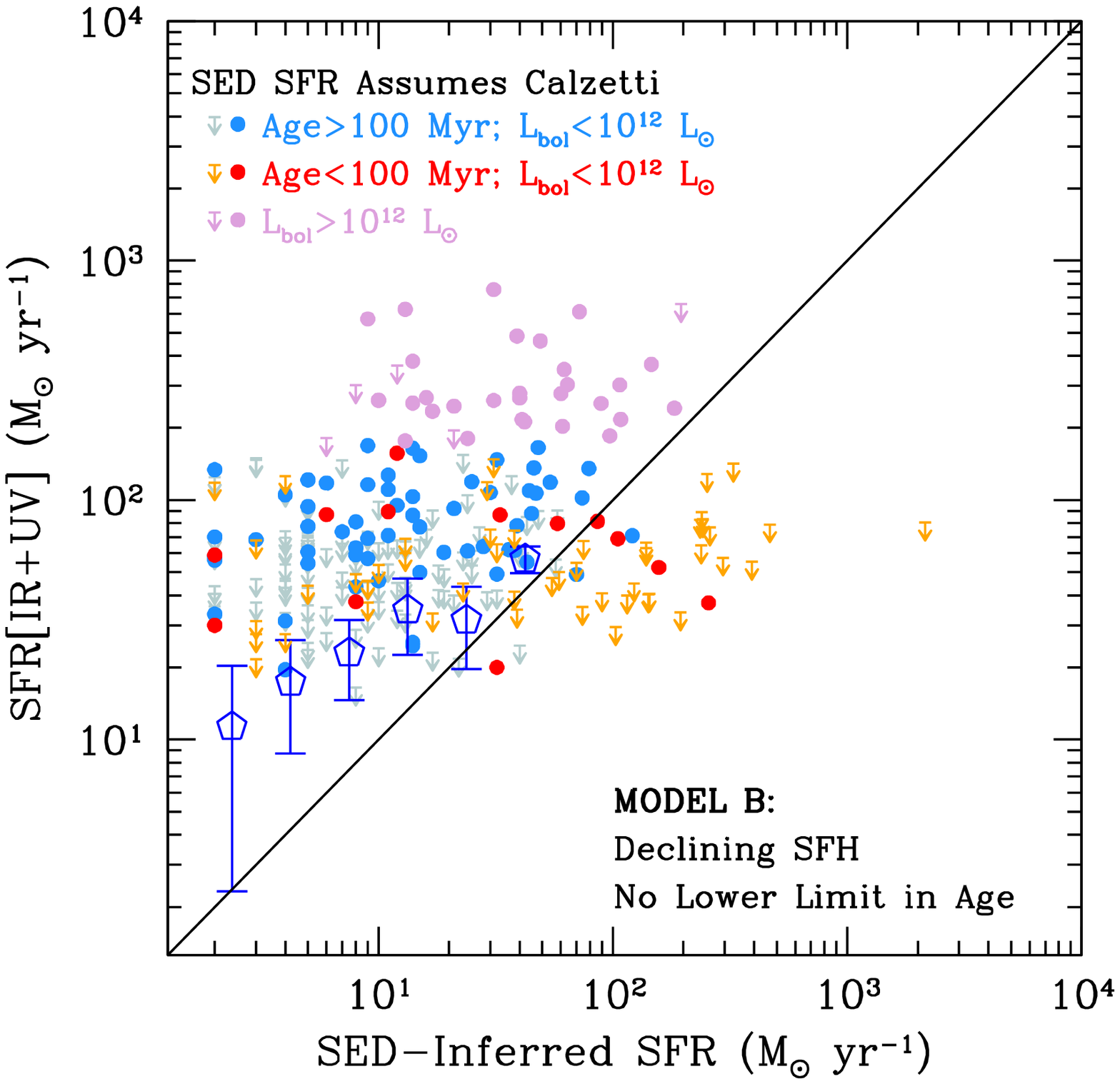}{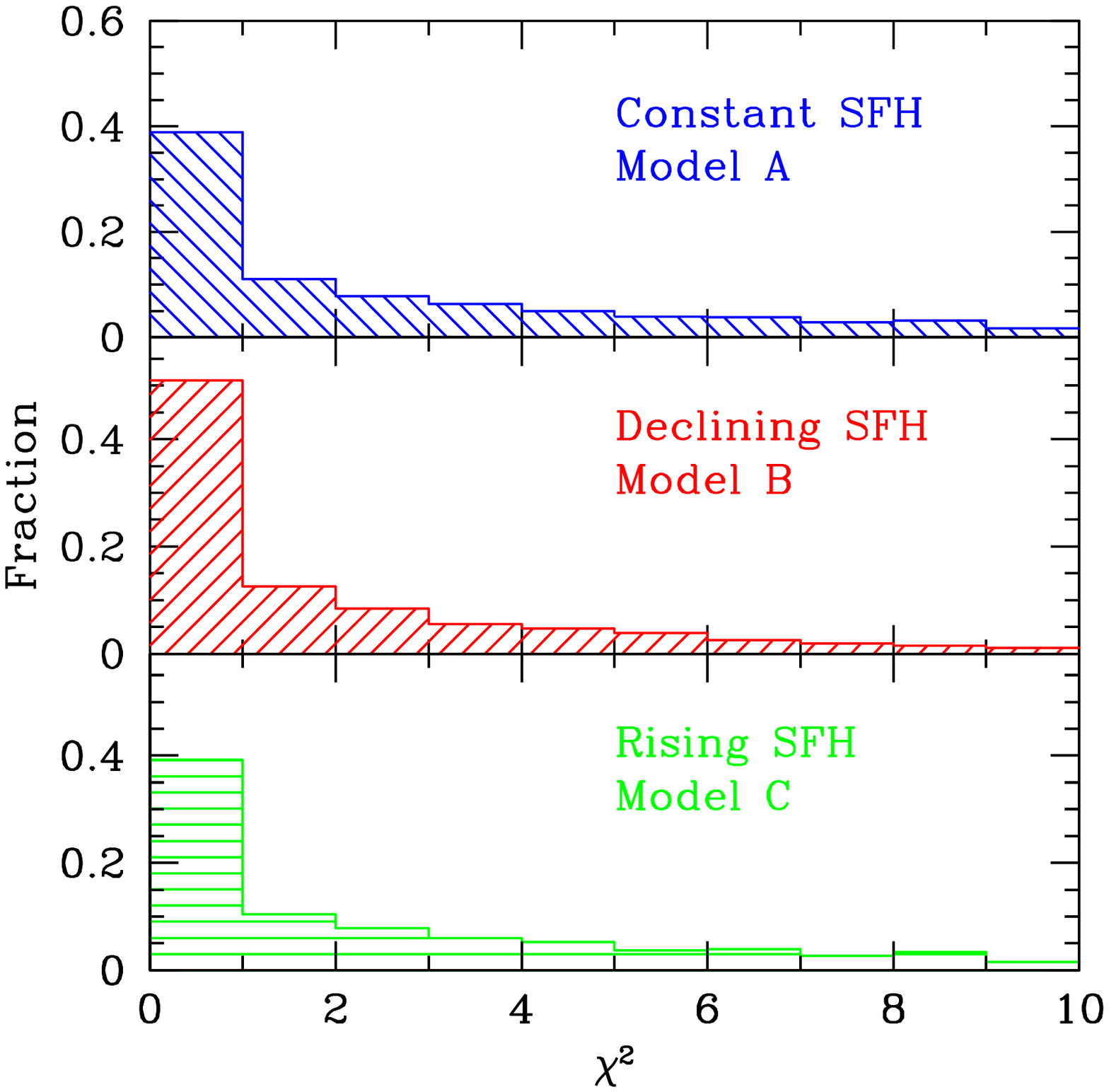}
\caption{({\em Left:}) Same as Figure~\ref{fig:sfrallage}, where we
  have assumed the best-fit exponentially declining star formation
  history.  ({\em Right:}) Histograms of the $\chi^2$ distributions of
  the best-fit SED models to the broadband photometry for 1959
  galaxies with spectroscopic redshifts $1.4\le z\le 3.7$, assuming
  declining, constant, and rising star formation histories
  (corresponding to Models A, B, and C, respectively, as listed in
  Table~\ref{tab:sfrrelations}).  The histograms have been normalized
  to the total number of galaxies (1959).}
\label{fig:declining}
\end{figure*}

We conclude this discussion by noting that the assumption of
exponentially rising star formation histories results in the same
qualitative behavior for the three sets of galaxies discussed above
(typical galaxies, ULIRGs, and young galaxies;
Figure~\ref{fig:sfrallage}).  The RMS around the best-fit relation
between SFR[SED] and SFR[IR+UV] is listed in
Table~\ref{tab:sfrrelations}.  In contrast, the assumption of
exponentially declining star formation histories results in SFRs[SED]
for typical galaxies that lie systematically below SFRs[IR+UV]
(Figure~\ref{fig:declining}).  A similar conclusion is reached when
comparing dust-corrected H$\alpha$ estimates of the SFR with those
computed from fitting exponentially declining models to the broadband
photometry of a subset of galaxies in our sample \citep{erb06c}.

The discrepancies that arise from adopting an exponentially declining
star formation history stem from two effects.  The first is that the
redness of the UV continuum is, to a greater extent, attributed to
older stars in the case when $t_{\rm age}/\tau_{\rm d} > 1$, relative
to the CSF case.  Hence, the $\ebmv$ for declining models will on
average be lower, implying lower SFRs.  The second effect is that the
ratio of the SFR to UV continuum is generally lower for galaxies with
smaller $\tau_{\rm d}$ (Appendix~\ref{sec:uvconv}; see also
\citealt{shapley05} for a discussion of these points), which also
results in smaller SFRs for a given UV luminosity.  \citet{shapley05}
discuss other reasons why extreme declining star formation histories
where $\tau_{\rm d}<100$\,Myr and $t_{\rm age}/\tau_{\rm d}\gg1$ are
implausible for the vast majority of galaxies in our sample.  In
particular, the correspondence between the H$\alpha$ and UV derived
SFRs for $z\sim 2$ galaxies \citep{erb06c, reddy10a}, as well as the
presence of O star features in the UV spectra of galaxies in our
sample---e.g., Ly$\alpha$ emission, and Wolf Rayet signatures
including broad HeII 1640\,\AA\, and P-Cygni wind features in the CIV
and SiIV lines \citep{shapley03, quider09}---imply that the UV
continuum is likely dominated by O stars, contrary to the expectation
if $t_{\rm age}/\tau_{\rm d}\gg1$.

The importance of the comparison to the MIPS+UV SFRs is underscored by
the fact that one cannot discriminate between these various star
formation histories, even the simple ones considered here, from the
broadband SEDs alone.  As the right panel of
Figure~\ref{fig:declining} shows, the $\chi^2$ distributions of the
best-fit SEDs for the different star formation histories are roughly
similar.  The MIPS data allow us to break this impasse, and they can
be used to demonstrate clearly that, on average, the exponentially
declining models yield SFRs that are statistically inconsistent with
those obtained from direct measures of the
SFRs.\footnote{\citet{wuyts11} show that restricting declining star
  formation histories to {\em e-}folding times $\ga 300$\,Myr result
  in a better agreement between SED-inferred SFRs and those obtained
  from combining UV and IR data.  Doing the same for our sample, we
  obtain a median value of $t/\tau \simeq 0.4$ which implies a
  behavior closer to that of the CSF models.}  Given this finding, we
focus the subsequent discussion on constant and rising star formation
histories.

\begin{figure}[tbp]
\plotone{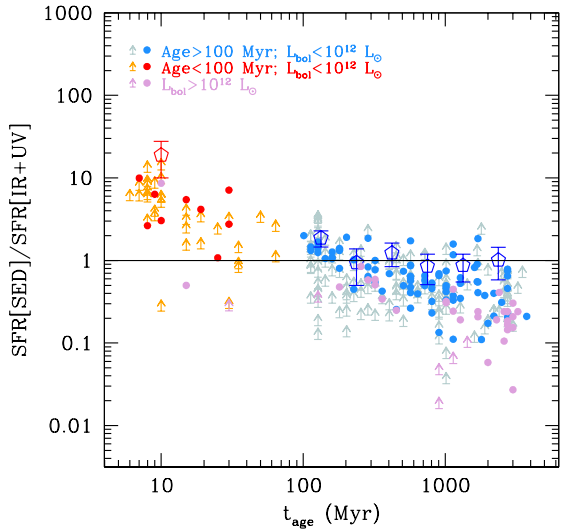}
\caption{Ratio of SED-inferred SFR and SFR derived from combining MIPS
  and UV data, vs. stellar population age, for Model C
  (Table~\ref{tab:sfrrelations}; rising star formation history with no
  age limit).  Symbols are the same as in Figure~\ref{fig:sfrallage}.
  The large red pentagon denotes the median value based on stacking
  the $24$\,$\mu$m data for the galaxies with $t_{\rm age}<100$\,Myr.
  For Model C, the SED-inferred SFRs of the young galaxies exceed by
  an order of magnitude those estimates from combining the MIPS and UV
  data.}
\label{fig:sfrvage}
\end{figure}

\section{RESOLVING THE CONFLICTING SFRs FOR YOUNG GALAXIES}
\label{sec:sfrsyoung}

\begin{figure*}[tbp]
\plottwo{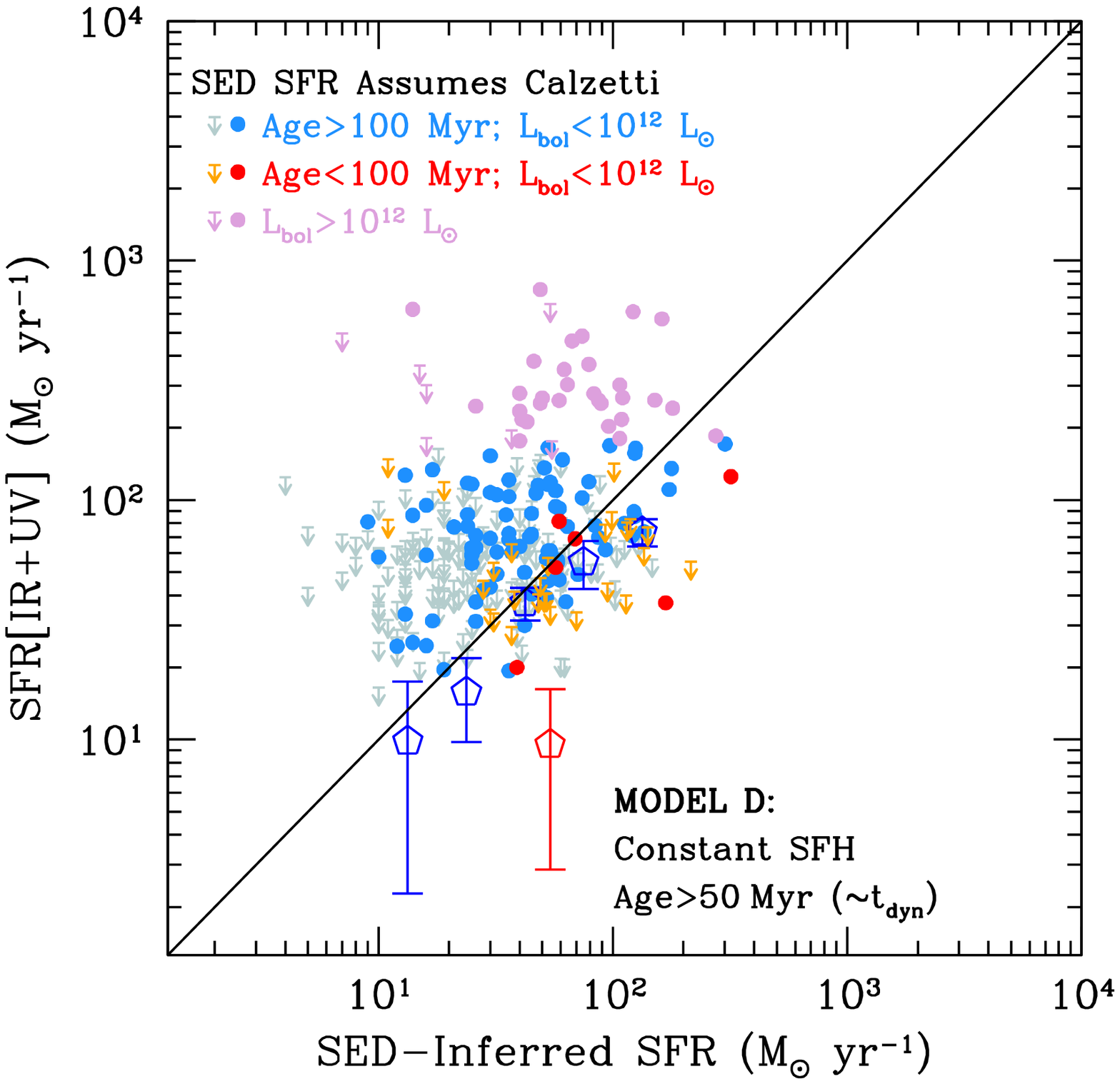}{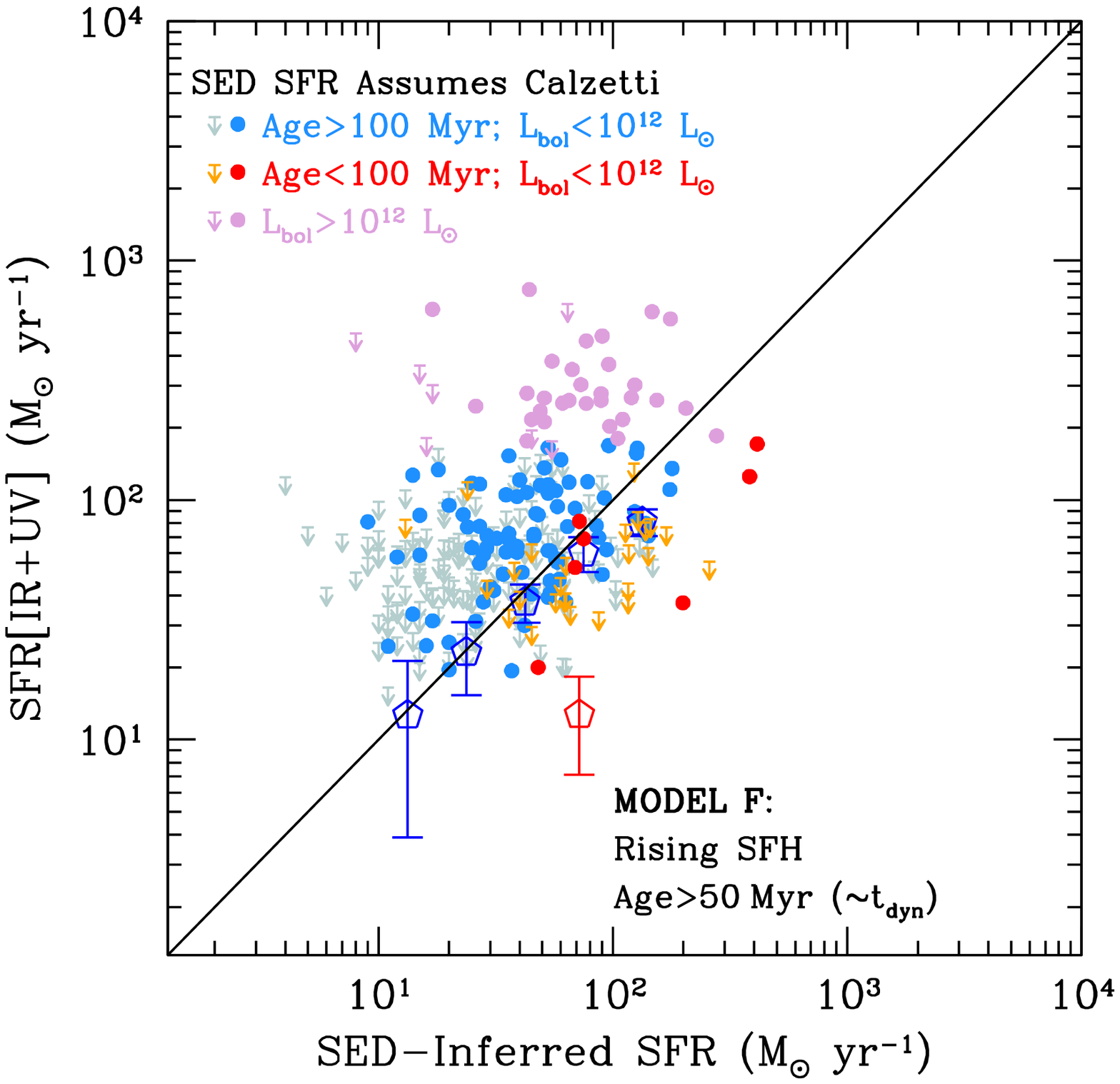}
\caption{Same as Figure~\ref{fig:sfrallage} where we have imposed a
  lower limit on the age of $t_{\rm age}=50$\,Myr in the SED modeling
  including constant (left) and rising (right) star
  formation histories.  The large red pentagon denotes the stacked
  results for galaxies with $50\le t_{\rm age}<100$\,Myr.}
\label{fig:sfragegt50}
\end{figure*}

As noted in Section~\ref{sec:youngsfrs}, all of the star formation
histories considered here predict SFRs that are substantially larger
than those computed from the MIPS and UV data for young galaxies with
$t_{\rm age}<100$\,Myr.  Figure~\ref{fig:sfrvage} shows that this
systematic difference is a strong function of age, being most severe
for galaxies with $t_{\rm age}\la 50$\,Myr.  Given that the true ages
are unlikely to be significantly smaller than the dynamical timescale
of $\simeq 50$\,Myr, we have imposed the restriction $t_{\rm
  age}>t_{\rm dyn}\simeq 50$\,Myr when fitting the SEDs.  Note that
the exact limit in age is unimportant given the relatively few
galaxies with $20\la t_{\rm age}\la 100$\,Myr.  Adopting an age
restriction, the comparison between SED and MIPS+UV SFRs is shown in
Figure~\ref{fig:sfragegt50}.  At face value, based on the individual
data points, the agreement between the SFRs for galaxies with $t_{\rm
  age}<100$\,Myr is better once we have restricted the age to be older
than $t_{\rm age}=50$\,Myr.  However, a MIPS stack for those galaxies
with $50<t_{\rm age}<100$\,Myr results in a formal non-detection at
$24$\,$\mu$m, inconsistent at the $3\sigma$ level with our expectation
based on the SFRs[SED] for these galaxies ({\em red pentagons} in
Figure~\ref{fig:sfragegt50}).

\begin{figure*}[tbp]
\plottwo{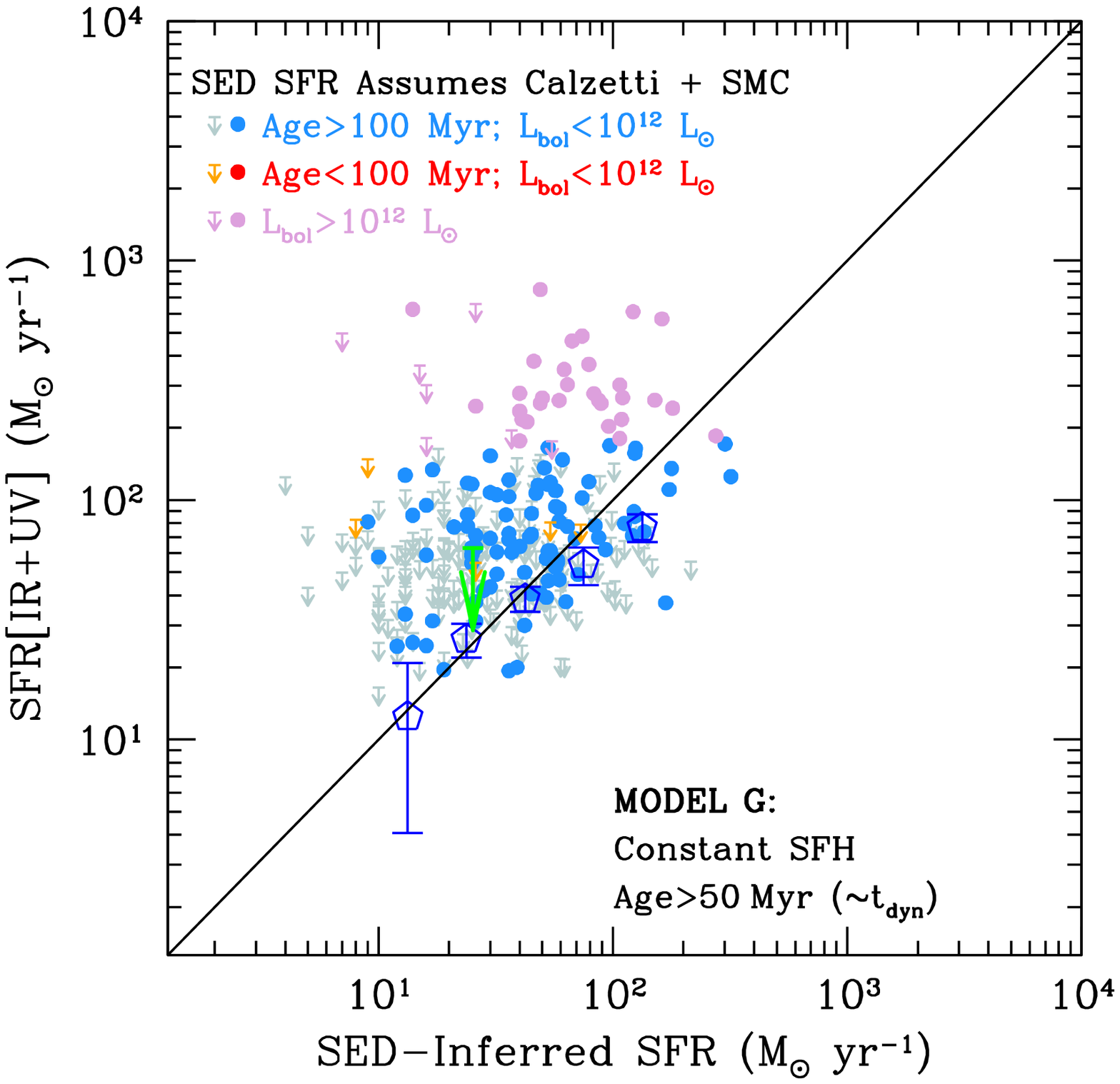}{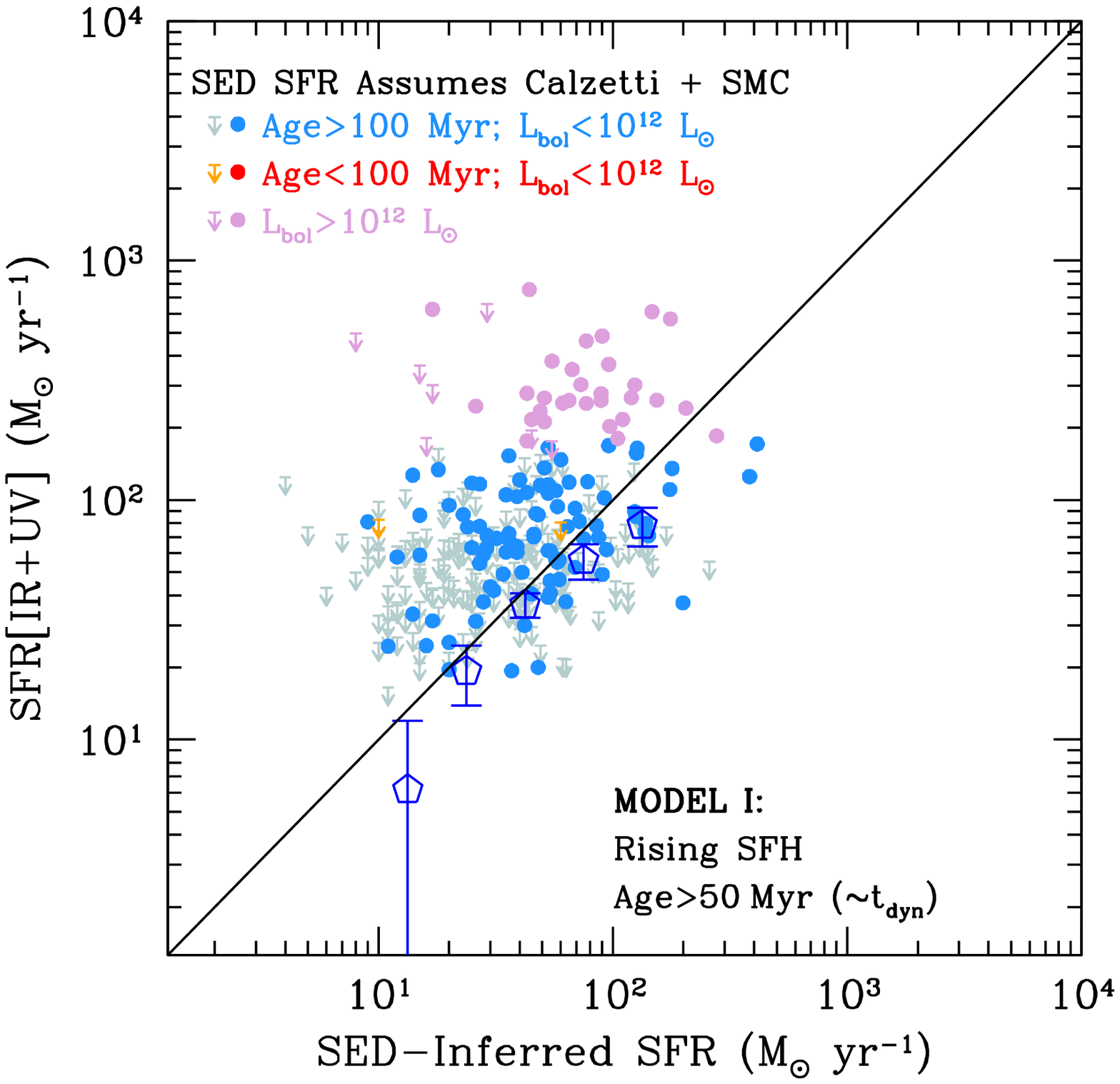}
\caption{Same as Figure~\ref{fig:sfragegt50} where we have imposed a
  lower limit on the age of $t_{\rm age}=50$\,Myr in the SED modeling,
  {\em and} we have remodeled those galaxies with Calzetti inferred
  ages $t_{\rm age}<100$\,Myr with an SMC attenuation curve.  Results
  are shown for the constant (left) and rising (right) star formation
  histories.  The green arrow in the left panel indicates the
  3\,$\sigma$ upper limit to SFR[IR+UV] for the 4 $24$\,$\mu$m
  undetected galaxies that are considered young under the assumption
  of the SMC attenuation curve.}
\label{fig:sfrsmc}
\end{figure*}

\subsection{Age Dependence of the UV Attenuation Curve}

One possible solution to this discrepancy comes from the correlation
between the dust obscuration, $\lir/\luv$, and UV slope ($\beta$) for
these young galaxies.  In particular, \citet{reddy06a, reddy10a} found
that such galaxies have redder UV slopes for a given dust attenuation
relative to older galaxies with $t_{\rm age}>100$\,Myr.  This effect
may stem from a geometrically different distribution of dust with
respect to the stars in a galaxy where, for young galaxies, this dust
may preferentially lie foreground to the stars.  Regardless, the
attenuation curve obtained for the young galaxies (on average) appears
to mimic that of the Small Magellanic Cloud (SMC) attenuation curve.
A steeper attenuation curve (e.g., SMC curve) will result in older
ages than those obtained in the Calzetti case; because not as much
dust is needed to redden the continuum for a foreground screen of dust
relative to a uniform distribution of dust and stars, the redness of
the UV slope will be attributed more to an older stellar population
than dust attenuation.  If we remodel those galaxies with
Calzetti-inferred ages $50<t_{\rm age}<100$\,Myr with the SMC
attenuation curve, then the fraction of galaxies that are still
considered to be ``young'' ($t_{\rm age}<100$\,Myr) is reduced by at
least a factor of nine.  In the CSF case, the number of galaxies with
$t_{\rm age}<100$\,Myr drops from 36 to just 4.  In the exponentially
rising case, this number drops from 36 to just 2.  The small fraction
of galaxies that are still considered young under the SMC assumption
have upper limits in SFR[IR+UV] that are consistent with SFR[SED] when
we assume an SMC attenuation curve (Figure~\ref{fig:sfrsmc}).  Hence,
if we assume an SMC attenuation curve, there is no longer a conflict
between the upper limit to SFR[IR+UV] (from the MIPS-nondetection) and
SFR[SED] for the young galaxies.\footnote{As we discuss in
  Section~\ref{sec:ages}, there are certain star formation histories
  (namely rising ones) where the age of galaxy becomes an ill-defined
  quantity, thus obfuscating the distinction between ``young'' and
  ``old'' galaxies as discussed here.  In these cases, it is useful to
  distinguish galaxies based on their stellar masses (or
  metallicities, if such measurements are available).  From the
  SED-fitting assuming a CSF history, the vast majority of those
  galaxies with ages $\la 100$\,Myr also have $M_{\ast}\la
  10^{10}$\,M$_{\odot}$.  As we show in Section~\ref{sec:ages}, the
  assumption of a rising star formation history does not significantly
  alter the stellar masses relative to those obtained under the CSF
  case and, as such, one can just as easily adopt the stellar mass
  threshold of $M_{\ast}\approx 10^{10}$\,M$_{\odot}$ to reach the
  same conclusions regarding the validity of the various dust
  attenuation curves for $z\sim 2$ galaxies.}

At first glance, the adoption of a lower limit in age that is
equivalent to the dynamical timescale, combined with a different
attenuation curve for the ``young'' galaxies, may appear to be a
contrived and inelegant solution to resolving the discrepancy between
SFR[IR+UV] and SFR[SED] for such galaxies.  However, our primary goal
is to derive SED parameters that are based on physically motivated
ages (i.e., dynamical time constraints) and are consistent with
observations of the dustiness and UV slopes for such galaxies
\citep{reddy06a, reddy10a}.  A consistent treatment of the
multi-wavelength data allows us to resolve the discrepancy between the
different measures of SFRs for young galaxies.

\begin{figure*}[tbp]
\plottwo{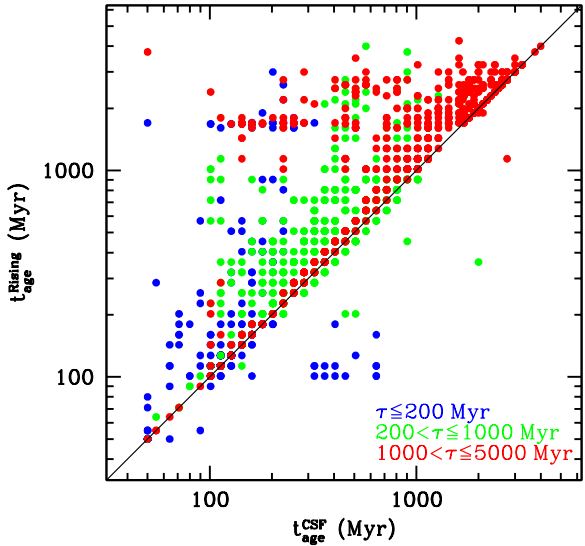}{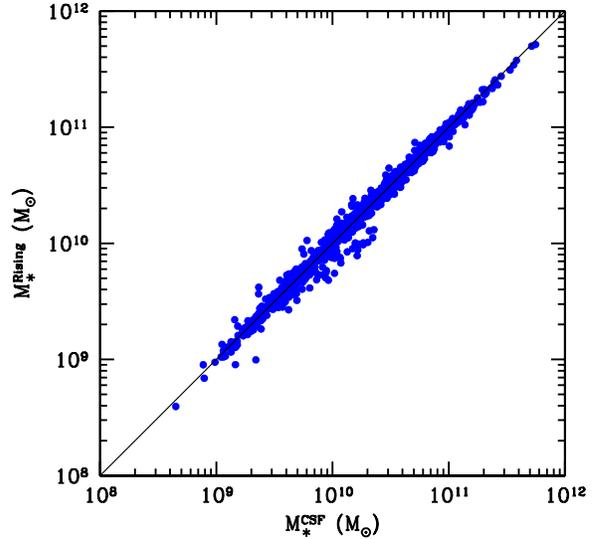}
\caption{Comparison of the ages (left) and stellar masses (right)
  between the constant (Model G) and rising star formation models
  (Model I), respectively.}
\label{fig:compareagemass}
\end{figure*}

\subsection{Summary of SFR Comparisons}

In the previous sections, we presented the comparison of the SFRs
determined from independent indicators of dust (MIPS $24$\,$\mu$m) to
those computed from fitting the broadband SEDs of galaxies at
redshifts $1.5\la z\la 2.6$.  Three sets of galaxies are identified in
this comparison: (1) ULIRGs with $\lbol \approx \lir >
10^{12}$\,L$_{\odot}$; (2) typical star-forming galaxies with $\lbol
\le 10^{12}$\,L$_{\odot}$ and ages $t_{\rm age}\ga 100$\,Myr; and (3)
``young'' galaxies with $\lbol \le 10^{12}$\,L$_{\odot}$ and ages
$t_{\rm age}\la 100$\,Myr.  For typical galaxies, an exponentially
declining star formation history yields SFRs that are inconsistent
with those obtained from combining the MIPS and UV data.  Assuming
constant or rising star formation histories yields SFRs that are in
reasonable agreement with the MIPS+UV determinations.  Alternatively,
none of the simple star formation histories considered here (constant,
declining, or rising) are able to reproduce the lower SFRs found for
``young'' galaxies.  We explore several possibilities for resolving
this discrepancy, including dynamical time constraints and systematics
in the UV attenuation curve.  A physically plausible solution --- and
one which is consistent with independent measurements of the age
dependence of the dust attenuation curve at $z\sim 2-3$
\citep{reddy06a, reddy10a} --- is to adopt a lower limit in age
corresponding to the dynamical timescale typical of galaxies in our
sample ($\simeq 50$\,Myr), and to assume a steeper (e.g., ``SMC'')
attenuation curve for the young galaxies (corresponding to Models G
and I in Table~\ref{tab:sfrrelations}).  Based on these findings, we
proceed in the next section with a discussion of the stellar masses
and ages derived from constant and rising star formation histories.

\section{INTERDEPENDENCIES OF THE SED PARAMETERS}
\label{sec:agemass}

Having just discussed the most plausible star formation histories that
are constrainable with the available data, in this section we focus on
the systematic variations in age and stellar mass with the overall
form of the star formation history (constant versus rising), the
typical uncertainties in ages and stellar masses, and the correlation
between SFR and stellar mass.

\subsection{Systematic Variations in Age and Stellar Mass}
\label{sec:compareagemass}

\begin{figure}[tbp]
\plotone{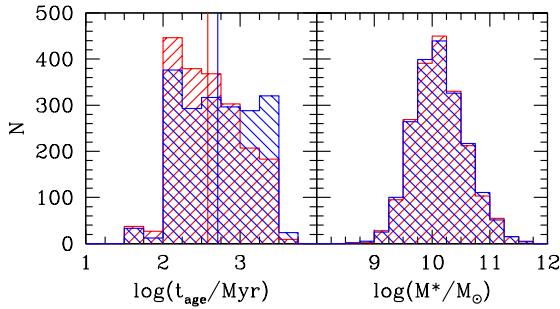}
\caption{Histograms of age and stellar mass for the constant (red) and
  rising (blue) star formation histories.  The vertical lines in the
  left panel indicate the mean age found when assuming the constant
  and rising star formation histories.}
\label{fig:agemasshist}
\end{figure}

The ages and stellar masses derived for 1959 objects with
spectroscopic redshifts $1.4\le z\le 3.7$, assuming constant and
exponentially rising star formation histories, are shown in
Figure~\ref{fig:compareagemass}.  Histograms of the age and stellar
mass distributions are presented in Figure~\ref{fig:agemasshist}.  The
stellar mass for any given galaxy derived with a rising star formation
history is essentially identical within the uncertainties (that stem
primarily from photometric errors and the degeneracy between the
parameters being fit) to those obtained with a CSF history.  The
primary difference, therefore, is in the distribution of best-fit
ages; ages derived with a rising star formation history are older by
$\approx .12$\,dex.  The fraction of galaxies in our spectroscopic
sample with derived ages older than 1\,Gyr increases from $20\%$ to
$32\%$ when we assume rising star formation histories versus CSF.

\begin{figure*}[tbp]
\plottwo{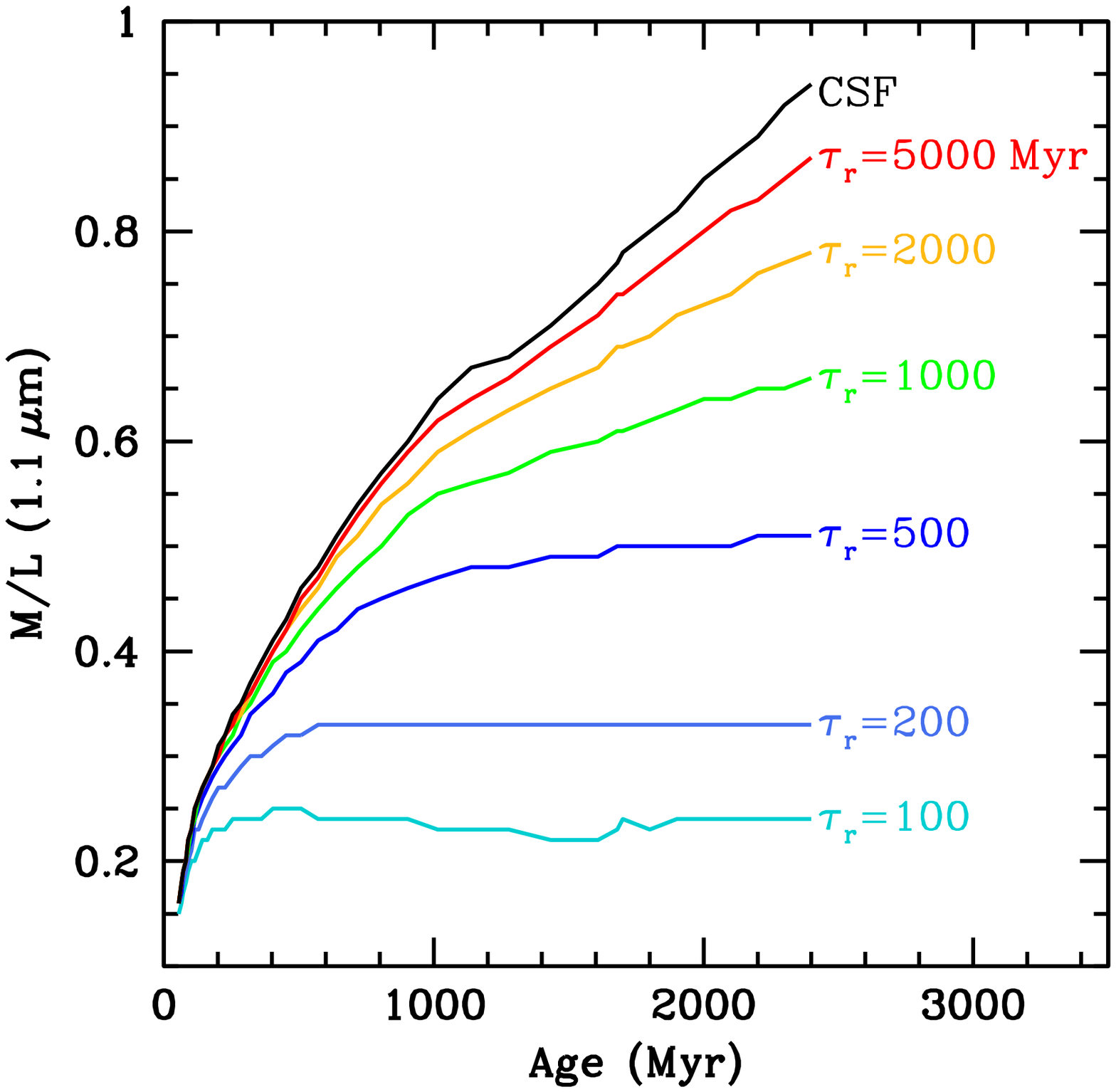}{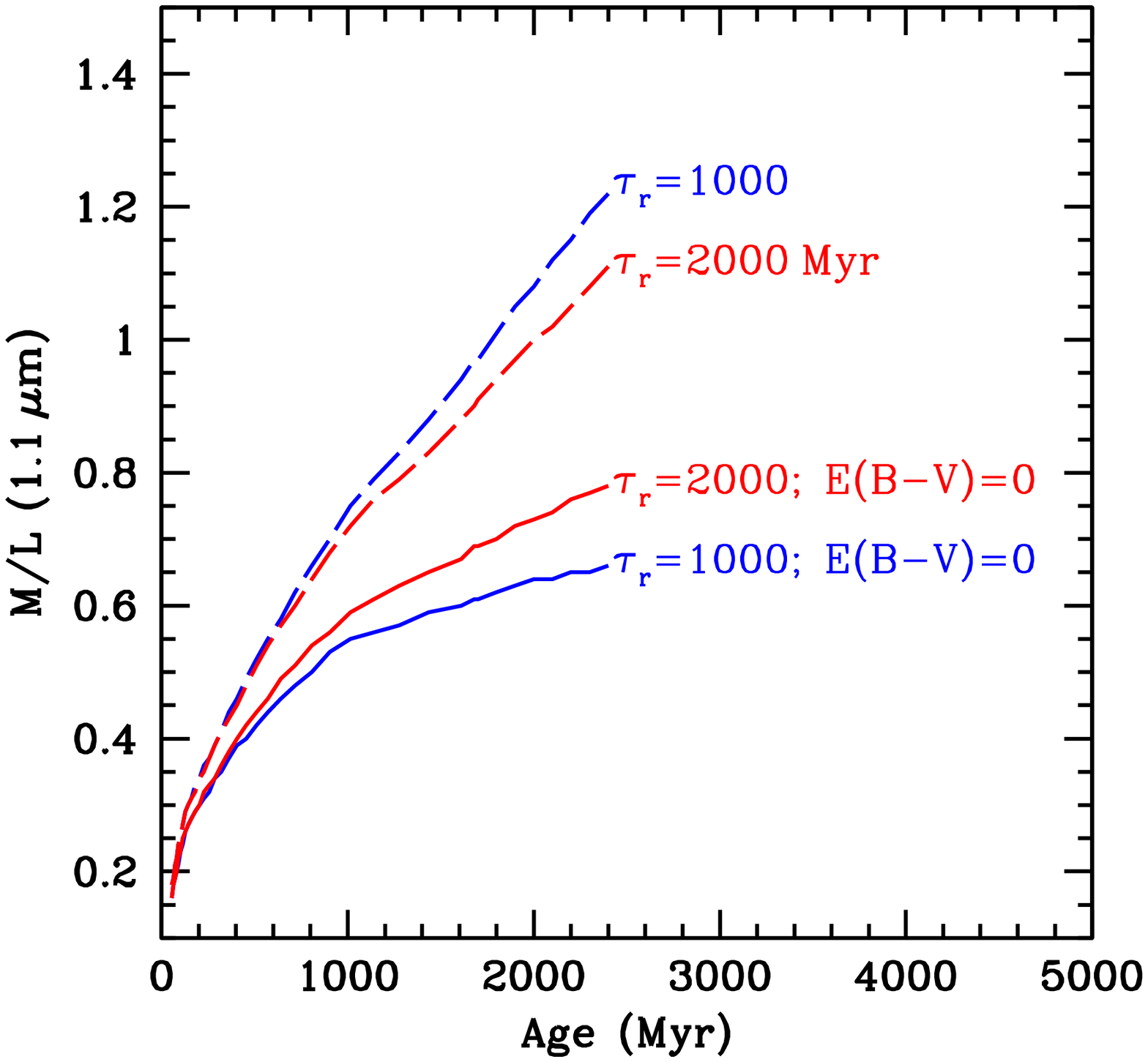}
\caption{{\em Left:} Stellar mass to light ratio at $1.1$\,$\mu$m (in
  solar units) as a function of age for CSF and exponentially rising
  star formation histories, with $\tau_{r}=100$ to $5000$\,Myr.  The
  curves assume $\ebmv=0$.  {\em Right:} Evolution in $M/L$ ratio
  including the effects of dust attenuation (dashed lines), where we
  have assumed the relationship between SFR and $\ebmv$ as determined
  from the SED modeling, compared to the cases with no dust
  attenuation, $\ebmv=0$ (solid lines).  For this example, we consider
  $\tau_{\rm r}=1000$ and $2000$\,Myr, and an initial SFR of
  $20$\,M$_{\odot}$\,yr$^{-1}$.  The difference between the curves for
  $\tau_{\rm r}=1000$\,Myr is larger than the difference between the
  $\tau_{\rm r}=2000$\,Myr curves because the SFR, and hence dust
  reddening, rises more rapidly for smaller values of $\tau_{\rm r}$.}
\label{fig:mlage}
\end{figure*}

We can better understand these trends by examining the evolution in
the stellar mass-to-light ratio ($M/L$) as a function of galaxy age
for the different star formation histories considered here.  At early
times, for ages $t_{\rm age}\la \tau_{\rm r}$, the light is dominated
by current star formation and the $M/L$ ratio is low.  As the stellar
mass builds up, older stars will begin to contribute significantly to
the light, causing the $M/L$ ratio to increase.  Neglecting dust
reddening, for $t_{\rm age}\gg\tau_{\rm r}$, the $M/L$ ratio
stabilizes to some constant value.  In reality, the $M/L$ ratio will
still increase (even though the ratio of stellar mass to total SFR may
remain roughly constant), because dust reddening will increase with
SFR, and hence the bolometric light will suffer an increasing amount
of obscuration (Figure~\ref{fig:mlage}).

Regardless, as shown in Figure~\ref{fig:mlage}, the $M/L$ ratio (at
rest-frame $1.1$\,$\mu$m, corresponding roughly to an observed
wavelength at $z\sim 2.3$ that coincides with IRAC channel 1)
increases more slowly with age for galaxies with smaller values of
$\tau_{\rm r}$.  Hence, for the same $M/L$ ratio, rising star
formation histories correspond to an older age than a CSF history.
Note that these trends hold on average, and there is still a fair
fraction of galaxies where the ages (and stellar masses) do not change
substantially between the CSF and rising star formation histories.  In
particular, approximately $55\%$ of galaxies in our sample have a
best-fit $\tau_{\rm r}=5000$\,Myr, and it is for these longer
exponential timescales that the $M/L$ ratio is not substantially
different than that obtained in the CSF case when $t_{\rm
  age}<\tau_{\rm r}$.  In any case, the stellar mass distributions do
not change substantially when we adopt a rising versus CSF history,
and in both cases, the mean stellar mass of galaxies in our
spectroscopically confirmed ($\rs<25.5$) sample at $z\sim 2.3$ is
$\langle M_{\ast}\rangle=1.6\times 10^{10}$\,M$_{\odot}$.  As we
discuss in Section~\ref{sec:uvmass}, the mean stellar mass is a strong
function of absolute magnitude, and including galaxies fainter than
our spectroscopic limit will result in a lower mean stellar mass.

\subsection{The ``Ages'' of Galaxies with Rising Star Formation Histories}
\label{sec:ages}

\begin{figure}[tbp]
\plotone{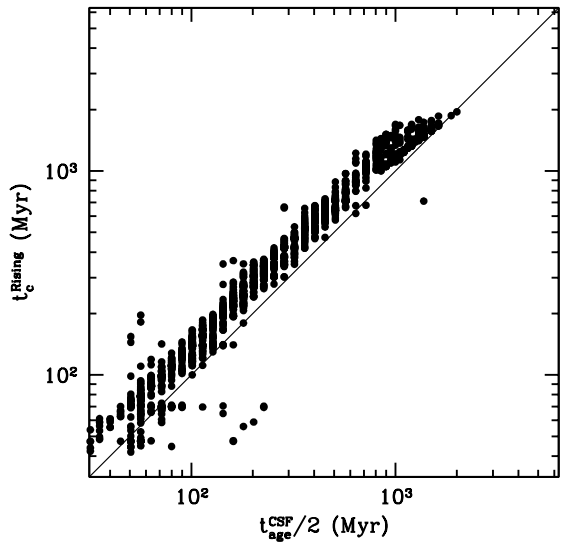}
\caption{Comparison between the characteristic timescales required to
  double the stellar mass to the present value with a rising ($t_{\rm
    c}^{Rising}$) and constant ($t_{\rm age}^{CSF}/2$) star formation
  histories for the full SED sample of 1959 galaxies.}
\label{fig:compareagetc}
\end{figure}

As just discussed, rising star formation histories require older ages
than CSF histories to achieve a given mass-to-light ratio.  In
general, the ages derived with either a constant or rising star
formation history will be a lower limit to the true age, because of
the possibility that there may be an underlying older stellar
population whose near-IR light is overwhelmed by the near-IR light
from current star formation (i.e., the ``outshining'' problem as
discussed above).  A further systematic effect pertains to ``ages''
derived under the assumption of different star formation histories.
In the simplest case for constant star formation, the age is simply
determined by the time required to build up the observed stellar mass
given the current rate of star formation.  Integrating
Equation~\ref{eq:sfrequation} yields the stellar mass after time $t$
for an exponentially rising star formation history:
\begin{equation}
M_{\ast}(t)=\Psi_{o}\tau_{\rm r}[{\rm exp}(t/\tau_{\rm r})-1].
\end{equation}
For simplicity, we have ignored the gas recycling fraction (i.e., the
fraction of gas released back into the interstellar medium (ISM) from
supernovae and stellar winds), as adopting this correction will simply
adjust downwards the stellar mass by some multiplicative factor close
to $\approx 0.7$ \citep{bruzual03}.  With this parameterization, the
stellar mass $M_{\ast}=0$\,M$_\odot$ at time $t=0$.  In this case, the
``age'' is well defined in a mathematical sense, and these are the
ages that we have referred to in the previous sections.  Note that the
specific SFR
\begin{equation}
\phi(t)\equiv \frac{\Psi(t)}{M_{\ast}(t)} = \frac{{\rm exp}(t/\tau_{\rm r})}{\tau_{\rm r}[{\rm exp}(t/\tau_{\rm r})-1]},
\label{eq:ssfr}
\end{equation}
is roughly constant as a function of time if $\exp(t/\tau_{\rm r})\gg
1$.  As long as this condition is satisfied, the age $t$ can be
arbitrarily old.  In that case, we would simply adjust the
normalization $\Psi_{\rm o}$ for a given $t$ in order to recover the
observed stellar mass.  As a simple example, if we assume that
$\tau_{\rm r}\approx 400$\,Myr and we consider ages of $t=1$ and
$2$\,Gyr, then we must set $\Psi_{\rm o}\approx 3.6$ and
$0.3$\,M$_{\odot}$\,yr$^{-1}$, respectively, in order for the current
stellar mass to be $M_{\ast}\approx 1.6\times 10^{10}$\,M$_{\odot}$.
Given the ambiguity of the derived ``age'' for a rising star formation
history, it is useful to define a characteristic timescale $t_{\rm c}$
over which the galaxy has doubled its stellar mass:
\begin{equation}
t_{\rm c}\equiv t_{\rm age}-t_{\rm 1/2} \approx \tau_{\rm r}\ln(2) =
\frac{1}{\phi}\ln(2),
\label{eq:massdouble}
\end{equation}
where $t_{\rm 1/2}$ indicates the time at which the galaxy had half of
its current stellar mass.  We show the comparison between $t_{\rm c}$
and $t_{\rm age}/2$ for the rising and constant star formation models,
respectively, in Figure~\ref{fig:compareagetc}.  The comparison shows
that there is a better correspondence between the times required to
double the stellar mass to the currently observed value with a rising
and CSF models.  Of course, for a given currently observed SFR, the
rising history (as compared to a CSF model) will implies a longer
amount of time to build up the current stellar mass given that the SFR
was lower in the past.  In Section~\ref{sec:discussion}, we discuss
the mass doubling time, derived ages, and implied formation redshifts,
in the context of direct observations of $z\ga 3$ galaxies.

\subsection{Uncertainties in Age and Stellar Mass with Simple 
Star Formation Histories}

As just discussed, the assumption of rising versus CSF histories
results in systematic differences in the stellar population ages.
Other systematics affecting ages and stellar masses include
uncertainties in the IMF, differences between stellar population
synthesis models (e.g., \citealt{bruzual03, maraston06}; CB11), and
the degree to which simple star formation histories can capture the
complexity of the ``real'' star formation history of a galaxy (e.g.,
\citealt{papovich01, shapley01, shapley05, marchesini09, muzzin09,
  maraston10, papovich11}).  While the uncertainties in the complexity
of the star formation history of any individual galaxy may be large,
if such variations are stochastic (e.g., such as having multiple
independent bursts of star formation), then large galaxy samples can
be used to average over these stochastic effects and yield insight
into the characteristic, or typical, star formation history.  In
addition to these systematic effects, there are also uncertainties
related to our ability to accurately constrain the stellar population
model given the flux errors and limited wavelength coverage of
photometry.  These effects result in typical fractional uncertainties
of $\sigma_{\rm age}/\langle t_{\rm age}\rangle = 0.5$ and
$\sigma_{\rm M^\ast}/\langle M^\ast\rangle = 0.4$, respectively, as
determined from Monte Carlo simulations \citep{shapley05, erb06b}.

\subsection{Relationship between Star Formation Rate and Stellar Mass}
\label{sec:sfrm}

\begin{figure*}[tbp]
\plottwo{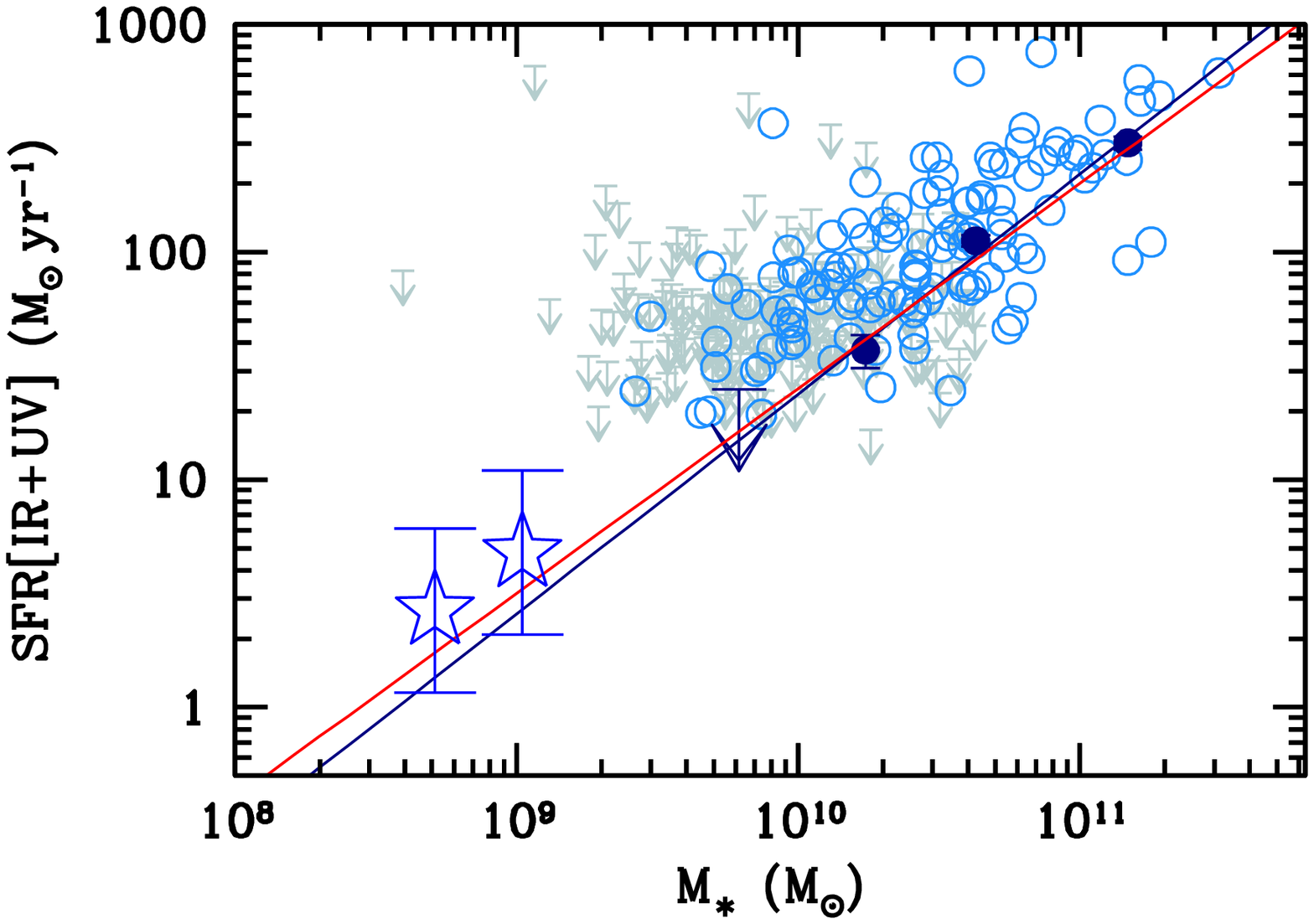}{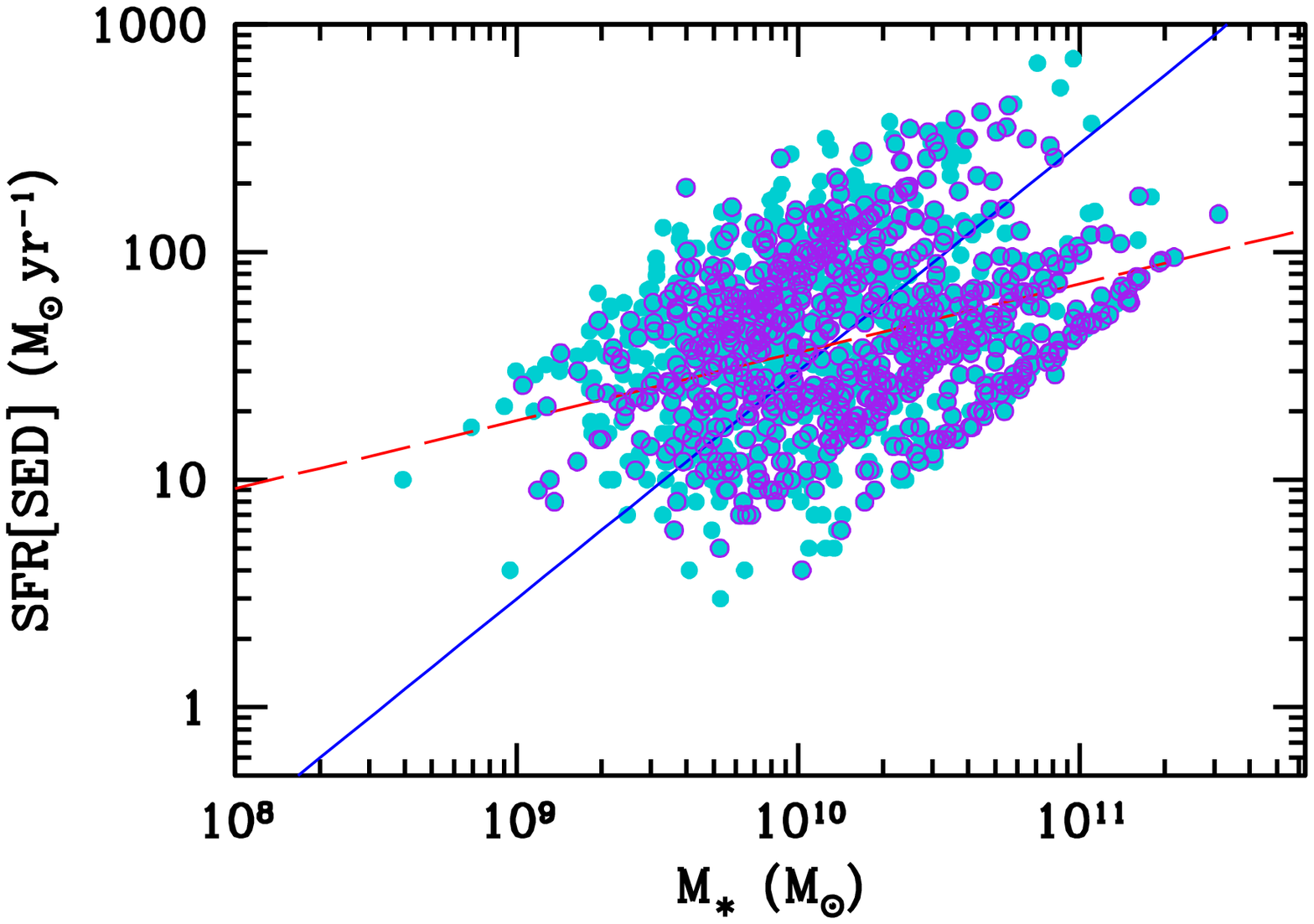}
\caption{{\em Left:} Relationship between SFR, determined from
  combining MIPS and UV data, and stellar mass as determined from SED
  fitting for 302 spectroscopically confirmed galaxies at $1.5\le z\le
  2.6$.  We have assumed rising star formation histories to derive the
  stellar masses (e.g., using Model I [Table~\ref{tab:sfrrelations}]).
  Upper limits and detections are indicated by the arrows and hollow
  circles, respectively, and the darker filled symbols and upper limit
  denote the stacked SFR in bins of stellar mass.  The best-fit linear
  correlation between SFR and M$_{\ast}$ for the spectroscopic sample,
  taking into account upper limits, is shown by the solid blue line.
  The large open stars are for photometrically selected faint $z\sim
  2$ galaxies.  The red line indicates the relationship found by
  \citet{daddi07a} at similar redshifts. {\em Right:} Same as left
  panel, where SFRs are determined from SED fitting of a larger sample
  of 1264 galaxies with spectroscopic redshifts in the same range as
  the MIPS sample ({\em cyan points}).  The purple points show the
  distribution for the subset of 705 galaxies that have at least two
  detections longward of the Balmer/4000\,\AA\, breaks.  The solid
  line shows the intrinsic correlation between SFR and $M_{\ast}$, and
  the dashed line indicates a linear least squares fit to the data.}
\label{fig:sfrm}
\end{figure*}

The relationship between SFR[IR+UV] and stellar mass for the 302
galaxies at $1.5\le z_{\rm spec}\le 2.6$ with MIPS $24$\,$\mu$m
observations (121 of which are detected at $24$\,$\mu$m) is shown in
the left panel of Figure~\ref{fig:sfrm}.  Rising star formation
histories are assumed; stellar masses derived from a CSF history will
be similar.  As found by several previous investigations at similar
redshifts (e.g., \citealt{reddy06a, papovich06, daddi07a}), we find a
relatively tight positive correlation between SFR and stellar mass
(though we discuss below potential biases), which has typically been
interpreted as an indication that most of the stellar mass at these
redshifts accumulated in a relatively ``smooth'' manner and not in
major mergers.  Taking into account the upper limits with a survival
analysis, we find an rms scatter about the best-fit linear relation
between the log of SFR and log of stellar mass of $\approx 0.37$\,dex,
and a slope of $0.97\pm0.05$.

Because the relationship between SFR and stellar mass has been studied
extensively over a large range of redshifts (e.g.,
\citealt{brinchmann04, reddy06a, noeske07, daddi07a, stark09,
  gonzalez10, sawicki11}), it is useful to examine how sample
selection may bias the measurement of this relationship.  The right
panel of Figure~\ref{fig:sfrm} shows the correlation between SFR and
$M_{\ast}$, where SFR is determined solely from the SED fitting (using
Model I), for a larger sample of 1244 galaxies with spectroscopic
redshifts in the same range as the MIPS sample.  For comparison, the
distribution of SFR versus stellar mass for the subset of 705 galaxies
with at least two detections longward of the Balmer and $4000$\,\AA\,
breaks is also shown.  We note that the distribution for this subset
is similar to that of the larger sample, implying that the inclusion
of galaxies with just a single point longward of the Balmer and
$4000$\,\AA\, breaks does not {\em systematically} bias the
distribution of SFR versus $M_{\ast}$.  In the SED fitting procedure,
both the SFR and stellar mass are determined by the normalization of
the model SED to the observed photometry
(Section~\ref{sec:sedmodeling}).  As such, SFR[SED] and $M_{\ast}$ are
highly correlated and exhibit an intrinsic slope of unity.  However, a
linear least squares fit to the data imply a much shallower slope of
$0.30$.  This bias is a result of the fact that our sample is selected
based on UV luminosity (e.g., and not stellar mass).  Hence, there
will be a Malmquist bias of selecting galaxies with larger SFRs at a
given stellar mass for the lowest stellar mass galaxies in our sample.
This bias is quantified via simulations that are presented in
Appendix~\ref{sec:malmquist}.  The effect of this bias on the best-fit
relation between stellar mass and near-IR magnitude is discussed in
Appendix~\ref{sec:massnir}.  Finally, Appendix~\ref{sec:sfrmbias}
examines the effect of these biases on the determination of mean
stellar mass from a UV selected sample.

A comparison of the left and right panels of Figure~\ref{fig:sfrm}
shows that the Malmquist bias is much less noticeable when the SFR is
determined independently of the stellar mass (which is {\em not} the
case when fitting the SEDs with stellar population models), primarily
because the SFRs[IR+UV] have additional scatter associated with them
(related to the dispersion in dust extinction at a given UV
luminosity) that is decoupled from the method we used to infer
$M_{\ast}$ from the SED fitting.  As we discuss below, the near-unity
slope of the SFR-$M_{\ast}$ relation, where SFRs are determined from
the MIPS+UV data, appears to hold to lower stellar masses, and it
implies that the specific SFR depends only weakly on stellar mass.  We
will return to the implication of this result in
Section~\ref{sec:discussion}.

Before proceeding, we comment briefly on a potential bias that may
exist at the bright, high-mass end of the SFR-$M_{\ast}$ relation.  As
noted in Section~\ref{sec:sfrcomparison}, the \citet{reddy10a}
conversion between rest-frame $8$\,$\mu$m and total IR luminosity
reproduces the average $\lir$ computed using stacked {\em
  Herschel}/PACS 100 and $160$\,$\mu$m data \citep{reddy12}.  This
conversion is also shown to over-predict $\lir$ for the most
bolometrically luminous galaxies in our sample.  Correcting for this
effect will lower SFR[IR+UV] by up to a factor of two, resulting in a
shallower slope of SFR-M$_{\ast}$ {\em at these high stellar masses}
where $M_{\ast}\ga 10^{11}$\,M$_{\odot}$.  On the other hand, at a
given high stellar mass, our UV selected sample will be biased against
the dustiest and hence most heavily star-forming galaxies.  This is
due the fact that such dusty galaxies either (a) have UV colors that
are too red to satisfy the color criteria or (b) are too faint to be
represented in the spectroscopic sample (e.g., \citealt{reddy08,
  reddy10a}).  Inclusion of these missing dusty galaxies would have
the opposite effect of {\em steepening} the slope of the SFR-M$_\ast$
relation at high stellar masses.  Regardless of these two competing
biases at the high stellar mass end, correcting for them is unlikely
to alter significantly the overall slope of the SFR-M$_\ast$ relation,
given that such high SFR galaxies constitute only a small fraction of
the number density of $L^{\ast}$ and fainter galaxies (e.g.,
\citealt{reddy08, reddy09, magnelli11}).

\subsection{Summary of Age and Stellar Mass Distributions, and
SFR-$M_{\ast}$ Relation}

Modeling the broadband photometry of spectroscopically confirmed
galaxies in the $z\sim 2-3$ sample allows us to examine the
distribution of ages and stellar masses.  We have shown that rising
star formation histories generally yield older ages for a given mass
to light ratio relative to the ages obtained with a CSF history.  More
generally, the ``age'' of galaxy is not well-defined for a rising star
formation history, in the sense that the age can be varied
simultaneously with the normalization of the star formation history to
yield the same value of current SFR and stellar mass, as long as
$\exp(t\tau_{\rm })\gg 1$.  Using SFRs that are derived independently
of the stellar masses, we find an intrinsic SFR versus stellar mass
correlation with roughly unity slope at $z\sim 2$.  So far, we have
concerned ourselves with the SED sample which, by construction, only
includes those galaxies that had at least one detection longward of
the $4000$\,\AA\, break.  However, there exist substantial numbers of
UV-selected galaxies that are not represented in the SED sample
because they are undetected longward of the break.  More generally, we
expect the SED sample to be biased to galaxies with larger stellar
masses at a given SFR relative to the UV sample as a whole.  In the
next section, we explore this bias by stacking the longer wavelength
data for galaxies in our sample, and we also extend the results on the
SFR-$M_{\ast}$ relation to UV faint galaxies at the same redshifts.

\begin{deluxetable*}{lccccc}
\tabletypesize{\footnotesize}
\tablewidth{0pc}
\tablecaption{Mass to Light Ratios}
\tablehead{
\colhead{Rest-frame $\lambda$} &
\colhead{} &
\colhead{} &
\colhead{} &
\colhead{} &
\colhead{} \\
\colhead{($\mu$m)} &
\colhead{$N$\tablenotemark{a}} &
\colhead{Spearman's $\sigma$\tablenotemark{b}} &
\colhead{RMS\tablenotemark{c}} &
\colhead{($\frac{M}{L}$)$_{\rm min}$\tablenotemark{d}} &
\colhead{($\frac{M}{L}$)$_{\rm max}$\tablenotemark{e}}}
\startdata
$0.50$ & 98 & 5.5 & 0.33 & 0.07 & 4.4 \\
$0.67$ & 491 & 14.5 & 0.32 & 0.06 & 4.4 \\
$1.1$ & 643 & 22.3 & 0.20 & 0.14 & 4.4 \\
$1.4$ & 673 & 23.0 & 0.18 & 0.18 & 4.4 \\
$1.8$ & 180 & 11.8 & 0.19 & 0.24 & 2.4 \\
$2.4$ & 190 & 10.9 & 0.23 & 0.21 & 4.4 
\enddata
\tablenotetext{a}{Number of spectroscopically confirmed galaxies.}
\tablenotetext{b}{Significance of correlation between log stellar mass and absolute magnitude
computed from Spearman's rank correlation test, in units of standard deviation.}
\tablenotetext{c}{RMS of data about best-fit linear relation between log stellar mass
and absolute magnitude.}
\tablenotetext{d}{Minimum observed mass to light ratio in our sample, in units of the Sun.}
\tablenotetext{e}{Maximum observed mass to light ratio in our sample, in units of the Sun.}
\label{tab:mlratios}
\end{deluxetable*}

\section{MASS-TO-LIGHT RATIOS OF $z\sim 2-3$ GALAXIES}
\label{sec:mlratio}

\subsection{Rest-frame Near-IR $M/L$ Ratios}

\begin{figure}[tbp]
\plotone{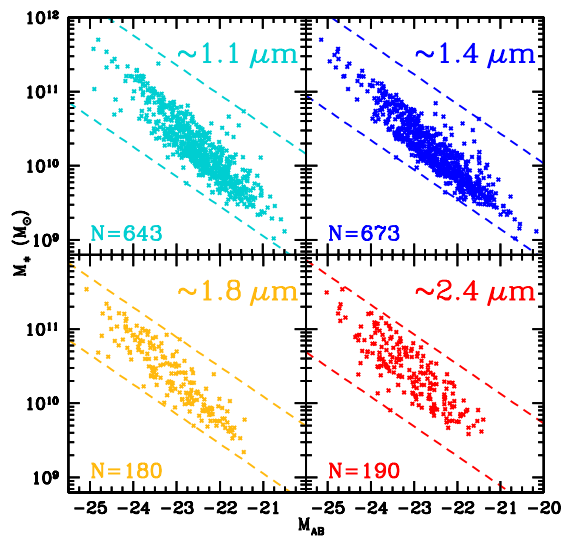}
\caption{Relationship between stellar mass and absolute magnitude
  measured in the four IRAC channels.  Approximate rest-frame
  wavelengths are indicated in each panel, along with the number of
  objects used to define the relationship.  Short dashed lines
  indicate the minimum and maximum $M/L$ ratio.}
\label{fig:masstraceirac}
\end{figure}

Because the data required to model the stellar populations do not
exist for every galaxy in our sample (e.g., those galaxies that did
not have near-IR or IRAC coverage, as well as those undetected at
these longer wavelengths), it is useful to establish some empirical
relationship between the luminosity of a galaxy and its stellar mass.
Given the stellar mass to light ratio ($M/L$), we can then estimate
the stellar mass of galaxy based simply on the flux at a given
wavelength.  The relation between stellar mass and rest-frame near-IR
light as probed by the IRAC observations is shown in
Figure~\ref{fig:masstraceirac}.  We have also computed these relations
at rest-frame optical wavelengths using the F160W and $\ks$-band data
As summarized in Table~\ref{tab:mlratios}, the dispersions between the
rest-frame near-IR luminosities and stellar masses are lower than
those found between rest-frame optical luminosity and stellar mass,
but we note the large variation in mass-to-light ratio even at
rest-frame wavelengths where the stellar emission peaks, around
$1.6$\,$\mu$m.  The current star formation in a galaxy may outshine
the light from the older stars at rest-frame optical and near-IR
wavelengths.  In these cases, the ages (and masses) typically reflect
those of the {\em current} star formation episode, though two
component models can be used to determine an upper limit to the hidden
stellar mass in such galaxies (e.g., \citealt{shapley05}).  For the
subsequent discussion, we focus on results using the IRAC
$3.6$\,$\mu$m data, noting that our results (e.g., derived stellar
masses) do not change substantially when we use data from the other
IRAC channels.

Note that there is a systematic trend towards lower $M/L$ at lower
stellar mass or fainter near-IR luminosity; i.e., the best-fit
relation between near-IR luminosity and stellar mass does not fall
onto a line of constant $M/L$.  At face value, this systematic trend
implies that galaxies with lower stellar masses have larger specific
SFRs ($\phi$), resulting in lower $M/L$ ratios.  However, as discussed
in Section~\ref{sec:sfrm} and Appendix~\ref{sec:malmquist}, there is
Malmquist bias of selecting galaxies with larger SFRs, and hence
larger $\phi$, for the lowest mass galaxies in our spectroscopic
sample.  This directly affects the best-fit relation between stellar
mass and near-IR luminosity, as discussed in
Appendix~\ref{sec:massnir}.  The basic conclusion from these different
best-fit relations is that there are sufficient biases induced by
Malmquist effects and not correcting for dust extinction that can
produce an erroneously strong trend between $M/L$ ratio and
luminosity.  For the subsequent discussion, we employ the relations
discussed in Appendix~\ref{sec:massnir} to convert the average near-IR
magnitude to stellar mass for galaxies of a given UV luminosity.

\subsection{Rest-frame UV Mass-to-light Ratio}

The scatter in the $M/L$ ratio increases towards shorter wavelengths,
as the current star formation dominates the emission.  This can
already be seen in the larger scatter in $M/L$ at optical wavelengths
relative to that found in the near-IR, and we would expect the maximum
dispersion to occur at UV wavelengths.  Here, we investigate the mean
and dispersion in $M/L$ ratio at UV wavelengths, with the aim of
quantifying the stellar masses and star formation histories of
UV faint galaxies.

\subsubsection{Trend Between Stellar Mass and UV Luminosity in the
Spectroscopic Sample}
\label{sec:uvmass}

\begin{figure}[tbp]
\plotone{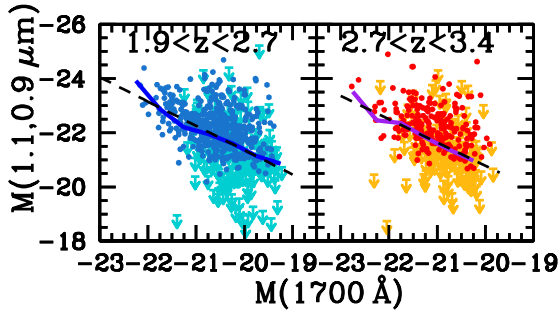}
\caption{Rest-frame $1.1$ and $0.9$\,$\mu$m absolute magnitudes (from
  IRAC channel 1 data) vs. UV absolute magnitude for 630
  spectroscopically confirmed galaxies at redshifts $1.4\le z<2.7$
  (left) and 344 spectroscopically confirmed galaxies at
  redshifts $2.7\le z<3.7$ (right).  Points and
  downward-pointing arrows denote galaxies detected and undetected (to
  $3\sigma$), respectively, at $3.6$\,$\mu$m, and the solid lines
  denote the stacked magnitudes as a function of UV luminosity
  (Table~\ref{tab:iracstack}).  The dashed lines show the best-fit
  linear relations obtained from the Buckley-James estimator, taking
  into account both IRAC detections and nondetections.}
\label{fig:uvch1}
\end{figure}

\begin{figure}[!t]
\plotone{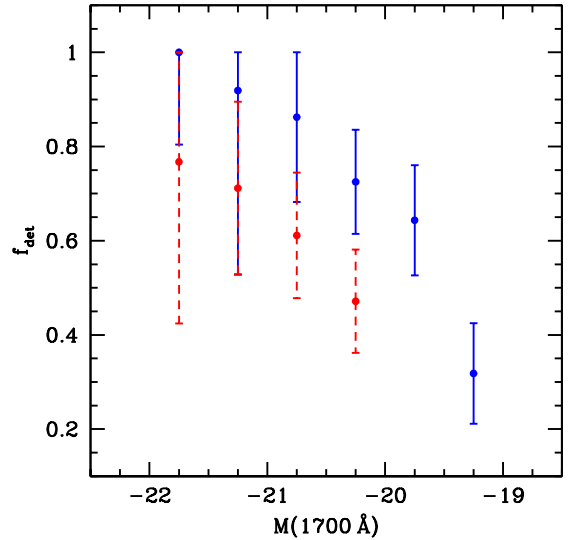}
\caption{Fraction of galaxies detected in IRAC channel 1 ($3.6$\,$\mu$m)
as a function of absolute magnitude at $1700$\,\AA, at redshifts
$1.4\le z<2.7$ (blue) and $2.7\le z<3.7$ (red).  Errorbars
reflect Poisson statistics.}
\label{fig:iracundet}
\end{figure}

Figure~\ref{fig:uvch1} shows the distribution of near-IR luminosity
(from the IRAC $3.6$\,$\mu$m data) with UV luminosity for 630 and 344
spectroscopically confirmed galaxies between redshifts $1.4\le z<2.7$
and $2.7\le z < 3.7$, respectively, with $3\sigma$ upper limits
indicated for those galaxies that are undetected at $3.6$\,$\mu$m.
The effect of the decreasing fraction of IRAC-detected galaxies with
fainter UV luminosities (Figure~\ref{fig:iracundet}) is evident when
examining the stacked IRAC fluxes, which are correspondingly fainter
for UV faint galaxies.  The procedure used to compute the stacked IRAC
fluxes is described in Appendix~\ref{sec:iracstack}.  The trend
between stacked IRAC flux and UV magnitude is essentially identical to
the trend inferred using the Buckley-James estimator on the individual
detections and non-detections.  The best-fit trends between near-IR 
and UV magnitude, taking into account IRAC non-detections are:
\begin{equation}
M_{1.1}(z\sim 2.3) = (0.90\pm0.08)M_{\rm UV}-(3.41\pm1.57),
\end{equation}
with an rms scatter of 1.01\,dex, and
\begin{equation}
M_{0.9}(z\sim 3.0) = (0.86\pm0.12)M_{\rm UV}-(3.62\pm2.45),
\end{equation}
with an rms scatter of 1.08\,dex.

As discussed in Appendix~\ref{sec:massnir}, the conversion between
near-IR magnitude and stellar mass depends on UV luminosity, simply
because current star formation can contribute to the near-IR
magnitude.  To properly account for this effect, we converted the
near-IR magnitude found for each bin of UV luminosity to a stellar
mass using the relation between near-IR magnitude and stellar mass
appropriate for that bin of UV luminosity (see
Appendix~\ref{sec:massnir}, Figure~\ref{fig:nirbestfit}, and
Table~\ref{tab:nirbestfit}).  In this way, we are able to account for
the star formation history of the average galaxy at a given UV
luminosity in estimating its stellar mass.

\begin{figure}[tbp]
\plotone{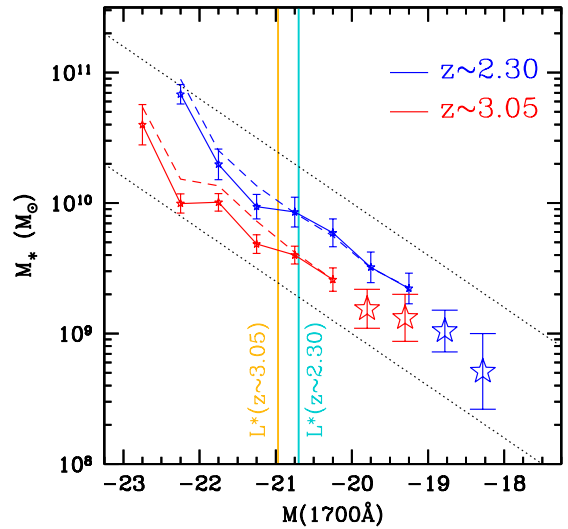}
\caption{Median stellar mass, inferred from stacking the IRAC data, in
  bins of UV absolute magnitude (points and solid lines).  The stacked
  IRAC magnitudes were converted to stellar mass assuming the best-fit
  linear relations between these two quantities in bins of UV
  magnitude (Appendix~\ref{sec:massnir}, Figure~\ref{fig:nirbestfit},
  Table~\ref{tab:nirbestfit}).  Results are also shown (dashed lines)
  where we have assumed a single best-fit linear relation for all
  galaxies in our sample.  The vertical lines denote the values of
  $L^{\ast}$ of the UV luminosity functions at $z\sim 2.30$ and $z\sim
  3.05$ \citep{reddy09}.  The open stars denote the median stellar
  mass in the bins of UV faint photometrically selected galaxies.  The
  conversion between near-IR magnitude and stellar mass for these
  faint bins is discussed in Appendix~\ref{sec:massnir}.  The dotted
  black lines indicate lines of constant $M/L$ ratio.}
\label{fig:nirmassuv}
\end{figure}

Figure~\ref{fig:nirmassuv} shows the resulting median stellar masses
of galaxies in different UV magnitude bins.  For comparison, the
dashed lines show the results at $z\sim 2.30$ and $z\sim 3.05$ if we
assume a single best-fit linear relation between $M_{\ast}$ and
$M_{1.1,0.9}$ for all galaxies (irrespective of UV luminosity).
Interestingly, it is primarily for galaxies brighter than
$L^{\ast}_{\rm UV}$ that the SFR is significant enough to bias the
$M_{\ast}$ versus near-IR relation towards higher masses.  Below
$L^{\ast}$, the discrepancy in stellar masses derived using relation
for all galaxies versus that derived using the relation only for UV
faint galaxies, is small.  This can be attributed to the fact that the
former relation is dominated by UV faint galaxies at fainter near-IR
magnitudes (Appendix~\ref{sec:massnir} and Figure~\ref{fig:ssfrcode}).
The relationship between UV luminosity and mass is similar to that
derived in \citet{sawicki11} for a sample of BX selected galaxies in
the {\em Hubble} Deep Field (HDF) North, once we have taken into
account differences in SED fitting by remodeling our galaxies using
the same templates \citep{bruzual03} used in that study.

\begin{deluxetable*}{lccc}
\tabletypesize{\footnotesize}
\tablewidth{0pc}
\tablecaption{Stacked IRAC Magnitudes for UV faint Sample}
\tablehead{
\colhead{Redshift Interval} &
\colhead{M(1700\AA) Range\tablenotemark{a}} &
\colhead{$3.6$\,$\mu$m\tablenotemark{b}} &
\colhead{$4.5$\,$\mu$m\tablenotemark{b}}}
\startdata
$1.4\le z<2.7$ & -19.03\,\,\,\,\,-18.53 & $-19.97\pm0.10$ (393) & $-20.09\pm0.09$ (260) \\
               & -18.53\,\,\,\,\,-18.03 & $-19.28\pm0.26$ (170) & $-19.24\pm0.34$ (112) \\
\\
$2.7\le z<3.7$ & -20.05\,\,\,\,\,-19.55 & $-20.54\pm0.08$ (386) & $-20.64\pm0.06$ (251) \\
               & -19.55\,\,\,\,\,-19.05 & $-20.29\pm0.13$ (230) & $-20.08\pm0.32$ (132) 
\enddata
\tablenotetext{a}{Absolute magnitude range assuming a mean redshift of $z=2.30$ and $z=3.05$.}
\tablenotetext{b}{Uncertainties in absolute magnitude reflect the stacked flux measurement
uncertainty combined in quadrature with the dispersion in absolute magnitude given the range
of redshifts of objects in each bin.  Parentheses indicate the number of galaxies in the stack.}
\label{tab:stackfaint}
\end{deluxetable*}

The results summarized in Figures~\ref{fig:uvch1} and
\ref{fig:nirmassuv} imply that even within the limited dynamic range
of UV luminosity probed with the spectroscopic sample, there is a
steep trend of mean stellar mass with UV luminosity.  The UV faintest
galaxies in our spectroscopic sample (around $0.16L^{\ast}$) have mean
stellar masses that are at least a factor of $\approx 40$ smaller than
those measured for UV-bright galaxies with $M_{\rm 1700}\simeq -22.5$.
This steep trend is only evident once galaxies undetected in the
near-IR are included (Figure~\ref{fig:uvch1}); the trend is much
shallower or non-existent when examining only those galaxies detected
in $\ks$-band or with IRAC \citep{shapley01, shapley05}.  Similar
trends between UV luminosity and stellar mass have been found for
dropout selected galaxies at higher redshifts \citep{stark09, lee11a},
and could be inferred from the fact that the SFR correlates with both
UV luminosity (Appendix~\ref{sec:malmquist}) and stellar mass
(Figure~\ref{fig:sfrm}).

\subsubsection{Stellar Masses of Photometrically Selected UV Faint
Galaxies}
\label{sec:massfaint}

To extend these spectroscopic results to fainter UV luminosities, we
have stacked the IRAC data for photometrically selected BXs and LBGs
with $\rs>25.5$, specifically in two bins with $25.5<\rs<26.5$ and
$26.5\le\rs<27.0$, using the same stacking method discussed in
Appendix~\ref{sec:iracstack}.  These photometrically selected faint
galaxies are likely to lie at the redshifts of interest, given the
strong correlation between contamination fraction and UV magnitude
\citep{reddy08, reddy09}.  Table~\ref{tab:stackfaint} lists the
stacked $3.6$ and $4.5$\,$\mu$m fluxes for these faint bins, along
with the number of objects contributing to the stacks.  The relatively
faint stacked IRAC magnitudes obtained for these galaxies suggests
that low redshift contaminants, if they exist, do not dominate the
signal; otherwise, we would have expected the stacked IRAC flux to be
brighter than the observed value.  The near-IR magnitudes are
converted to stellar masses as described in
Appendix~\ref{sec:massnir}.  These faint stacks imply that the trend
between stellar mass and UV luminosity derived from the spectroscopic
sample extends to fainter UV magnitudes and lower masses
(Figure~\ref{fig:nirmassuv}).

Having computed their mean stellar masses, we examine the UV faint
galaxies in the context of the SFR-$M_{\ast}$ relation.  The
unattenuated SFRs implied by the mean UV luminosities of the two bins
of photometrically selected faint galaxies at $z\sim 2.3$ are $\approx
2$ and $1$\,M$_{\odot}$\,yr$^{-1}$.  The average UV slopes for these
two bins, based on the relationship between $\beta$ and $M_{\rm UV}$
at $z\sim 2.5$ \citep{bouwens11b}, are $\langle\beta\rangle\approx
-1.8$ and $-1.9$.  Converting these $\beta$ to dust obscurations with
the \citet{meurer99} relation and applying them to the unobscured (UV)
SFRs then implies bolometric SFRs of $\approx 5$ and
$3$\,M$_{\odot}$\,yr$^{-1}$.  The dispersion in $\beta$ for a given
$M_{\rm UV}$ results in a factor of $\approx 2$ uncertainty in these
SFRs.  Adding the intrinsic scatter in the \citet{meurer99} relation
increases the SFR uncertainties to a factor of $\approx 2.3$.
Figure~\ref{fig:sfrm} includes the values of inferred bolometric SFR
for the faint samples analyzed here.

Given the biases discussed in Sections~\ref{sec:sfrm},
\ref{sec:mlratio}, and Appendices~\ref{sec:malmquist} and
\ref{sec:massnir}, it is prudent to determine whether the location of
the UV faint galaxies on the SFR-$M_{\ast}$ plane may be biased with
respect to all galaxies that lie within the same bins of stellar mass.
We investigate such biases using simulations that are presented in
Appendix~\ref{sec:sfrmbias}.  The first main result of our simulations
is that Malmquist bias can result in overestimated mean SFRs in bins
of stellar mass.  One may circumvent this bias by stacking in bins of
SFR (or UV luminosity), at the expense of probing the intrinsic
SFR-$M_{\ast}$ relation over a narrower range of $M_{\ast}$ where the
flux-limited sample is complete.  The critical point is that the
median stellar mass in bins of SFR may not directly translate to the
median SFR in that same bin of stellar mass.  Secondly, in an ideal
situation, it is desirable to perform full SED fitting for all
galaxies in a sample, in order to better constrain stellar masses that
take into account the star formation history and current SFR.  In our
case, a substantial fraction of galaxies in our sample are faint (and
low mass) and are undetected longward of the Balmer and $4000$\,\AA\,
breaks.  One can estimate stellar masses for such objects by assuming
some $M/L$ ratio.  However, as we have shown, a proper treatment must
take into account the UV luminosity dependence of the near-IR $M/L$
ratio when estimating stellar masses for undetected objects (or
objects detected in stack).  Having accounted for these biases, we
find that, within the uncertainties of the faint stacks, the
SFR-$M_{\ast}$ relation exhibits a close to unity slope to stellar
masses as low as $M_{\ast}\simeq 5\times 10^{8}$\,M$_{\odot}$.
Progress in quantifying any possible evolution of the SFR-$M_{\ast}$
relation at low masses should be made with future spectroscopic
observations of larger samples of UV faint galaxies, combined with
information from deep near-IR selected samples.

\section{DISCUSSION}
\label{sec:discussion}

A principal aspect of our analysis is the comparison of SED-inferred
SFRs to those determined from direct measurements of dusty star
formation at $z\sim 2$.  An important conclusion of our analysis is
that exponentially declining star formation histories yield SFRs that
are inconsistent with those obtained by combining mid-IR and UV data.
While both the constant and rising star formation histories yield SFRs
that consistent with the independent measurements, as we discuss
below, there is additional evidence that suggests that {\em on
  average} galaxies at $z\ga 2$ have SFRs that may be increasing with
time.

\subsection{The Typical Star Formation History of High-redshift Galaxies}

\begin{figure}[tbp]
\plotone{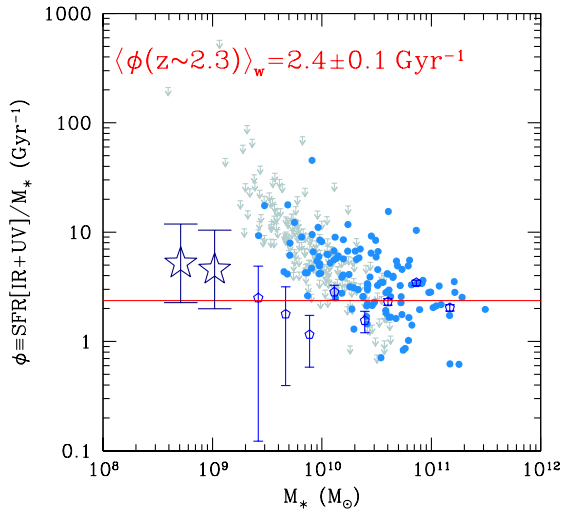}
\caption{Specific SFR, $\phi \equiv {\rm SFR}/M_{\ast}$, as a function
  of stellar mass for the MIPS sample of galaxies with redshifts
  $1.5\le z\le 2.6$.  Detections and non-detections are indicated by
  the filled symbols and downward arrows, respectively.  The mean
  specific SFR in bins of stellar mass for the spectroscopic sample
  and photometrically selected UV faint sample are indicated by the
  small open pentagons and large open stars, respectively.  The
  weighted mean specific SFR over all stellar masses is
  $\langle\phi\rangle=2.4\pm0.1$\,Gyr$^{-1}$, denoted by the solid
  line.}
\label{fig:ssfrm}
\end{figure}

The SED determined SFRs, and those computed from combining mid-IR and
UV data, vary roughly linearly with stellar mass
(Figure~\ref{fig:sfrm}).  Hence, as shown in Figure~\ref{fig:ssfrm},
the specific star formation $\phi \equiv {\rm SFR}/M_{\ast}$ of
galaxies with $1.5\le z_{\rm spec}\le 2.6$ is roughly constant over
$\approx 2.4$ orders of magnitude in stellar mass up to
$\log[M_{\ast}/{\rm M}_{\odot}]=11.0$, with a weighted mean of
$\phi(z\sim 2)=2.4\pm0.1$\,Gyr$^{-1}$, as measured from SFRs[IR+UV].
SFRs[SED] for galaxies with $2.7\le z\le 3.7$ imply $\phi(z\sim3) =
2.3\pm0.1$\,Gyr$^{-1}$.  Similar values of $\phi$ are obtained if we
assume a CSF model for deriving the SFRs[SED] and/or stellar masses.

\begin{figure}[tbp]
\plotone{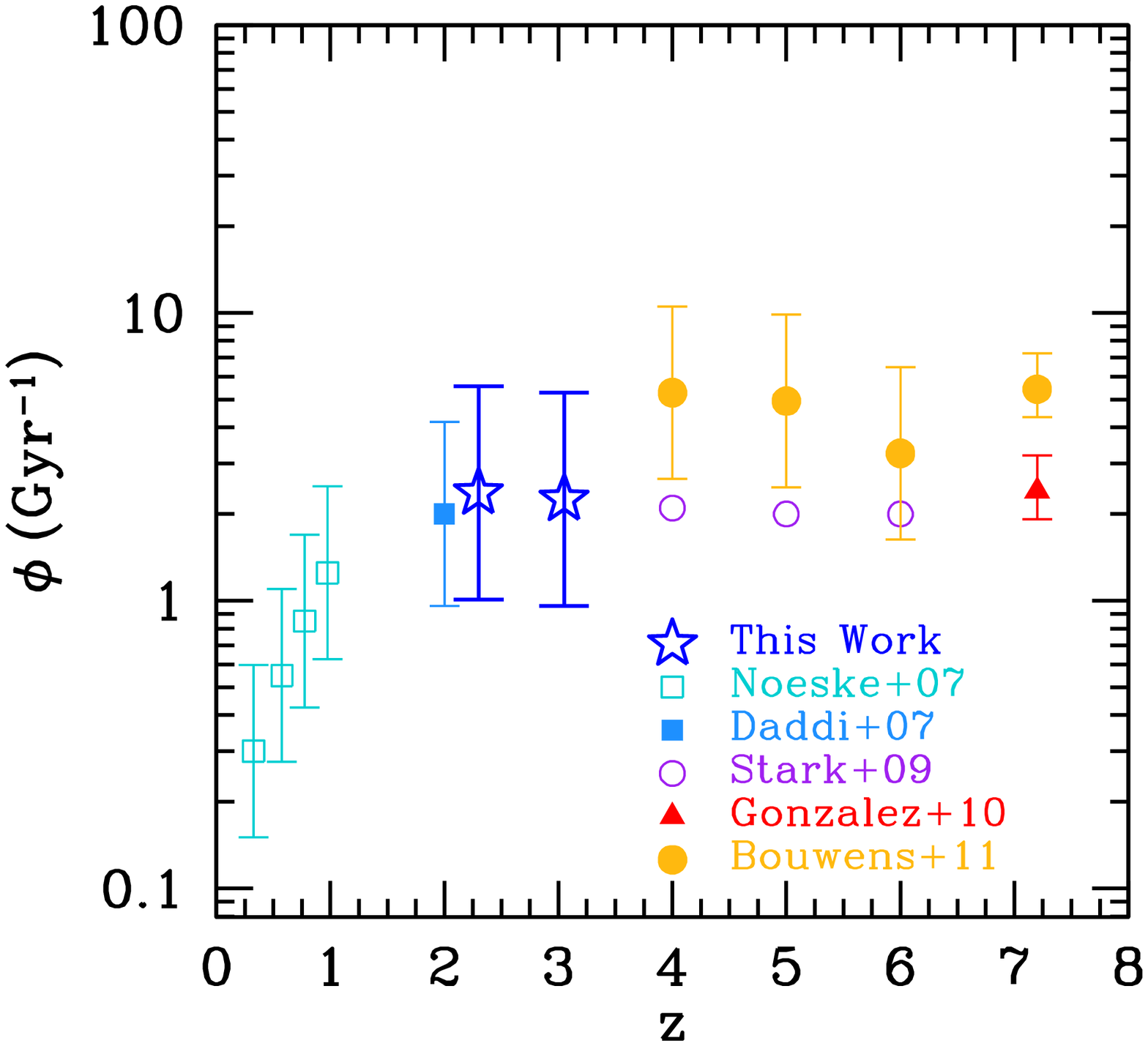}
\caption{Evolution of specific SFR $\phi$ as function of redshift for
  galaxies of a stellar mass $M_{\ast}\simeq 5\times
  10^{9}$\,M$_{\odot}$.  Data are taken from the compilation of
  \citet{gonzalez10}, which includes values from \citet{noeske07},
  \citet{daddi07a}, \citet{stark09}, and \citet{gonzalez10}.  We also
  show the dust corrected values from \citet{bouwens11b}.  The rms
  scatter in the specific SFR at a given mass is taken to be
  $0.3$\,dex for the \citet{noeske07} points, $0.32$\,dex for the
  \citet{daddi07a} point, $\approx 0.3$\,dex for the dust-corrected
  values at $4.0\la z\la 6$ from \citet{bouwens11b}, and $\approx
  0.10$\,dex for the \citet{gonzalez10} point at $z\sim 7$.  For our
  determination at $z\sim 2.3$, we compute a scatter of $\approx
  0.37$\,dex from a survival analysis that includes MIPS $24$\,$\mu$m
  non-detections (see text; note that this is the scatter of
  individual galaxies about the mean specific SFR, not the error in
  the mean specific SFR).  A similar value of the scatter is adopted
  for the $z\sim 3$ determination.}
\label{fig:ssfrevol}
\end{figure}

Several determinations of the specific SFR at low and high redshift,
nominally derived at a fixed stellar mass of $5\times
10^{9}$\,M$_{\odot}$, are compiled in
Figure~\ref{fig:ssfrevol}.\footnote{The difference in the mean stellar
  mass, and hence mean specific SFR, that results from using CB11
  versus \citet{bruzual03} models (the latter have been used in other
  studies; e.g., \citealt{gonzalez10}) is negligible compared to the
  intrinsic dispersion in specific SFR.}  The average specific SFRs
derived at $z\sim 2$ and $z\sim 3$ are similar to those measured at
redshifts $z>3$ (e.g., \citealt{daddi07a, stark09, gonzalez10});
including dust corrections to these higher redshift points results in
a roughly constant specific SFR at $z\ga 4$ \citep{bouwens11b}, and
one which is about a factor of $2-3$ larger than the values we find at
$z\sim 2-3$.  A note of caution regarding these higher redshift ($z\ga
3$) results is that the relatively young age of the universe implies a
limited dynamic range in $M/L$ ratio, translating to a limited range
in the possible {\em simple} star formation histories.  As such, it is
perhaps not at all surprising to find SFR\,$\propto M_{\ast}$, given
that the young stars that dominate the UV continuum also contribute
significantly to the rest-optical flux, the latter of which is used to
constrain the stellar mass (i.e., the ``outshining'' problem as
discussed earlier).  This degeneracy may also lead to an artificial
tightening of the scatter in the SFR-$M_{\ast}$ relation at high
redshift.  The advantage of the method employed here, for $z\sim 2$
galaxies, is that we have (a) determined SFRs largely independent of
the SED modeling (and hence $M_{\ast}$) by incorporating the MIPS
$24$\,$\mu$m data, and (b) the stellar masses are constrained with
IRAC data that probe the peak of the stellar emission at
$1.6$\,$\mu$m.

While independent estimates of SFRs are not practical at $z\ga 3$
given the general faintness of galaxies and their expectation of
having lower dust content, we note that: (a) the mean specific SFR at
$z\sim 2$ is similar to that derived at higher redshift, despite the
different methods used to measure SFRs at $z\sim 2$ versus $z\ga 3$
and (b) there is well-defined correlation between SFR and $M_{\ast}$
as $z\la 2$ \citep{noeske07}, where SED modeling independent
measurements of SFRs {\em can} be obtained and where rest-frame
near-IR constraints on stellar mass are accessible.  By extension,
these lower redshift results hint that there may indeed be an
intrinsic correlation between SFR and $M^{\ast}$ at the highest
redshifts.  Regardless, we note that if the stellar masses are
severely biased for large numbers of galaxies at $z\ga 3$ (due to
``contamination'' of the rest-optical light from the most massive
stars), then correcting for this effect result in stellar mass
densities at $z\ga 4$ that would be inconsistent with the integrated
star formation history inferred at these early cosmic times (e.g.,
\citealt{reddy09, bouwens11b}).  Such an inconsistency would be made
even worse if we then also corrected the stellar mass determinations
at $z\ga 4$ for the effect of strong emission line contribution to the
broadband optical fluxes \citep{schaerer09, schaerer10}.  With this in
mind, we find that even with the \citet{bouwens11b} dust corrections,
the overall picture does not change significantly and the specific
star formation rates appear to have remained approximately constant,
within a factor of a few, over the $2.5$\,Gyr interval from $z=7$ to
2.  As such, the average SFRs would be roughly proportional to stellar
mass, implying that both increase exponentially with time.

\subsubsection{Expectations from a Constant Star Formation History}
\label{sec:csfexpectation}

\begin{figure}[tbp]
\plotone{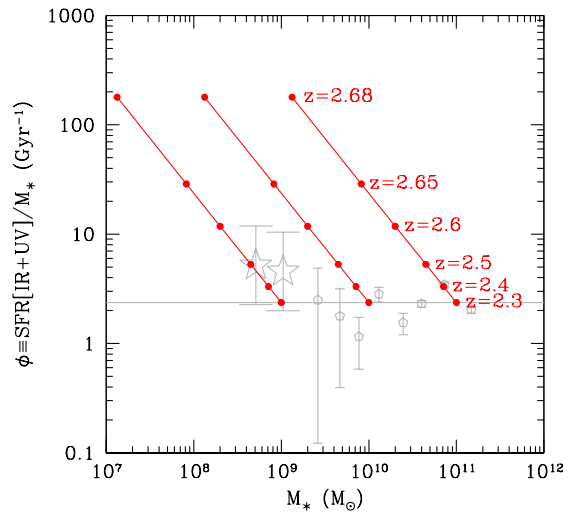}
\caption{Same as Figure~\ref{fig:ssfrm}, where median stacks of the
  specific SFR are shown by the grey points (individual points have
  been suppressed for clarity).  The connected red points show the
  evolution in $\phi$ and $M_{\ast}$ of a galaxy with
  $M_{\ast}=10^{9}$, $10^{10}$, and $10^{11}$\,M$_\odot$ and
  $\phi=2.4$\,Gyr$^{-1}$ at $z=2.3$, assuming a CSF history.}
\label{fig:ssfrmevol}
\end{figure}

Given this suggestion, it is useful to place the $z\sim 2-3$ galaxies
in the context of higher redshift samples by evolving their SFRs
backward in time.  For a CSF history, the specific SFR will be larger
at earlier times.  For example, a typical galaxy at redshift $z=2.3$
with $\phi\approx 2.4$\,Gyr$^{-1}$ would have had $\phi\ga
30$\,Gyr$^{-1}$ at $z\ga2.65$, $\approx 390$\,Myr earlier.  The upper
redshift $z=2.65$ is close the upper boundary where the $24$\,$\mu$m
data are still sensitive to the rest-frame $8$\,$\mu$m emission, and
the difference in time between $z=2.3$ and $z=2.65$ of $\approx
390$\,Myr is similar to the mean ``age'' determined for our
spectroscopic sample of $z\sim 2$ galaxies
(Figure~\ref{fig:agemasshist}).  Extending to higher redshift, the
specific star formation would increase by more than an order of
magnitude in the $\approx 80$\,Myr between $z=2.60$ and $z=2.68$,
reaching a value of $\phi\approx 180$\,Gyr$^{-1}$ at this latter
redshift (Figure~\ref{fig:ssfrmevol}).  Our sample, which encompasses
this higher redshift ($z=2.68$), includes very few galaxies with
$\phi$ as large as the one predicted with a CSF history.  The lack of
galaxies with such high specific SFRs is unlikely to be a selection
effect because the SFR would remain unchanged and such galaxies should
still be selectable via their UV emission.  If anything, the higher
redshift progenitors of galaxies at $z\sim 2$ would be more easily
detectable via their UV emission at the same SFR given that the
extinction per unit bolometric SFR decreases with increasing redshift
\citep{reddy10a}.  Hence, there appears to be a true paucity of
galaxies with specific SFRs $\phi\ga 100$\,Gyr$^{-1}$ relative to what
we would have predicted if the majority of galaxies at $z=2.3$ had CSF
(or declining\footnote{Aside from the disagreement between SED[IR+UV]
  and SED[SFR] with a declining star formation history, such histories
  also require larger SFRs at earlier times.  The tight correlation
  between SFR and dust attenuation would then imply an
  anti-correlation between dust attenuation and stellar mass for any
  {\em individual} galaxy.  If such an anti-correlation existed, we
  would have expected to see a significant evolution in the
  SFR-$M_{\ast}$ relation at $z>2$, yet this is not observed (e.g.,
  \citealt{gonzalez10, bouwens11b}).}) histories.  Stated another way,
the relatively small scatter in the SFR at a given stellar mass {\em
  over the entire redshift range $1.5\la z\la 2.6$} precludes the
possibility that there exists a large population of galaxies with
specific SFRs that are substantially larger than the mean values
observed at these epochs.\footnote{It is still possible that there may
  exist very low mass galaxies with high specific SFRs (e.g.,
  \citealt{vanderwel11}) that do not enter our selection, but may
  eventually merge to produce those galaxies seen in our sample.  On
  the other hand, there may also exist a heretofore undetected
  population of low mass galaxies with low specific SFRs that merge in
  the same way as the high specific SFR galaxies.  In this case, we
  might expect that the mean specific SFR in a given bin of low
  stellar mass may not be substantially different than that of a bin
  of higher stellar mass.  Our IRAC stacking analysis supports this
  expectation to a mass of $M_{\ast}\simeq 5\times 10^{8}$\,M$_\odot$.
  Undoubtedly, future observations should help to clarify the
  situation for galaxies with masses lower than our current mass
  threshold.}  These mean values of $\phi$ are roughly independent of
stellar mass with $\langle\phi\rangle\approx 2.4$\,Gyr$^{-1}$ (i.e.,
the lower mass progenitors of $z=2.3$ galaxies retain roughly the same
specific SFR, contrary to our expectation if galaxies were evolving
with a constant SFR).

\begin{figure*}[!t]
\plottwo{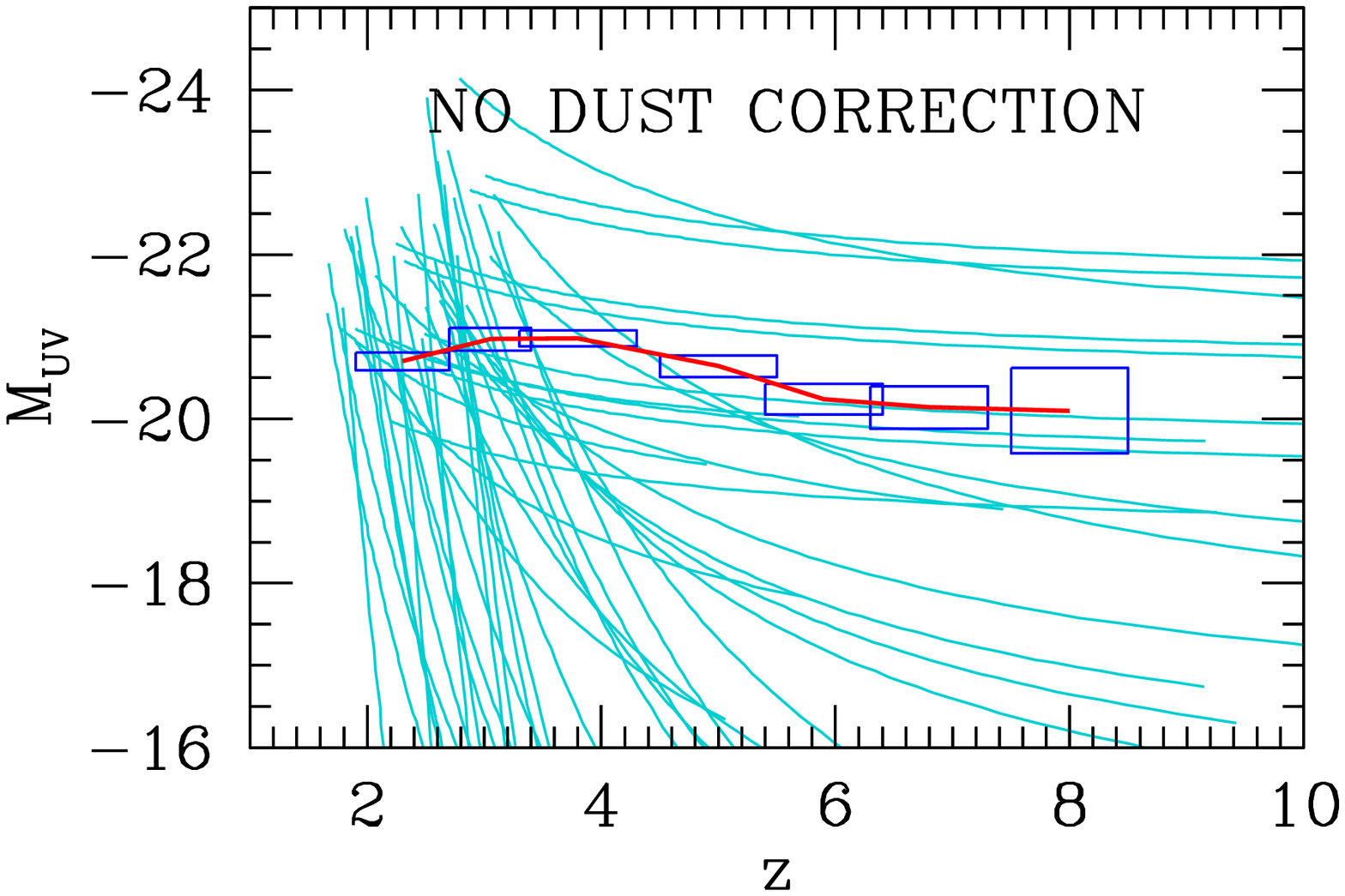}{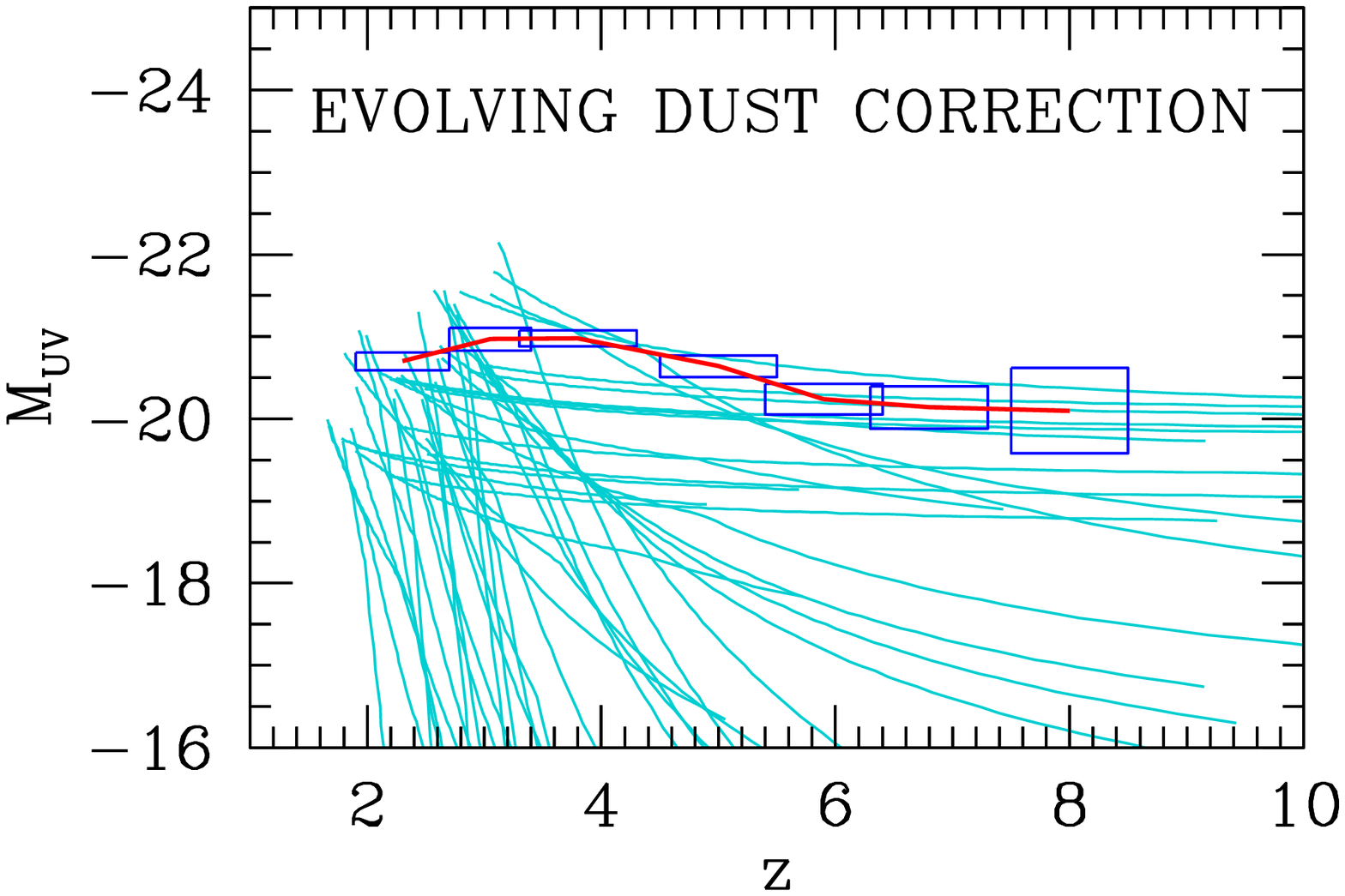}
\caption{Intrinsic (left) and observed (right) absolute UV magnitude
  as a function of redshift for 45 galaxies chosen randomly from our
  SED sample.  The curves assume a star formation history where the
  specific SFR $\phi$ is constant.  The intrinsic UV magnitude is
  derived directly from the bolometric SFR without any correction for
  dust.  The observed UV magnitude is computed under the assumption
  that the dust correction depends on the observed UV magnitude
  \citep{reddy09, bouwens11b}.  The rectangles and red line indicate
  the values and uncertainties in the characteristic absolute
  magnitude $M_{\rm UV}^{\ast}$ derived from the $z\sim 3.8-8.0$
  luminosity functions of \citet{bouwens11a} assuming $\delta z\approx
  1$; and the $z\sim 2-3$ luminosity functions of \citet{reddy09}
  assuming $\delta z=0.8$ and $0.7$ at $z\sim 2.30$ and $z\sim 3.05$,
  respectively.}
\label{fig:zevolplot}
\end{figure*}

\subsubsection{Expectations from a Rising Star Formation History and
Implications for the Evolution of $M_{\rm UV}$ with Redshift}

As discussed in Section~\ref{sec:ages}, a rising star formation
history where the specific SFR $\phi$ is roughly constant implies
``ages'' that can be arbitrarily old if we vary the normalization of
the star formation history.  Given such a star formation history, we
can estimate the redshift evolution in UV luminosity for each of the
galaxies in our sample.  For clarity, we show this evolution for a
subset of 45 galaxies drawn randomly from our SED sample of 1959
galaxies, assuming ages that are either $2$\,Gyr or $2.5$\,Gyr,
depending on the redshift of the galaxy (so as not to violate the age
of the universe at each redshift), in Figure~\ref{fig:zevolplot}.  The
observed UV magnitudes are calculated assuming that the dust
extinction varies with total SFR (or intrinsic UV luminosity;
e.g. \citealt{reddy08, bouwens09, reddy10a}).  We adopted the relation
between UV luminosity and bolometric luminosity in order to apply this
correction (see Appendix~\ref{sec:malmquist}; adopting the
\citealt{bouwens09} or \citealt{bouwens11b} relations between UV
slope, $\beta$, and $M_{\rm UV}$ at $z\sim 2.5$ yields similar
results).  The correction is renormalized to yield the same dust
reddening as indicated by the best-fit value of $\ebmv$ from the SED
fitting.\footnote{There is evidence of a mild evolution in $\beta$ at
  a given $M_{\rm UV}$ as a function of redshift, in the sense that
  higher redshift galaxies have slightly bluer UV slopes than lower
  redshift galaxies with the same absolute UV magnitude
  \citep{bouwens11b}.  For simplicity, we have not corrected for this
  redshift dependence.  Doing so would shift the UV magnitudes
  brighter at earlier times relative to what is shown in the right
  panel of Figure~\ref{fig:zevolplot}.}  For comparison, we also show
the maximum possible UV luminosity (i.e., assuming no dust reddening)
as a function of redshift in the left panel of
Figure~\ref{fig:zevolplot}.  Hence, the curves shown in the {\em left}
and {\em right} panels of Figure~\ref{fig:zevolplot} should encompass
the full range of possible $M_{\rm UV}$ versus redshift tracks for
each object.  Also shown are the characteristic magnitudes $M_{\rm
  UV}^{\ast}$ of the UV luminosity functions at $z\sim 2-3$ from
\citet{reddy09} and $z\sim 4-8$ from \citet{bouwens11a}.

In the case of rising SFRs, the redshift evolution in $M_{\rm UV}$ for
$z\sim 2-3$ galaxies with $\rs < 25.5$ implies that a substantial
fraction of them would have been fainter than $L^{\ast}$ (i.e., having
sub-$L^{\ast}$ luminosities) at $z\ga 4$.\footnote{This conclusion is
  even stronger for those galaxies that may have undergone merging,
  since the individual components would each have an SFR, and hence UV
  luminosity, lower than that of the merged galaxy.}  Hence, the
progenitors of $z\sim 2-3$ galaxies are likely to have appeared as
sub-$L^{\ast}$ galaxies in dropout samples at $z\ga 4$.  Further, the
mean specific SFR of $\langle \phi\rangle \simeq 2.4$\,Gyr$^{-1}$ at
$z\sim 2$ implies a mass doubling time of $\approx 300$\,Myr
(Equation~\ref{eq:massdouble}).  Hence, a typical star-forming galaxy
at $z=2.3$ has doubled its stellar mass since $z=2.8$.  Similarly, a
$z=2.3$ galaxy would have had a fifth of its current stellar mass (and
an intrinsic UV luminosity that is $\approx 1.25$\,mag fainter) at
$z=3.0$ (e.g., Figures~\ref{fig:nirmassuv} and \ref{fig:zevolplot}).

At face value, the average rising star formation history
characteristic of typical star-forming galaxies at $z\sim 2$ would
imply a significant evolution in $L^{\ast}$ of the UV luminosity
function at higher redshifts.  While this is true above $z\sim 4$, in
the sense that $L^{\ast}$ becomes progressively brighter with
decreasing redshift from $z\sim 7$ to $z\sim 4$ (e.g.,
\citealt{bouwens11a}), recent studies suggest little evolution in
$L^{\ast}$ over the redshift range $z\sim 2-4$ \citep{reddy09}.  How
can we reconcile the rising star formation histories of $z\sim 2$
galaxies with the fact that $L^{\ast}$ does not evolve significantly
between $z\sim 2-4$?  There are two possible explanations.  First,
observations indicate the emergence of a population of massive and
passively evolving galaxies over these redshifts \citep{vandokkum06},
implying that some fraction of UV-bright galaxies eventually ``turn
off'' by $z\sim 2$ \citep{stark09}.

Second, there is a non-monotonic relationship between bolometric and
UV luminosity for $z\sim 2$ galaxies, in the sense that as the
bolometric SFR of galaxy increases, dust attenuation will also
increase, and eventually there is a point at which the observed UV
luminosity saturates \citet{reddy10a}.  In effect, dust obscuration
limits the maximum observable UV luminosity of galaxy which, for
$z\sim 2$ galaxies, occurs around the value of $L^{\ast}$.  Therefore,
as a galaxy's SFR increases with time, there comes a point at which
dust attenuation becomes large enough that its observed UV luminosity
no longer increases with increasing SFR.  It is likely that some
combination of aforementioned effects (i.e., the fading of some
fraction of UV-bright galaxies between $z\sim 4$ and $z\sim 2$, and
the limit to the observed UV luminosity due to dust attenuation; e.g.
see also \citealt{bouwens09}) conspire to produce a relatively
constant value of $L^{\ast}$ between $z\sim 2-4$, even though the
average star formation history may be rising during these epochs.

\begin{figure*}[!t]
\plottwo{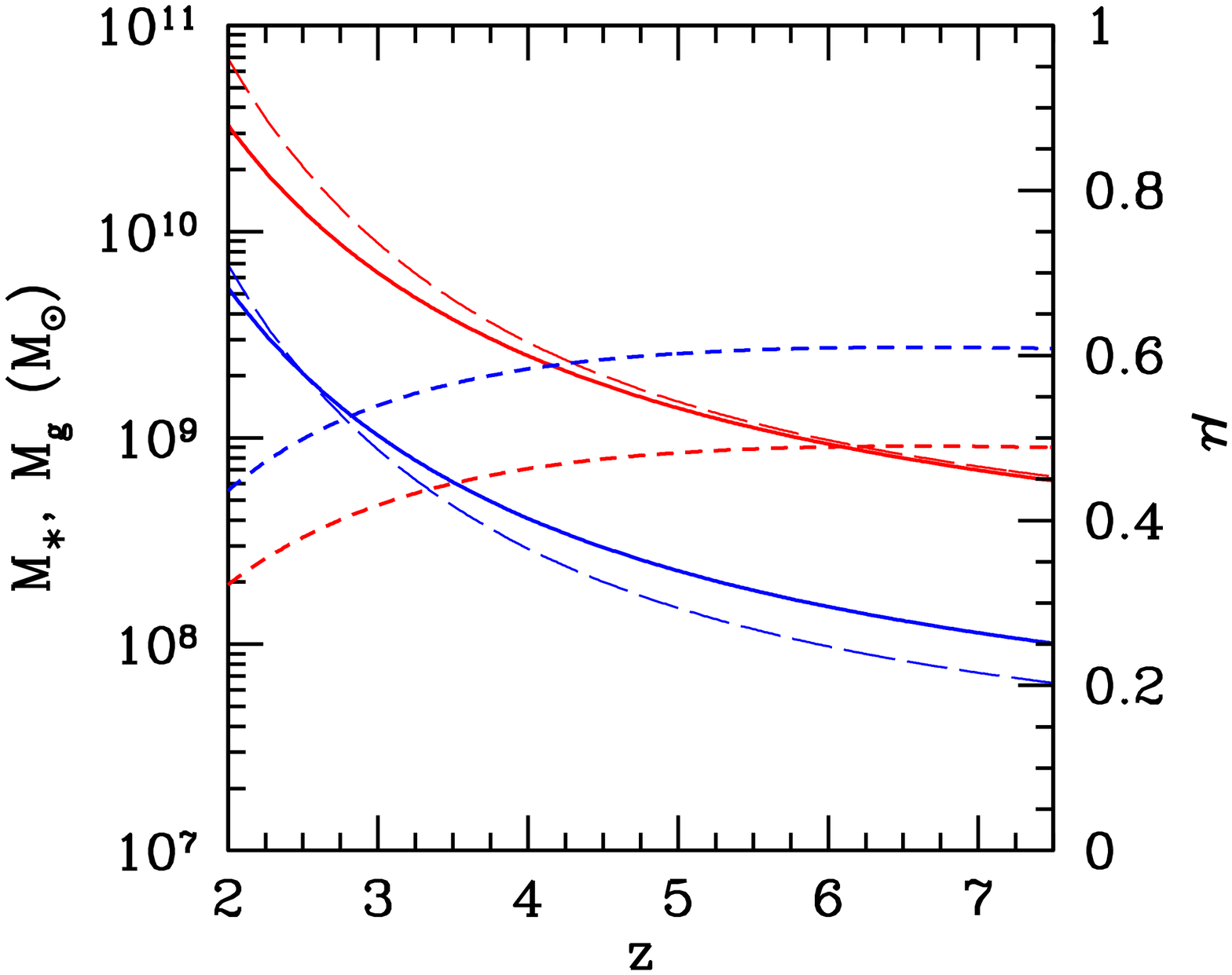}{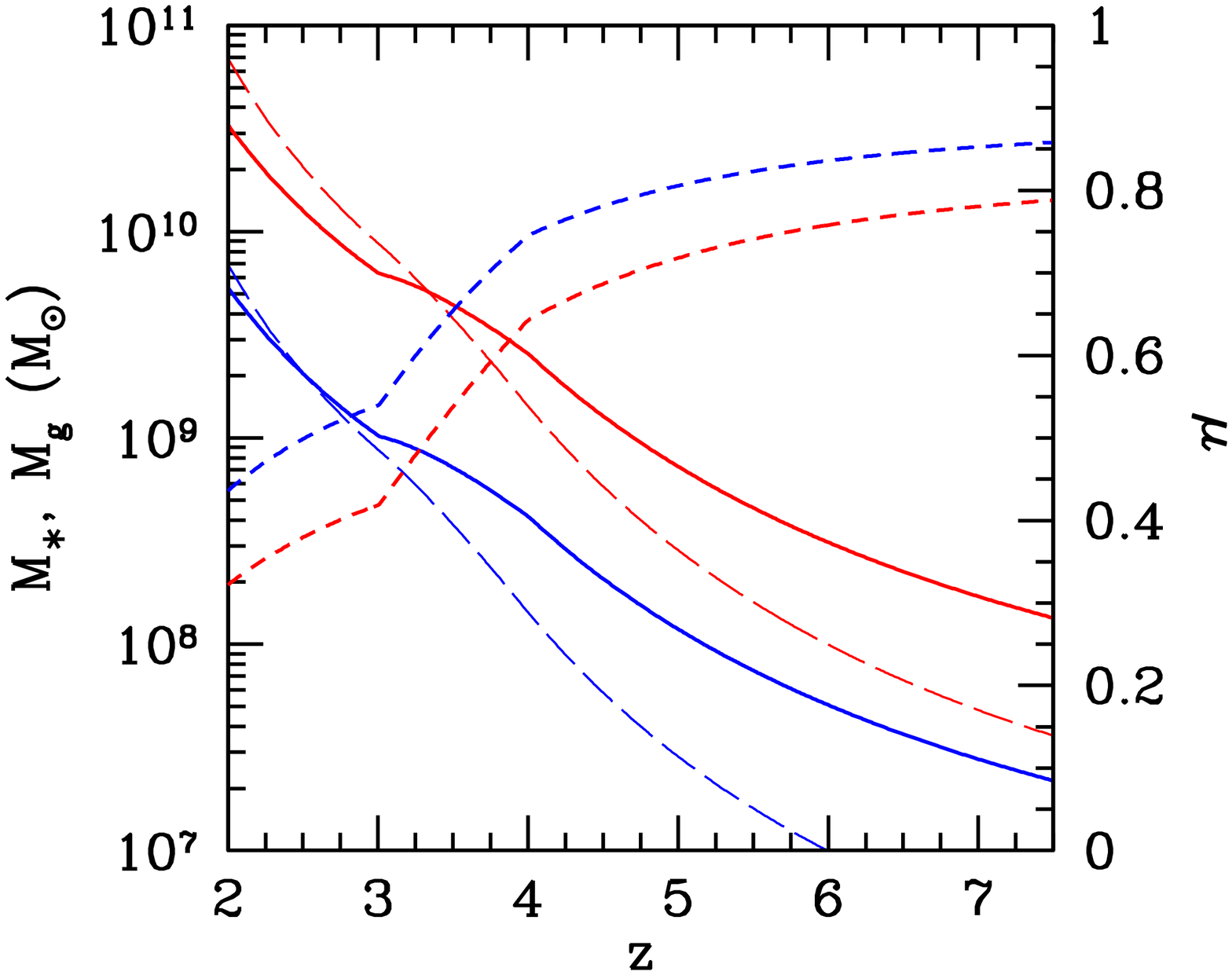}
\caption{{\em Left:} Redshift evolution in gas (solid lines) and
  stellar (long dashed lines) masses for galaxies with a final stellar
  mass of $\log M_{\ast}=9.5$ and 10.5 at $z\sim 2.3$, assuming a
  constant specific SFR of $\phi=2.4$\,Gyr$^{-1}$.  Also indicated
  (short dashed lines) is the evolution in gas mass fraction for these
  two examples.  {\em Right:} Same as left panel where we have assumed
  $\phi\simeq 2.4$\,Gyr$^{-1}$ at $z\la 3.0$ based on our sample, and
  $\phi\simeq 5.0$\,Gyr$^{-1}$ at $z\ga 4$ based on the dust-corrected
  specific SFRs presented by \citet{bouwens11b}.  We have used a
  spline fit to interpolate the specific SFRs over the redshift range
  $3.0\la z\la 4.0$.}
\label{fig:gasevol}
\end{figure*}

\subsection{Implications for Cold Gas Accretion at High Redshift}

In \citet{reddy06a}, we examined the relationship between specific
SFR, $\phi$, and cold gas mass fraction, $\mu$:
\begin{equation}
\phi = C\frac{\mu^{1.4}}{1-\mu},
\end{equation}
where we assumed the Kennicutt-Schmidt relation \citep{kennicutt98}
between SFR surface density and cold gas mass surface density.  The
multiplicative factor $C$ in the equation above depends on the
constant of proportionality between SFR and cold gas mass surface
density and the size of the star-forming region.  We have defined the
cold gas mass fraction $\mu$ as
\begin{equation}
\mu \equiv \frac{M_{\rm g}}{M_{\rm g}+M_{\ast}}.
\end{equation}
We revisit this relationship in light of our new determinations of the
specific SFRs of $z\sim 2-3$ galaxies.  The \citet{kennicutt98}
relation between SFR and cold gas mass surface density reduces to
\begin{equation}
M_{\rm g}\simeq 5.2\times 10^{8} \left[\frac{\Psi}{\rm M_{\odot}\,yr^{-1}}\right]^{\frac{1}{N}}\left[\frac{r_{\rm 1/2}}{\rm kpc}\right]^{2-\frac{2}{N}}\,{\rm M_\odot},
\label{eq:gasmass}
\end{equation}
where $M_{\rm g}$ is the cold gas mass is solar masses, $\Psi$ is the
SFR, $r_{\rm 1/2}$ is the half-light radius characterizing the size of
the star-forming region, and the power law index of the
Kennicutt-Schmidt relation is $N=1.4$ (below, we explore how changing
the value of $N$ affects our results).  \citet{law12} have quantified
the sizes and their redshift evolution of a subset of the UV selected
galaxies in the SED sample that have {\em HST}/WFC3 IR imaging.  These
authors find characteristic half-light radii of $r_{\rm 1/2}\approx
1.3$ and $1.8$\,kpc for $z\sim 2$ galaxies in stellar mass bins
centered at $\log (M_{\ast}/M_{\odot})=9.5$ and $10.5$, respectively.
These sizes are measured in the rest-frame optical and are roughly
$20\%$ smaller than those measured in the rest-UV, where the latter
may be more appropriate for the size of the star-forming regions.
\citet{law12} also find $r_{\rm 1/2}\propto (1+z)^\gamma$ where
$\gamma\approx -1.07$, measured from $z=3.6$ to $z=1.5$.  This size
scaling with a power law close to unity is similar to that inferred
from higher redshift ($z\ga 4$) dropout samples \citep{oesch10}, and
is indistinguishable within the errors from the $r_{\rm 1/2}\propto
H(z)^{-1}$ scaling proposed by \citet{ferguson04}.  In the subsequent
discussion, we assume the results of \citet{law12} for the redshift
evolution in the characteristic half-light radii measured in the
rest-frame UV.

Adopting Equation~\ref{eq:gasmass}, we show in
Figure~\ref{fig:gasevol} the evolution of the cold gas masses for a
typical star-forming galaxy with a constant specific SFR of
$\phi=2.4$\,Gyr$^{-1}$ (and, hence, an exponentially rising star
formation history with $\tau_{\rm r} = \phi^{-1}\simeq
400$\,Myr\footnote{This exponentially rising timescale is similar to
  the $\tau_{\rm r}\simeq 0.5$\,Gyr inferred at $z\ga 5$ using the
  models of \citet{bouche10}; and is also similar to that inferred by
  \citet{papovich11} to describe the average star formation history of
  galaxies with a constant comoving number density of $2\times
  10^{-4}$\,Mpc$^{-3}$ at $z\ga 4$.}) and (final) stellar mass at
$z\sim 2.3$ of $\log (M_{\ast}^{z=2.3}/M_{\odot})=9.5$ and $10.5$
(denoted by the solid lines in the left panel of
Figure~\ref{fig:gasevol}).  Also shown (in the {\em right} panel of
Figure~\ref{fig:gasevol}) is the inferred evolution assuming a median
specific SFR that evolves with redshift: at $z\sim 2-3$ we assume
$\phi\simeq 2.4$\,Gyr$^{-1}$, based on our sample; at $z\ga 4$ we
assume $\phi\simeq 5.0$\,Gyr$^{-1}$, based on the dust-corrected
specific SFRs found by \citet{bouwens11b}.  We have adopted a spline
fit to these values in order to infer the specific SFR over the range
$3.0\la z\la 4.0$.  In addition, the stellar masses are evolved
backwards in time assuming a gas recycling rate from stellar remnants
as computed from the CB11 stellar population synthesis models for an
exponentially rising star formation history with $\tau_{\rm r}\simeq
400$\,Myr.

There are several important implications of these results.
Qualitatively, the inferred evolution assuming a constant and evolving
median specific SFR ({\em left} and {\em right} panels, respectively,
of Figure~\ref{fig:gasevol}) are similar.  If we assume that the
Kennicutt-Schmidt relation applies at high redshift, then a rising SFR
must be accompanied by an increase in cold gas surface density.
Hence, the cold gas mass should increase as $\approx M_{\ast}^{5/7}$,
assuming a Schmidt law index of $N=1.4$.  The cold gas mass {\em
  fraction}, $\mu$, will therefore evolve with time more slowly than
if the SFR is constant or declining exponentially.  The results of
Figure~\ref{fig:gasevol} indicate that the characteristic cold gas
mass fraction (assuming $\phi=2.4$\,Gyr$^{-1}$) is relatively constant
above $z\sim 4$, and declines more rapidly at lower redshift.  The
average stellar mass of galaxies in our {\em spectroscopically
  confirmed} sample at $1.4\le z\le 2.7$ is $\langle M_{\ast}\rangle =
1.6\times 10^{10}$\,M$_{\odot}$ (see
Section~\ref{sec:compareagemass}).  For this stellar mass, the cold
gas fraction is inferred to be between $50\% - 60\%$ at $z\ga 4$
assuming $\phi=2.4$\,Gyr$^{-1}$ (or $\ga 60\%$ at $z\ga 4$ assuming
the dust corrected $\phi$ of \citealt{bouwens11b}), and decline to
$\approx 40\% -50\%$ at $z\approx 2.8$.  More generally, on an object
by object basis, we find a large range in cold gas fraction at $z\sim
2$, with typical values of $\mu\simeq 0.4-0.6$.  This is quite similar
to the average cold gas fractions of $\mu\simeq 0.3-0.6$ inferred for
star-forming galaxies with similar stellar masses as represented in
our spectroscopic sample (Figure~\ref{fig:agemasshist}) at $z\sim 2$
based either on dust-corrected H$\alpha$ (via the Kennicutt-Schmidt
relation; \citealt{erb06b}) or CO measurements (e.g.,
\citealt{daddi10, tacconi10,riechers10}; see also
\citealt{papovich11}).

\begin{figure}[tbp]
\plotone{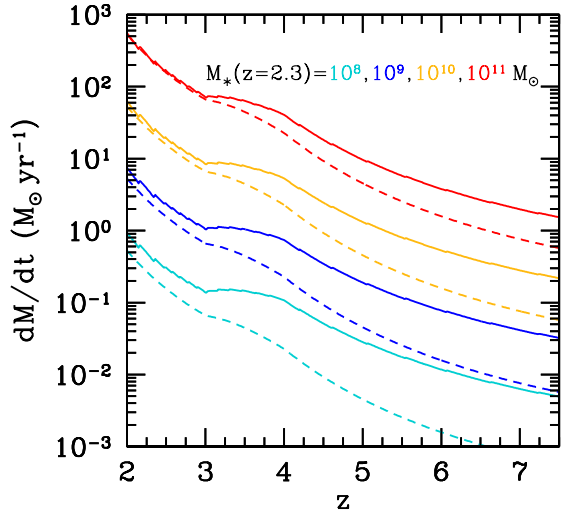}
\caption{Redshift evolution in the SFR (dashed lines) and net cold gas
  accretion rate (solid lines) for galaxies with final stellar masses
  at $z\sim 2.3$ $M_{\ast}^{z=2.3}=10^{8}$, $10^{9}$, $10^{10}$, and
  $10^{11}$\,M$_{\odot}$.  The curves assume a specific SFR that
  evolves from $\phi\simeq 2.4$\,Gyr$^{-1}$ at $z\sim 2-3$ to
  $\phi\simeq 5.0$\,Gyr$^{-1}$ at $z\ga 4$ and a gas recycling rate
  from stellar remnants computed from the CB11 stellar population
  synthesis models for a exponentially rising star formation history
  with $\tau_{\rm r}\simeq 400$\,Myr.}
\label{fig:gasrate}
\end{figure}

As we have just discussed, the mean cold gas fraction is likely higher
at $z\sim 4$ than at $z\sim 2$, and the mean fraction may not evolve
strongly above $z\sim 4$ for typical star-forming galaxies at these
higher redshifts (e.g., stronger evolution may be expected for
galaxies with larger stellar masses).  To illustrate these issues
further, we show in Figure~\ref{fig:gasrate} the net cold gas
accretion rates (as inferred from the specific SFRs) and SFRs for
galaxies that have $M_{\ast}^{z=2.3} = 10^{8}$, $10^{9}$, $10^{10}$,
and $10^{11}$\,M$_{\odot}$ by $z\sim 2.3$, adopting the evolution in
$\phi$ that was assumed in the {\em left} panel of
Figure~\ref{fig:gasevol}.  We computed the net cold gas accretion rate
as $\dot{M}^{\rm cold}_{\rm acc}$, which is related simply to: (a) the
amount of cold gas consumed to form stars in a time $\delta t$ (i.e.,
the SFR); (b) the time derivative of the cold gas mass (computed from
Equation~\ref{eq:gasmass}); and (c) the rate at which the mass formed
in stars is released back into the ISM as computed from the CB11
stellar population synthesis models ($\dot{R}$).  For a Salpeter IMF,
the CB11 models indicate that the fraction of gas returned to the ISM
to the stellar mass reaches a value of $\approx 30\%$ at 2\,Gyr,
assuming an exponentially rising star formation history with
$\tau_{\rm r}\simeq 400$\,Myr.  Hence, we can write the net cold gas
accretion rate as
\begin{equation}
\dot{M}^{\rm cold}_{\rm acc} = \dot{M}_{\rm g} + \Psi - \dot{R}
\label{eq:gasacc}
\end{equation}
(see also Equation~8 of \citealt{papovich11}).

Above $z\sim 4$, the net cold gas accretion rate outpaces the SFR by
at least a factor of two, with the difference between cold gas
accretion rate and SFR being largest for lower mass galaxies at any
given redshift above $z\sim 3-4$.  It is evident that the SFR becomes
comparable to the net cold gas accretion rate at a redshift which is
dependent upon stellar mass (i.e., the redshift where the dashed and
solid lines come together in Figure~\ref{fig:gasrate}).  In
particular, $\Psi \approx \dot{M}^{\rm cold}_{\rm acc}$ at higher
redshifts for galaxies with larger stellar masses.

Above $z\sim 2-3$, the accretion rate outpaces the SFR and the {\em
  average} cold gas fraction remains consistently high (e.g., {\em
  right} panel of Figure~\ref{fig:gasevol}; \citealt{bouche10}; see
also Figure~4 of \citealt{papovich11}).  Of course, the actual
dispersion in $\mu$ at a given redshift (even above $z\sim 4$) may
still be quite large.  Nonetheless, the steep UV luminosity function
at $z\ga 4$ (e.g., \citealt{bouwens09}) implies that less UV luminous
and hence lower stellar mass galaxies (e.g., \citealt{stark09, lee11a,
  lee11b}; see also Figure~\ref{fig:uvch1}) should dominate, and hence
the number-weighted mean cold gas fraction at these redshifts should
be quite high with $\mu\ga 60\%$ (Figure~\ref{fig:gasevol}).  Finally,
we note that the difference in $\dot{M}^{\rm cold}_{\rm acc}$ and SFR
will be slightly larger at $z\ga 4$ had we assumed a shallower slope
of the \citet{schmidt59} relation of $N\simeq 1.1-1.2$ (e.g., as
suggested by \citealt{genzel10}), with a further result being that the
point at which $\Psi \approx \dot{M}^{\rm cold}_{\rm acc}$ shifts to
lower redshifts for galaxies of a given stellar mass.

The physical picture implied by the redshift evolution in cold gas
accretion is that {\em on average} galaxies must be continually
accreting gas to support their increasing SFRs.  Indeed, the gas
consumption timescales assuming a CSF history for $z\sim 2$ UV
selected galaxies are on average a factor of two lower than median
stellar ages, implying that significant gas accretion must be
occurring at $z\sim 2$ \citep{erb08}, while at the same time, such
galaxies are inferred from clustering measurements to have long duty
cycles (e.g., \citealt{conroy08}).  A high rate of gas accretion is
also required to explain the increasing SFRs (and SFR density)
contributed by galaxies of a fixed number density at $z\ga 4$
\citep{papovich11}.

We conclude with a caution against over-interpreting the physical
implications of the redshift at which $\Psi \approx \dot{M}^{\rm
  cold}_{\rm acc}$.  One might speculate that below this
redshift--which for typical galaxies with $M_{\ast}\ga
10^{9}$\,M$_{\odot}$ occurs around $z\sim 2$
(Figure~\ref{fig:gasrate})---the SFR declines primarily in response to
a lower cold gas accretion rate.  While this may be the case---and
would explain naturally the overall decline in the cosmic SFR density
at the same redshifts ($z\sim 2$)---as we discuss in the next section,
the cold gas accretion computed above evidently constitutes only a
small fraction of the total baryons accreted onto halos.  Therefore,
some other mechanism(s) must be responsible for regulating the large
number of baryons that never end up in stars at high redshift.  We
discuss this further in Section~\ref{sec:totacc}.

\subsection{A Consideration of Gas Inflows, Outflows, and
the Expectations from Simulations}
\label{sec:totacc}

\begin{figure}[tbp]
\plotone{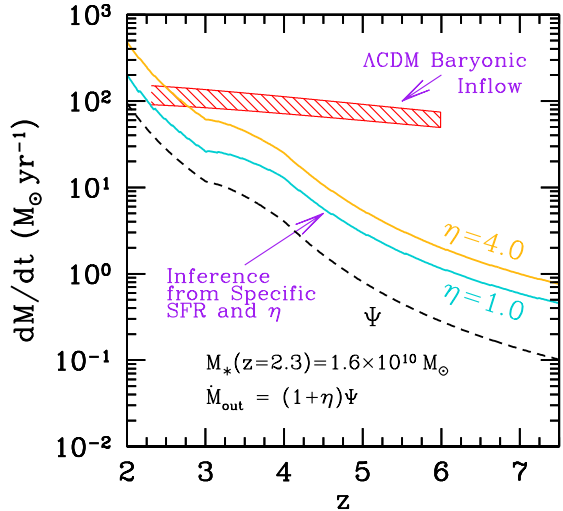}
\caption{SFR (dashed line) and inferred cold gas inflow rates (solid
  lines) for different values of the mass-loading factor $\eta$, for a
  galaxy with a final stellar mass of $M_{\ast}=1.6\times
  10^{10}$\,M$_{\odot}$ at $z=2.3$.  The curves assume a specific SFR
  that evolves from $\phi\simeq 2.4$\,Gyr$^{-1}$ at $z\sim 2-3$ to
  $\phi\simeq 5.0$\,Gyr$^{-1}$ at $z\ga 4$ and a gas recycling rate
  from stellar remnants computed from the CB11 stellar population
  synthesis models for a exponentially rising star formation history
  with $\tau_{\rm r}\simeq 400$\,Myr.  The hashed red region
  denotes the inferred baryonic accretion rate (dark matter accretion
  times the baryonic fraction of $\Omega_{\rm b}/\Omega_{\rm
    m}=0.165$) for
  halos with $\log[M_{\rm h}/{\rm M}_{\odot}]\simeq 12.0-12.2$ at
  $z=2.3$.}
\label{fig:inrate}
\end{figure}

So far, we have not distinguished between gas inflows and outflows,
and have only referred to the net cold gas accretion rate as inferred
from the specific SFRs of $z\ga 2$ galaxies.  We can estimate the cold
gas inflow rate by making some assumptions of the gas outflow rate.
Taking the net cold gas accretion rate as the difference between
inflow ($\dot{M}_{\rm in}$) and outflow rates ($\dot{M}_{\rm out}$),
and assuming that $\dot{M}_{\rm out} = \eta_{\rm out}\Psi$, where
$\eta_{\rm out}$ is commonly referred to as the ``mass-loading
factor,''\,\footnote{Hydrodynamical simulations suggest that under the
  assumption of momentum-driven winds, the instantaneous mass outflow
  rate $\eta_{\rm inst}\propto v_{\rm c}^{-1}$, where $v_{\rm c}$ is
  the circular velocity, and hence $\eta_{\rm inst}\propto
  M_{\ast}^{-1/3}$ (e.g., \citealt{finlator08, dave11}).  In this
  case, $\eta_{\rm out}$ could be thought of as the mass-loading
  factor averaged over the star formation history of the galaxy
  \citep{finlator08}.}  we can rewrite Equation~\ref{eq:gasacc}:
\begin{equation}
\dot{M}_{\rm in} = \dot{M}_{\rm g} + (1+\eta_{\rm out})\Psi - \dot{R}.
\end{equation}
Direct measurements of $\eta_{\rm out}$ in high-redshift galaxies are
sparse, given the difficulty of measuring the covering fraction and
terminal velocity of the gas and its radial extent, as well as the HI
column density (due to saturation of the typical absorption lines used
to make such measurements).  \citet{pettini02} estimated the mass
outflow rate of the low-ionization and HI gas of the lensed Lyman
Break galaxy cB58, finding $\eta_{\rm out}\simeq 1.75$.  This is
likely a lower limit because a significant fraction of the outflowing
gas may be at higher velocity and/or in a more highly ionized state.
Alternatively, \citet{erb08} finds that $\eta_{\rm out}\simeq 1$ is
required to reproduce the mass-metallicity relation at $z\sim 2$.
From the theoretical standpoint, hydrodynamical simulations find mean
mass-loading factors of at least a factor of a few for galaxies with
stellar masses at the low end of those found in our sample
($M_{\ast}\sim 10^{9}$\,M$_{\odot}$), and being less than unity for
the most massive galaxies \citep{oppenheimer06, oppenheimer08,
  finlator08, oppenheimer11, dave11}.  To encompass the range of
observed and expected mass outflow rates, we consider the effect of
changing $\eta_{\rm out}$ on the cold gas inflow rate inferred for a
typical galaxy in our sample with $M^{z=2.3}_{\ast}\simeq 1.6\times
10^{10}$\,M$_{\odot}$ , as shown in Figure~\ref{fig:inrate}.

Figure~\ref{fig:inrate} suggests that the total baryonic accretion
rate must be quite large relative to the SFR if $\eta_{\rm out}$ is at
least a factor of a few.  It is useful to compare these inferred
inflow rates to the {\em total} gas accretion expected for the dark
matter halos that host galaxies in our sample.  Based on clustering
analysis, the correlation length of galaxies in our spectroscopic
sample (with luminosities around $L^{\ast}$ of the UV luminosity
function; \citealt{reddy09}) is $6.5\pm0.5$\,$h^{-1}$\,Mpc (comoving),
which corresponds to dark matter halos with masses $\log [M_{\rm
    h}/{\rm M}_{\odot}]>11.8$, with an average halo mass of $\langle
\log [M_{\rm h}/{\rm M}_{\odot}]\rangle = 12.2$ and median halo mass
of $\log [M_{\rm h}/{\rm M}_{\odot}] = 12.0$ (Trainor \& Steidel, in
preparation).  The total baryonic accretion rate, which is a
generic and well-determined quantity from $\Lambda$CDM cosmological
simulations, can be approximated as
\begin{equation}
\dot{M}_{\rm b}\simeq 6.6 \left[\frac{M_{\rm h}}{10^{12}\,{\rm M}_{\odot}}\right]^{1.15} (1+z)^{2.25} f_{0.165}\,\,\,{\rm M}_{\odot}\,{\rm yr}^{-1}
\label{eq:baryonacc}
\end{equation}
\citep{dekel09}, where $f_{0.165}$ is the fraction of baryons to total
matter (baryons plus dark matter) in units of the cosmological value,
$\Omega_{\rm b}/\Omega_{\rm m} = 0.165\pm 0.009$
\citep{komatsu09}.\footnote{The fraction, $\epsilon$, of the baryonic
  accretion rate that is in a cold state ($T\la 10^{4}$\,K) is model
  dependent and highly uncertain, with suggested values of
  $\epsilon\simeq 0.4-0.8$ (e.g., \citealt{bouche10, faucher11}).
  Adopting these typical values would lower the total cold baryonic
  accretion rate by a factor of $\la 2$ and does not change any of the
  subsequent conclusions.}  Employing Equation~\ref{eq:baryonacc}, we
show in Figure~\ref{fig:inrate} the total baryonic accretion rate (red
shaded region) based on hydrodynamical simulations for the halos that
are inferred to host galaxies in our sample.

The expected baryonic accretion rates from the simulations are
relatively constant with redshift, owing to the fact that the larger
accretion rates expected for higher mass halos at lower redshift are
counteracted by cosmological expansion (which tends to decrease the
accretion rate at a given halo mass).  At $z\sim 2-3$, larger values
of the mass-loading factor ($\eta \ga 1$) yield a total cold gas
inflow rate (as inferred from the specific SFR) that is similar to the
baryonic gas accretion expected for halos with $\log [M_{\rm h}/{\rm
    M}_{\odot}]\simeq 12.0-12.2$.  Strikingly, however, at $z\ga 3.5$
there is a large discrepancy between the baryonic gas accretion rate
expected for the halos and the cold gas accretion rate inferred from
the specific SFR with reasonable values of $\eta$ (cyan and orange
lines in Figure~\ref{fig:inrate}).  For instance, there is a factor of
$\sim 40$ difference between the two expectations at $z\sim 6$.

While the discrepancies at $z\ga 3.5$ could be resolved by adopting a
mean mass-loading factor that is redshift dependent and substantially
larger than $\eta=4.0$, several other possibilities have been
suggested in the literature.  At face value, the hydrodynamical
simulations suggest that the SFR for a galaxy in a given halo should
be roughly constant with time if the fraction of accreting baryons
that turns into stars is roughly constant with redshift.  Could the
galaxies in our sample truly be forming stars at a constant rate?  As
argued in Section~\ref{sec:csfexpectation}, this is unlikely given the
fact that we do not observe galaxies at the higher redshift end of our
selection function with specific SFRs that are substantially larger
than the median specific SFR.  On the other hand, the specific SFRs
measured at $z\ga 4$ are still subject to several systematic
uncertainties, including the overestimation of stellar mass due to
emission line contribution to the broadband near-IR luminosity (e.g.,
\citealt{schaerer10}).  A star formation history that rises more
slowly than an exponential (i.e., as inferred from mean specific SFRs
that increase towards higher redshift) can bring the inferences from
the specific SFR closer to the expectations from the simulations.
Alternatively, an IMF that becomes progressively more bottom-light
towards higher redshift could also lead to elevated specific SFRs at
early times, though such IMF evolution would result in an integrated
SFR density that disagrees with estimates of the stellar mass density
at $z\ga 4$ \citep{reddy09}.

From the theoretical side, \citet{krumholz11} suggest that star
formation at $z\ga 2$ may be suppressed relative to the baryonic
accretion rate due to limited free-fall time and a lack of metals in
halos of mass $M_{\rm h}\la 10^{11}$\,M$_{\odot}$.  This suppression
results in a departure between the gas accretion rate and SFR as
predicted from the \citet{kennicutt98} relation.  However, we note
that the redshift and mass dependence of the baryonic accretion rate
implies that the halo of an $L^{\ast}$ galaxy at $z\sim 2.3$ with
$M_{\rm h}\approx 10^{12}$\,M$_{\odot}$, would have had $M_{\rm
  h}\simeq 1.6\times 10^{11}$\,M$_{\odot}$ at $z\sim 6$, on average.
This mass lies just at the threshold of where metallicity effects are
inferred (from the simulations) to suppress star formation
\citep{krumholz11}.  Yet, even for halos as massive as $M_{\rm
  h}\approx 10^{11}$\,M$_{\odot}$ at $z\sim 6$, we infer gas inflow
rates that are a factor of $40-200$ times smaller than those inferred
from the simulations (Figure~\ref{fig:inrate}).\footnote{It is
  possible some fraction of the $M_{\rm h}\approx
  10^{12}$\,M$_{\odot}$ halos at $z\sim 2.3$ had progenitor halos that
  are less massive than $M_{\rm h}\sim 10^{11}$\,M$_{\odot}$ at $z\sim
  6$, where metallicity effects would be noticeable, if such low mass
  halos merge in a way that the time-averaged baryonic accretion rate
  evolution occurs more rapidly than that assumed by
  Equation~\ref{eq:baryonacc}.}

Another possibility suggested by \citep{krumholz11} is that the bulk
of the stellar mass growth at high redshift occurs from the accretion
of stars formed in other galaxies, in which case the observed specific
SFRs would be biased lower than the intrinsic specific SFR derived by
considering only {\em in situ} star formation.  Nonetheless, the
actual SFR of galaxies within such halos apparently constitutes only a
small fraction of the baryons accreted onto those halos.  It is
possible that some combination of the aforementioned effects may
result in the very efficient suppression of star formation in even
relatively massive halos ($M_{\rm h}\approx 10^{11}$\,M$_{\odot}$) at
$z\ga 2$, but we note that such mechanisms must become increasingly
more important with increasing redshift.  Irrespective of the exact
mechanism, what is perhaps quite striking from Figure~\ref{fig:inrate}
is that star formation in general may have been an extremely
inefficient process even at early cosmic times when galaxies were
first assembling.

\begin{figure}[tbp]
\plotone{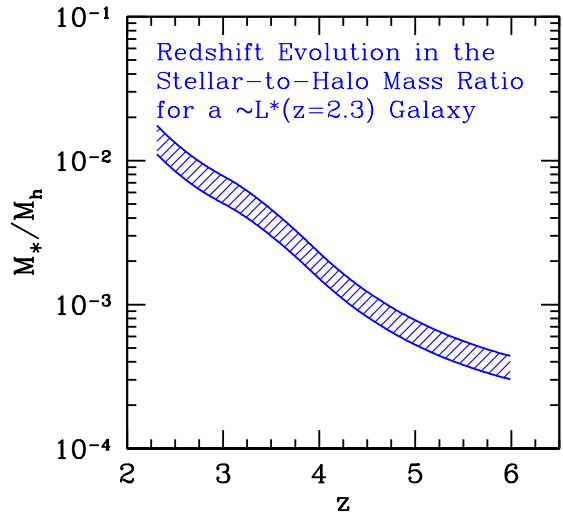}
\caption{Stellar-to-halo mass ratio as a function of redshift for a
  typical star-forming galaxy with a stellar mass at $z=2.3$ of
  $M_{\ast}^{z=2.3}=1.6\times 10^{10}$\,M$_{\odot}$.  Clustering
  results are used to estimate the corresponding average and median
  dark matter halo masses (Trainor \& Steidel, in preparation).  The
  stellar masses are evolved backwards in time using the same
  assumptions as in Figure~\ref{fig:gasevol}.  The median and average
  halo masses are evolved to higher redshift assuming the baryonic
  rate accretion formula specified by Equation~\ref{eq:baryonacc}.}
\label{fig:smhm}
\end{figure}

The relative inefficiency of star formation at early times can be
envisioned also by examining the stellar-to-halo mass ratio
($M_{\ast}/M_{\rm h}$), as shown in Figure~\ref{fig:smhm}.  At
$z=2.3$, the mean stellar mass of galaxies in our sample, and their
median and average halo masses as inferred from clustering, imply
$M_{\ast}/M_{\rm h}\simeq 0.10-0.16$.  Evolving the stellar masses
backwards in time assuming the typical star formation history (e.g.,
as in Figure~\ref{fig:gasevol}), and evolving the halo masses toward
higher redshift assuming the baryonic accretion rate formula
(Equation~\ref{eq:baryonacc}), then implies a stellar-to-halo mass
ratio that is an order of magnitude smaller at $z\sim 4.5$ than at
$z=2.3$, and a factor of $40$ lower at $z\sim 6$ than at $z=2.3$.  The
results imply again that on average a very small fraction the baryons
accreted onto halos at early times actually ends up forming the
stellar mass of a galaxy at similar epochs.\footnote{We have been
  careful to emphasize that only a small fraction of baryons accreted
  in the early history of a typical galaxy contributes to star
  formation {\em at those epochs}.  This does not necessarily imply
  that such baryons are forever precluded from forming stars at some
  later epoch (e.g., at $z\la 2$), particularly if the gas is
  continually cycled into and out of galaxies.}  These results agree
qualitatively with the lower $M_{\ast}/M_{\rm h}$ expected at higher
redshifts, and for lower mass halos at a given redshift, from $N$-body
simulations (e.g., \citealt{moster10}).

\subsection{Summary and Implications}

In this section, we have discussed the typical star formation
histories of high-redshift galaxies based on (a) an analysis of the
median specific SFR as a function of stellar mass at $z\sim 2$,
combined with (b) the observation that the specific SFR does not
evolve strongly with redshift above $z\sim 4$, at least for galaxies
for which such measurements have been obtained.  These two pieces of
evidence lead us to the conclusion that on average the SFRs of $z\ga
2$ galaxies are likely rising monotonically with time.  In this case,
the ``age'' of the galaxy is an ill-defined quantity and in principle
could be arbitrarily old depending on the normalization of the star
formation history.  As such, it is possible that a substantial
fraction of the UV faint sub-$L^{\ast}$ galaxies at $z\ga 4$ evolve
into UV-bright galaxies (i.e., ones that we would spectroscopically
confirm) at $z\sim 2$.  Studies of the clustering properties of
sub-L$^\ast$ galaxies combined with halo abundance matching will
undoubtedly help elucidate the probable progenitors of $z\sim 2$
galaxies.

We proceed to examine the evolution in gas masses and gas accretion
rates implied by rising star formation histories assuming that the
Kennicutt-Schmidt relation holds at high redshift.  The results imply
that the gas masses will increase with increasing SFR, and that the
gas mass fraction will evolve more slowly than in a scenario in which
galaxies are forming stars at a constant or declining rate.  We find
in general that mean cold gas fraction remains quite high at $\ga
60\%$ at $z\ga 4$.  A further consequence of the slow evolution of gas
mass fraction at $z\ga 4$ is that the average gas phase metallicity
may not evolve with time as strongly at $z\ga 4$ relative to that
observed at $z\la 2$.  This is because newly accreted gas will
continually dilute the overall gas-phase metallicity.  At face value,
the fundamental metallicity relation (FMR) derived in
\citet{mannucci10} implies an $\approx 0.5$\,dex increase in mean
oxygen abundance from $z=7$ to 4.  However this relation has only been
calibrated up to $z\sim 2.5$ and for larger SFRs at higher redshifts,
than the SFRs considered here at $z\ga 4$.  Direct metallicity
measurements for such high-redshift galaxies, while unobtainable at
the present time, will prove crucial in connecting the metallicity
evolution with that expected from the average increasing star
formation rate between $z=7$ and 2.

At these early times, the {\em net} gas accretion rate exceeds the SFR
by at least a factor of two, where the former is determined by
relating the specific SFR to the cold gas mass via the
\citet{kennicutt98} relation.  For the progenitors of typical
star-forming galaxies in our $z\sim 2$ sample, the SFR eventually
equals the net gas accretion rate around $z\sim 2-3$, at which point
the gas fraction decreases below $50\%$.  These results are consistent
with measurements of the gas mass fractions of high-redshift galaxies,
and the fact that such galaxies must be accreting substantial amounts
of gas in order to sustain their SFRs.

We proceed to compare the gas accretion rates inferred from the
specific SFR---combined with some assumption for the mass-loading
factor of outflowing winds---with the accretion rates expected for the
dark matter halos that are inferred from clustering to host galaxies
in our sample.  We find that the two methods of inferring gas
accretion rates agree at $z\la 3$, but they become increasingly
divergent above $z\ga 4$, such that less than one percent of the
baryons accreted by a $\log[M_{\rm h}/{\rm M}_{\odot}]=11.2$ halo at
$z\sim 6$ would end up in stars at that epoch.  While several
possibilities are discussed to explain this discrepancy, the results
imply that star formation must have been extremely inefficient at
early cosmic times.

\section{CONCLUSIONS}

We use a large spectroscopic sample of $L^{\ast}$ galaxies at
redshifts $1.4\la z\la 3.7$, and a photometrically selected sample of
fainter galaxies at the same redshifts, to constrain the average star
formation history of typical star-forming galaxies between redshift
$z=7$ and $z=2$.  Our analysis takes advantage of rest-frame UV
spectroscopy, and optical, ground-based near-IR, {\em Hubble Space
  Telescope} WFC3/IR, and {\em Spitzer Space Telescope} IRAC and MIPS
$24$\,$\mu$m imaging, in order to measure the stellar populations and
dust obscured star formation of galaxies in our sample.  We have taken
into account a number of systematic effects in computing these
quantities, including the contribution of line emission to the
broadband photometry, dynamical time constraints on galaxy ages, and
the stellar population dependence of the UV attenuation curve.  

With these data, we perform detailed comparisons between SED-inferred
SFRs (SFRs[SED]) and those computed by combining deep UV and MIPS
$24$\,$\mu$m data (SFRs[IR+UV]), for 302 spectroscopically confirmed
galaxies at redshifts $1.5\la z\la 2.6$, where the MIPS $24$\,$\mu$m
observations are sensitive to the dust emission features at rest-frame
$8$\,$\mu$m; such comparisons allow us to rule out some simple star
formation histories.  The uncertainties in stellar masses and ages are
discussed.  We proceed to examine the correlation between SFR[IR+UV]
and stellar mass, as well as SFR[SED] and stellar mass for a larger
sample of 1959 spectroscopically confirmed galaxies, taking into
account several systematic effects including Malmquist bias and
differences in the conversion from near-IR luminosity to stellar mass.
These results are extended to lower SFRs and stellar masses via a
photometrically selected sample of 563 UV faint ($\rs>25.5$) galaxies
over the same redshifts.  This information is used to constrain the
average star formation history of high-redshift galaxies, which in
turn yields information on the gas mass evolution and gas mass
accretion rates for galaxies of different stellar masses.

The main conclusions are:

\begin{itemize}

\item{SED modeling that assumes exponentially declining star formation
  histories results in SFRs[SED] that are on average a factor of
  $5-10\times$ lower than SFRs[IR+UV].  On the other hand, modeling
  typical star-forming galaxies ($\lbol\la 10^{12}$\,L$_{\odot}$; ages
  $\la 100$\,Myr) with constant or rising star formation histories and
  a \citet{calzetti00} attenuation curve results in SFRs[SED] that are
  in good agreement with SFRs[IR+UV].  This suggests that most
  $L^{\ast}$ galaxies at $z\sim 2-3$ did not follow a declining star
  formation history prior to the epoch at which they are observed.
  The traditional use of exponentially declining models to describe
  the star formation histories of high-redshift galaxies is called
  into question by the analysis presented in this paper.}

\item{Assuming standard modeling assumptions (\citealt{calzetti00}
  attenuation curve and no age constraint) results in SFRs[SED] that
  are systematically larger than SFRs[IR+UV] for galaxies inferred to
  have younger stellar populations ($\la 100$\,Myr).  However, by
  building a coherent picture for the dust attenuation (SMC-like) and
  dynamical timescale of younger galaxies, we are able to reconcile
  the SFRs[SED] and SFRs[IR+UV] for such galaxies.}

\item{Rising star formation histories yield stellar masses that are
  comparable to (and ages that are typically older than) those
  obtained for a constant star formation history, owing to the slower
  evolution in $M/L$ ratio for rising histories.  More generally, the
  ``age'' of a galaxy is an ill-defined quantity for rising histories,
  as the observed $M/L$ ratio can be obtained by simultaneously
  varying the age of a galaxy and the normalization of its star
  formation history.}

\item{The scatter in the $M/L$ ratios of galaxies at a given
  luminosity decreases at longer wavelengths.  Nonetheless, the
  scatter, which is induced partly by the current star formation, is
  still relatively large ($0.20$\,dex) at wavelengths where the
  stellar emission peaks ($\ga 1$\,$\mu$m).  The luminosity dependence
  of the $M/L$ ratio at $1.1$\,$\mu$m is partly a result of sampling a
  finite range of UV luminosity in a flux limited sample like ours,
  combined with the effects of dust obscuration.  The systematic
  variation of the stellar mass versus near-IR magnitude relation with
  UV luminosity can result in biased measurements of the masses of
  UV-faint galaxies.}

\item{Taking into account IRAC upper limits and the aforementioned UV
  luminosity dependence of the conversion between near-IR magnitude
  and stellar mass, we find a strong correlation between UV luminosity
  and stellar mass, even within the limited dynamic range of our
  spectroscopic sample.  An IRAC stacking of photometrically selected
  UV faint galaxies to $\rs \sim 27.0$ shows that this trend extends
  to UV faint galaxies.  In particular, $0.1L^{\ast}$ galaxies at
  $z\sim 2$ have stellar masses that are a factor of $\approx
  140\times$ smaller than those of the UV brightest ($\sim 5L^{\ast}$)
  galaxies in our sample.}

\item{Combining the UV-faint data with our spectroscopic sample, and
  taking into account systematics in the calculation of stellar
  masses, we find a close to unity slope of the SFR-$M_{\ast}$
  relation over $2.5$ orders of magnitude in stellar mass and $2.8$
  orders of magnitude in SFR at $z\sim 2$.  This relation implies that
  the specific SFR (SFR/$M_{\ast}$) is roughly constant for stellar
  masses $M_{\ast}\ga 5\times 10^{8}$\,M$_{\odot}$ and SFRs $\ga
  2$\,M$_{\odot}$\,yr$^{-1}$.}

\item{The constant specific SFR with mass at $z\sim 2$, combined with
  the fact that this median SFR is similar to that found at $z\ga 4$,
  implies that the SFRs must be increasing with (and roughly
  proportional to) the stellar mass.  A constant star formation
  history would imply much larger specific SFRs at slightly higher
  redshifts, and such objects are not observed at those higher
  redshifts that are probed within our sample.  If galaxies on average
  have SFRs that increase with cosmic time, then it suggests that the
  UV-bright galaxies observed at $z\sim 2-3$ had progenitors that
  would be detected as sub$-L^{\ast}$ galaxies in $z\ga 4$ dropout
  samples.}

\item{If we assume that the Kennicutt-Schmidt relation
  \citet{kennicutt98} holds at high redshift, then an average rising
  star formation history implies a gas mass that increases with time
  and a gas mass fraction that evolves slowly at $z\ga 4$.  The
  specific SFRs at $z\ga 2$ indicate a net cold gas accretion rate
  that outpaces the SFR at high redshift, until $z\sim 2-3$, when the
  SFR becomes comparable to the net cold gas accretion rate for
  typical star-forming galaxies.  We have inferred the cold gas inflow
  rate from the net cold gas accretion rate with some assumption for
  the mass-loading factor of galactic outflows.  The inflow rates
  derived in this manner diverge increasingly from the total gas
  inflow rates expected for the dark matter halos that host our
  galaxies (as inferred from clustering analysis; Trainor \& Steidel,
  in prep.).  At face value, the observations imply that less than one
  percent of the baryons accreted onto halos of mass $M_{\rm h}\simeq
  1.6\times 10^{11}$\,M$_{\odot}$ at $z\sim 6$ end up contributing to
  star formation at that epoch, thus highlighting the inefficiency of
  star formation even at early cosmic times when galaxies were growing
  rapidly.}

\end{itemize}

There are two important points to make.  First, taken together, the
results of the SFR comparisons (Section~\ref{sec:sfrcomparison}), the
near-unity slope of the SFR-$M^{\ast}$ relation, and the roughly
similar specific SFRs at $z\ga 2$, point to a picture in which the
average SFRs of high-redshift galaxies increase with cosmic time.
This behavior produces naturally the observed rise in the SFR density
between $z=7$ and $z=4$ \citep{papovich11}.  Second, an analysis of
the specific SFRs indicate that these rising histories cannot be
sustained indefinitely, and once the SFR exceeds the gas accretion
rate, we expect the gas mass to decrease leading to an eventual
decrease in SFR.  However, our analysis underscores the fact that the
net cold gas accretion inferred from the specific SFRs is only a small
constituent of the total baryons accreted onto halos at $z\ga 2$, and
that in general star formation must have been very inefficient at
early times.

We conclude by noting several future avenues of investigation to
confirm more robustly our findings.  We have emphasized a few of the
biases (e.g., Malmquist bias and systematics in the conversion between
near-IR luminosity and stellar mass) that are important for analyses
of the SFR-$M^{\ast}$ (and specific SFR versus $M^{\ast}$) relations.
The presence of a Malmquist bias, however, does not preclude the
possibility of there being a real evolution in the SFR-$M^{\ast}$
relation at faint luminosities and low stellar masses.  Hence,
significant progress in evaluating possible evolution can be achieved
with deep near-IR photometry (e.g., with {\em Hubble Space
  Telescope}'s WFC3/IR Camera) that in turn enables SED-fitting for
individual UV-faint and low mass galaxies at high redshift.  As
emphasized above, there is a substantial dispersion in the star
formation histories of individual galaxies; future deeper studies will
enable us to quantify how the dispersion in SFRs may change as a
function of $M^{\ast}$, particularly for UV-faint galaxies at $z\ga 3$
which, from clustering studies, may exhibit a more episodic star
formation history \citep{lee11b}.  Further, as recently highlighted in
\citet{schaerer09, schaerer10}, the expected high equivalent width
emission lines (e.g., H$\alpha$) of $z\ga 4$ galaxies can result in
factors of $2-3\times$ overestimates of their stellar masses from
broadband SED fitting.  Correcting for the presence of those emission
lines may potentially result in a stronger evolution of specific SFR
than has been found in other studies of $z\ga 4$ galaxies
\citep{stark09, gonzalez10, bouwens11b}, and may partially resolve the
discrepancies between current observations and numerical simulations
regarding the evolution of specific SFRs at $z\ga 4$ (e.g.,
\citealt{bouche10, dutton10, dave11, weinmann11, krumholz11}; see
discussion in \citealt{bouwens11b}).  The forthcoming deep near-IR
imaging surveys, along with deep imaging and spectroscopic campaigns
with the next generation of ground-based (TMT and GMT) and space-based
facilities ({\em JWST}), promise substantial progress in understanding
star formation and the buildup of stellar mass at early cosmic times.

\acknowledgements

N.A.R. thanks Romeel Dav\'{e}, Claude-Andr$\acute{\rm e}$
Faucher-Gigu$\grave{\rm e}$re, and Kristian Finlator for useful
discussions and comments.  We acknowledge the referee for useful
suggestions that improved the clarity of the manuscript.  We thank the
staff of the Keck and Palomar Observatories for their help in
obtaining the data presented here.  Support for N.A.R. was provided by
NASA through Hubble Fellowship grant HST-HF-01223.01 awarded by the
Space Telescope Science Institute, which is operated by the
Association of Universities for Research in Astronomy, Inc., for NASA,
under contract NAS 5-26555.  Additional support has been provided by
research funding for the {\em Spitzer Space Telescope} Legacy Science
Program, provided by NASA through contracts 1224666 and 1287778,
issued by the Jet Propulsion Laboratory, California Institute of
Technology.  N.A.R. also acknowledges the Beatrice Tinsley Visiting
Scholar program administered by the Astronomy Department at the
University of Texas at Austin, where part of this research was
conducted.  C.C.S. has been supported by NSF grants AST-0606912 and
AST-0908805, with additional support from the John D. and Catherine T.
MacArthur Foundation and the Peter and Patricia Gruber Foundation.


\begin{thebibliography}{100}
\expandafter\ifx\csname natexlab\endcsname\relax\def\natexlab#1{#1}\fi

\bibitem[{{Adelberger} {et~al.}(1998){Adelberger}, {Steidel},
    {Giavalisco}, {Dickinson}, {Pettini}, \& {Kellogg}}]{adelberger98}
  {Adelberger}, K.~L., {Steidel}, C.~C., {Giavalisco}, M.,
  et~al. 1998, \apj, 505, 18

\bibitem[{{Adelberger} {et~al.}(2004){Adelberger}, {Steidel}, {Shapley},
  {Hunt}, {Erb}, {Reddy}, \& {Pettini}}]{adelberger04}
{Adelberger}, K.~L., {Steidel}, C.~C., {Shapley}, A.~E., et~al. 2004, \apj, 607, 226

\bibitem[{{Bell} \& {de Jong}(2000)}]{bell00}
{Bell}, E.~F. \& {de Jong}, R.~S. 2000, \mnras, 312, 497

\bibitem[{{Bertin} \& {Arnouts}(1996)}]{bertin96}
{Bertin}, E. \& {Arnouts}, S. 1996, \aaps, 117, 393

\bibitem[{{Bouch{\'e}} {et~al.}(2010){Bouch{\'e}}, {Dekel}, {Genzel}, {Genel},
  {Cresci}, {F{\"o}rster Schreiber}, {Shapiro}, {Davies}, \&
  {Tacconi}}]{bouche10}
{Bouch{\'e}}, N., {Dekel}, A., {Genzel}, R., et~al. 2010, \apj, 718, 1001

\bibitem[{{Bouwens} {et~al.}(2009){Bouwens}, {Illingworth}, {Franx}, {Chary},
  {Meurer}, {Conselice}, {Ford}, {Giavalisco}, \& {van Dokkum}}]{bouwens09}
{Bouwens}, R.~J., {Illingworth}, G.~D., {Franx}, M., et~al. 2009, \apj, 705, 936

\bibitem[{{Bouwens} {et~al.}(2011{\natexlab{a}}){Bouwens},
    {Illingworth}, {Oesch}, {Franx}, {Labbe}, {Trenti}, {van Dokkum},
    {Carollo}, {Gonzalez}, \& {Magee}}]{bouwens11b} {Bouwens}, R.~J.,
  {Illingworth}, G.~D., {Oesch}, P.~A., et~al. 2011{\natexlab{a}},
  ArXiv e-prints

\bibitem[{{Bouwens} {et~al.}(2011{\natexlab{b}}){Bouwens}, {Illingworth},
  {Oesch}, {Labb{\'e}}, {Trenti}, {van Dokkum}, {Franx}, {Stiavelli},
  {Carollo}, {Magee}, \& {Gonzalez}}]{bouwens11a}
{Bouwens}, R.~J., {Illingworth}, G.~D., {Oesch}, P.~A., et~al. 2011{\natexlab{b}}, \apj, 737, 90

\bibitem[{{Brinchmann} {et~al.}(2004){Brinchmann}, {Charlot}, {White},
  {Tremonti}, {Kauffmann}, {Heckman}, \& {Brinkmann}}]{brinchmann04}
{Brinchmann}, J., {Charlot}, S., {White}, S.~D.~M., et~al. 2004, \mnras, 351, 1151

\bibitem[{{Bruzual} \& {Charlot}(2003)}]{bruzual03}
{Bruzual}, G. \& {Charlot}, S. 2003, \mnras, 344, 1000

\bibitem[{{Calzetti} {et~al.}(2000){Calzetti}, {Armus}, {Bohlin}, {Kinney},
  {Koornneef}, \& {Storchi-Bergmann}}]{calzetti00}
{Calzetti}, D., {Armus}, L., {Bohlin}, R.~C., et~al. 2000, \apj, 533, 682

\bibitem[{{Conroy} {et~al.}(2008){Conroy}, {Shapley}, {Tinker}, {Santos}, \&
  {Lemson}}]{conroy08}
{Conroy}, C., {Shapley}, A.~E., {Tinker}, J.~L., {Santos}, M.~R., \& {Lemson},
  G. 2008, \apj, 679, 1192

\bibitem[{{Daddi} {et~al.}(2010){Daddi}, {Bournaud}, {Walter}, {Dannerbauer},
  {Carilli}, {Dickinson}, {Elbaz}, {Morrison}, {Riechers}, {Onodera}, {Salmi},
  {Krips}, \& {Stern}}]{daddi10}
{Daddi}, E., {Bournaud}, F., {Walter}, F., et~al. 2010, \apj, 713, 686

\bibitem[{{Daddi} {et~al.}(2007){Daddi}, {Dickinson}, {Morrison}, {Chary},
  {Cimatti}, {Elbaz}, {Frayer}, {Renzini}, {Pope}, {Alexander}, {Bauer},
  {Giavalisco}, {Huynh}, {Kurk}, \& {Mignoli}}]{daddi07a}
{Daddi}, E., {Dickinson}, M., {Morrison}, G., et~al. 2007, \apj, 670, 156

\bibitem[{{Dav{\'e}} {et~al.}(2011){Dav{\'e}}, {Oppenheimer}, \&
  {Finlator}}]{dave11}
{Dav{\'e}}, R., {Oppenheimer}, B.~D., \& {Finlator}, K. 2011, \mnras, 415, 11

\bibitem[{{Dekel} {et~al.}(2009){Dekel}, {Birnboim}, {Engel}, {Freundlich},
  {Goerdt}, {Mumcuoglu}, {Neistein}, {Pichon}, {Teyssier}, \&
  {Zinger}}]{dekel09}
{Dekel}, A., {Birnboim}, Y., {Engel}, G., et~al. 2009, \nat, 457, 451

\bibitem[{{Dutton} {et~al.}(2010){Dutton}, {van den Bosch}, \&
  {Dekel}}]{dutton10}
{Dutton}, A.~A., {van den Bosch}, F.~C., \& {Dekel}, A. 2010, \mnras, 405, 1690

\bibitem[{{Erb}(2008)}]{erb08}
{Erb}, D.~K. 2008, \apj, 674, 151

\bibitem[{{Erb} {et~al.}(2006{\natexlab{a}}){Erb}, {Steidel}, {Shapley},
  {Pettini}, {Reddy}, \& {Adelberger}}]{erb06c}
{Erb}, D.~K., {Steidel}, C.~C., {Shapley}, A.~E., et~al. 2006{\natexlab{a}}, \apj, 647, 128

\bibitem[{{Erb} {et~al.}(2006{\natexlab{b}}){Erb}, {Steidel}, {Shapley},
  {Pettini}, {Reddy}, \& {Adelberger}}]{erb06b}
---. 2006{\natexlab{b}}, \apj, 646, 107

\bibitem[{{Faucher-Gigu{\`e}re} {et~al.}(2011){Faucher-Gigu{\`e}re}, {Kere{\v
  s}}, \& {Ma}}]{faucher11}
{Faucher-Gigu{\`e}re}, C.-A., {Kere{\v s}}, D., \& {Ma}, C.-P. 2011, \mnras,
  1399

\bibitem[{{Ferguson} {et~al.}(2004){Ferguson}, {Dickinson}, {Giavalisco},
  {Kretchmer}, {Ravindranath}, {Idzi}, {Taylor}, {Conselice}, {Fall},
  {Gardner}, {Livio}, {Madau}, {Moustakas}, {Papovich}, {Somerville},
  {Spinrad}, \& {Stern}}]{ferguson04}
{Ferguson}, H.~C., {Dickinson}, M., {Giavalisco}, M., et~al. 2004,
  \apjl, 600, L107

\bibitem[{{Finlator} \& {Dav{\'e}}(2008)}]{finlator08}
{Finlator}, K. \& {Dav{\'e}}, R. 2008, \mnras, 385, 2181

\bibitem[{{Finlator} {et~al.}(2011){Finlator}, {Oppenheimer}, \&
  {Dav{\'e}}}]{finlator11}
{Finlator}, K., {Oppenheimer}, B.~D., \& {Dav{\'e}}, R. 2011, \mnras, 410, 1703

\bibitem[{{Genzel} {et~al.}(2010){Genzel}, {Tacconi}, {Gracia-Carpio},
  {Sternberg}, {Cooper}, {Shapiro}, {Bolatto}, {Bouch{\'e}}, {Bournaud},
  {Burkert}, {Combes}, {Comerford}, {Cox}, {Davis}, {Schreiber},
  {Garcia-Burillo}, {Lutz}, {Naab}, {Neri}, {Omont}, {Shapley}, \&
  {Weiner}}]{genzel10}
{Genzel}, R., {Tacconi}, L.~J., {Gracia-Carpio}, J., et~al. 2010, \mnras, 407, 2091

\bibitem[{{Goldader} {et~al.}(2002){Goldader}, {Meurer}, {Heckman}, {Seibert},
  {Sanders}, {Calzetti}, \& {Steidel}}]{goldader02}
{Goldader}, J.~D., {Meurer}, G., {Heckman}, T.~M., et~al. 2002, \apj, 568, 651

\bibitem[{{Gonzalez} {et~al.}(2011){Gonzalez}, {Bouwens}, {Labbe},
  {Illingworth}, {Oesch}, {Franx}, \& {Magee}}]{gonzalez11}
{Gonzalez}, V., {Bouwens}, R., {Labbe}, I., et~al. 2011, ArXiv e-prints

\bibitem[{{Gonz{\'a}lez} {et~al.}(2010){Gonz{\'a}lez}, {Labb{\'e}}, {Bouwens},
  {Illingworth}, {Franx}, {Kriek}, \& {Brammer}}]{gonzalez10}
{Gonz{\'a}lez}, V., {Labb{\'e}}, I., {Bouwens}, R.~J., et~al. 2010, \apj, 713, 115

\bibitem[{{Gordon} {et~al.}(2001){Gordon}, {Misselt}, {Witt}, \&
  {Clayton}}]{gordon01}
{Gordon}, K.~D., {Misselt}, K.~A., {Witt}, A.~N., \& {Clayton}, G.~C. 2001,
  \apj, 551, 269

\bibitem[{{Isobe} {et~al.}(1986){Isobe}, {Feigelson}, \& {Nelson}}]{isobe86}
{Isobe}, T., {Feigelson}, E.~D., \& {Nelson}, P.~I. 1986, \apj, 306, 490

\bibitem[{{Kennicutt}(1998)}]{kennicutt98}
{Kennicutt}, R.~C. 1998, \araa, 36, 189

\bibitem[{{Koekemoer} {et~al.}(2002){Koekemoer}, {Fruchter}, {Hook}, \&
  {Hack}}]{koekemoer02}
{Koekemoer}, A.~M., {Fruchter}, A.~S., {Hook}, R.~N., \& {Hack}, W. 2002, in
  The 2002 HST Calibration Workshop : Hubble after the Installation of the ACS
  and the NICMOS Cooling System, ed. {S.~Arribas, A.~Koekemoer, \&
  B.~Whitmore}, 337--+

\bibitem[{{Komatsu} {et~al.}(2009){Komatsu}, {Dunkley}, {Nolta}, {Bennett},
  {Gold}, {Hinshaw}, {Jarosik}, {Larson}, {Limon}, {Page}, {Spergel},
  {Halpern}, {Hill}, {Kogut}, {Meyer}, {Tucker}, {Weiland}, {Wollack}, \&
  {Wright}}]{komatsu09}
{Komatsu}, E., {Dunkley}, J., {Nolta}, M.~R., et~al. 2009,
  \apjs, 180, 330

\bibitem[{{Kriek} {et~al.}(2010){Kriek}, {Labb{\'e}}, {Conroy}, {Whitaker},
  {van Dokkum}, {Brammer}, {Franx}, {Illingworth}, {Marchesini}, {Muzzin},
  {Quadri}, \& {Rudnick}}]{kriek10}
{Kriek}, M., {Labb{\'e}}, I., {Conroy}, C., et~al. 2010, \apjl, 722, L64

\bibitem[{{Krumholz} \& {Dekel}(2011)}]{krumholz11}
{Krumholz}, M.~R. \& {Dekel}, A. 2011, ArXiv e-prints

\bibitem[{{Law} {et~al.}(2009){Law}, {Steidel}, {Erb}, {Larkin}, {Pettini},
  {Shapley}, \& {Wright}}]{law09}
{Law}, D.~R., {Steidel}, C.~C., {Erb}, D.~K., et~al. 2009, \apj, 697, 2057

\bibitem[{{Law} {et~al.}(2012){Law}, {Steidel}, {Shapley}, {Nagy}, {Reddy}, \&
  {Erb}}]{law12}
{Law}, D.~R., {Steidel}, C.~C., {Shapley}, A.~E., et~al. 2012, \apj, 745, 85

\bibitem[{{Lee} {et~al.}(2011{\natexlab{a}}){Lee}, {Dey}, {Reddy}, {Brown},
  {Gonzalez}, {Jannuzi}, {Cooper}, {Fan}, {Bian}, {Glikman}, {Stern},
  {Brodwin}, \& {Cooray}}]{lee11a}
{Lee}, K.-S., {Dey}, A., {Reddy}, N., et~al. 2011{\natexlab{a}}, \apj, 733, 99

\bibitem[{{Lee} {et~al.}(2011{\natexlab{b}}){Lee}, {Ferguson}, {Wiklind},
  {Dahlen}, {Dickinson}, {Giavalisco}, {Grogin}, {Papovich}, {Messias}, {Guo},
  \& {Lin}}]{lee11b}
{Lee}, K.-S., {Ferguson}, H.~C., {Wiklind}, T., et~al. 2011{\natexlab{b}}, ArXiv e-prints

\bibitem[{{Lee} {et~al.}(2009){Lee}, {Giavalisco}, {Conroy}, {Wechsler},
  {Ferguson}, {Somerville}, {Dickinson}, \& {Urry}}]{lee09}
{Lee}, K.-S., {Giavalisco}, M., {Conroy}, C., et~al. 2009, \apj,
  695, 368

\bibitem[{{Madau}(1995)}]{madau95}
{Madau}, P. 1995, \apj, 441, 18

\bibitem[{{Madau} {et~al.}(1998){Madau}, {Pozzetti}, \& {Dickinson}}]{madau98}
{Madau}, P., {Pozzetti}, L., \& {Dickinson}, M. 1998, \apj, 498, 106

\bibitem[{{Magdis} {et~al.}(2010){Magdis}, {Elbaz}, {Daddi}, {Morrison},
  {Dickinson}, {Rigopoulou}, {Gobat}, \& {Hwang}}]{magdis10a}
{Magdis}, G.~E., {Elbaz}, D., {Daddi}, E., et~al. 2010, \apj, 714, 1740

\bibitem[{{Magnelli} {et~al.}(2011){Magnelli}, {Elbaz}, {Chary}, {Dickinson},
  {Le Borgne}, {Frayer}, \& {Willmer}}]{magnelli11}
{Magnelli}, B., {Elbaz}, D., {Chary}, R.~R., et~al. 2011, \aap, 528, A35+

\bibitem[{{Makovoz} \& {Marleau}(2005)}]{makovoz05}
{Makovoz}, D. \& {Marleau}, F.~R. 2005, astro-ph/0507007

\bibitem[{{Mannucci} {et~al.}(2010){Mannucci}, {Cresci}, {Maiolino}, {Marconi},
  \& {Gnerucci}}]{mannucci10}
{Mannucci}, F., {Cresci}, G., {Maiolino}, R., {Marconi}, A., \& {Gnerucci}, A.
  2010, \mnras, 408, 2115

\bibitem[{{Maraston} {et~al.}(2006){Maraston}, {Daddi}, {Renzini}, {Cimatti},
  {Dickinson}, {Papovich}, {Pasquali}, \& {Pirzkal}}]{maraston06}
{Maraston}, C., {Daddi}, E., {Renzini}, A., et~al. 2006, \apj, 652, 85

\bibitem[{{Maraston} {et~al.}(2010){Maraston}, {Pforr}, {Renzini}, {Daddi},
  {Dickinson}, {Cimatti}, \& {Tonini}}]{maraston10}
{Maraston}, C., {Pforr}, J., {Renzini}, A., et~al. 2010, \mnras, 407, 830

\bibitem[{{Marchesini} {et~al.}(2009){Marchesini}, {van Dokkum}, {F{\"o}rster
  Schreiber}, {Franx}, {Labb{\'e}}, \& {Wuyts}}]{marchesini09}
{Marchesini}, D., {van Dokkum}, P.~G., {F{\"o}rster Schreiber}, N.~M., et~al. 2009, \apj, 701, 1765

\bibitem[{{Marigo} \& {Girardi}(2007)}]{marigo07}
{Marigo}, P. \& {Girardi}, L. 2007, \aap, 469, 239

\bibitem[{{Melbourne} {et~al.}(2012){Melbourne}, {Williams}, {Dalcanton},
  {Rosenfield}, {Girardi}, {Marigo}, {Weisz}, {Dolphin}, {Boyer}, {Olsen},
  {Skillman}, \& {Seth}}]{melbourne12}
{Melbourne}, J., {Williams}, B.~F., {Dalcanton}, J.~J., et~al. 2012, \apj, 748, 47

\bibitem[{{Meurer} {et~al.}(1999){Meurer}, {Heckman}, \& {Calzetti}}]{meurer99}
{Meurer}, G.~R., {Heckman}, T.~M., \& {Calzetti}, D. 1999, \apj, 521, 64

\bibitem[{{Misselt} {et~al.}(2001){Misselt}, {Gordon}, {Clayton}, \&
  {Wolff}}]{misselt01}
{Misselt}, K.~A., {Gordon}, K.~D., {Clayton}, G.~C., \& {Wolff}, M.~J. 2001,
  \apj, 551, 277

\bibitem[{{Moster} {et~al.}(2010){Moster}, {Somerville}, {Maulbetsch}, {van den
  Bosch}, {Macci{\`o}}, {Naab}, \& {Oser}}]{moster10}
{Moster}, B.~P., {Somerville}, R.~S., {Maulbetsch}, C., et~al. 2010, \apj, 710, 903

\bibitem[{{Muzzin} {et~al.}(2009){Muzzin}, {Marchesini}, {van Dokkum},
  {Labb{\'e}}, {Kriek}, \& {Franx}}]{muzzin09}
{Muzzin}, A., {Marchesini}, D., {van Dokkum}, P.~G., et~al. 2009, \apj, 701, 1839

\bibitem[{{Nagamine} {et~al.}(2000){Nagamine}, {Cen}, \&
  {Ostriker}}]{nagamine00}
{Nagamine}, K., {Cen}, R., \& {Ostriker}, J.~P. 2000, \apj, 541, 25

\bibitem[{{Nandra} {et~al.}(2002){Nandra}, {Mushotzky}, {Arnaud}, {Steidel},
  {Adelberger}, {Gardner}, {Teplitz}, \& {Windhorst}}]{nandra02}
{Nandra}, K., {Mushotzky}, R.~F., {Arnaud}, K., et~al. 2002, \apj,
  576, 625

\bibitem[{{Noeske} {et~al.}(2007){Noeske}, {Weiner}, {Faber}, {Papovich},
  {Koo}, {Somerville}, {Bundy}, {Conselice}, {Newman}, {Schiminovich}, {Le
  Floc'h}, {Coil}, {Rieke}, {Lotz}, {Primack}, {Barmby}, {Cooper}, {Davis},
  {Ellis}, {Fazio}, {Guhathakurta}, {Huang}, {Kassin}, {Martin}, {Phillips},
  {Rich}, {Small}, {Willmer}, \& {Wilson}}]{noeske07}
{Noeske}, K.~G., {Weiner}, B.~J., {Faber}, S.~M., et~al. 2007, \apjl, 660, L43

\bibitem[{{Noll} {et~al.}(2009){Noll}, {Burgarella}, {Giovannoli}, {Buat},
  {Marcillac}, \& {Mu{\~n}oz-Mateos}}]{noll09}
{Noll}, S., {Burgarella}, D., {Giovannoli}, E., et~al. 2009, \aap, 507, 1793

\bibitem[{{Oesch} {et~al.}(2010){Oesch}, {Bouwens}, {Carollo}, {Illingworth},
  {Trenti}, {Stiavelli}, {Magee}, {Labb{\'e}}, \& {Franx}}]{oesch10}
{Oesch}, P.~A., {Bouwens}, R.~J., {Carollo}, C.~M., et~al. 2010, \apjl, 709, L21

\bibitem[{{Oke} {et~al.}(1995){Oke}, {Cohen}, {Carr}, {Cromer}, {Dingizian},
  {Harris}, {Labrecque}, {Lucinio}, {Schaal}, {Epps}, \& {Miller}}]{oke95}
{Oke}, J.~B., {Cohen}, J.~G., {Carr}, M., et~al. 1995, \pasp, 107, 375

\bibitem[{{Oppenheimer} \& {Dav{\'e}}(2006)}]{oppenheimer06}
{Oppenheimer}, B.~D. \& {Dav{\'e}}, R. 2006, \mnras, 373, 1265

\bibitem[{{Oppenheimer} \& {Dav{\'e}}(2008)}]{oppenheimer08}
---. 2008, \mnras, 642

\bibitem[{{Oppenheimer} {et~al.}(2011){Oppenheimer}, {Dav{\'e}}, {Katz},
  {Kollmeier}, \& {Weinberg}}]{oppenheimer11}
{Oppenheimer}, B.~D., {Dav{\'e}}, R., {Katz}, N., {Kollmeier}, J.~A., \&
  {Weinberg}, D.~H. 2011, \mnras, 2017

\bibitem[{{Pannella} {et~al.}(2009){Pannella}, {Carilli}, {Daddi}, {McCracken},
  {Owen}, {Renzini}, {Strazzullo}, {Civano}, {Koekemoer}, {Schinnerer},
  {Scoville}, {Smol{\v c}i{\'c}}, {Taniguchi}, {Aussel}, {Kneib}, {Ilbert},
  {Mellier}, {Salvato}, {Thompson}, \& {Willott}}]{pannella09}
{Pannella}, M., {Carilli}, C.~L., {Daddi}, E., et~al. 2009, \apjl, 698, L116

\bibitem[{{Papovich} {et~al.}(2001){Papovich}, {Dickinson}, \&
  {Ferguson}}]{papovich01}
{Papovich}, C., {Dickinson}, M., \& {Ferguson}, H.~C. 2001, \apj, 559, 620

\bibitem[{{Papovich} {et~al.}(2011){Papovich}, {Finkelstein}, {Ferguson},
  {Lotz}, \& {Giavalisco}}]{papovich11}
{Papovich}, C., {Finkelstein}, S.~L., {Ferguson}, H.~C., {Lotz}, J.~M., \&
  {Giavalisco}, M. 2011, \mnras, 412, 1123

\bibitem[{{Papovich} {et~al.}(2006){Papovich}, {Moustakas}, {Dickinson}, {Le
  Floc'h}, {Rieke}, {Daddi}, {Alexander}, {Bauer}, {Brandt}, {Dahlen}, {Egami},
  {Eisenhardt}, {Elbaz}, {Ferguson}, {Giavalisco}, {Lucas}, {Mobasher},
  {P{\'e}rez-Gonz{\'a}lez}, {Stutz}, {Rieke}, \& {Yan}}]{papovich06}
{Papovich}, C., {Moustakas}, L.~A., {Dickinson}, M., et~al. 2006, \apj, 640, 92

\bibitem[{{Pettini} {et~al.}(2002){Pettini}, {Rix}, {Steidel}, {Adelberger},
  {Hunt}, \& {Shapley}}]{pettini02}
{Pettini}, M., {Rix}, S.~A., {Steidel}, C.~C., et~al. 2002, \apj, 569, 742

\bibitem[{{Quider} {et~al.}(2009){Quider}, {Pettini}, {Shapley}, \&
  {Steidel}}]{quider09}
{Quider}, A.~M., {Pettini}, M., {Shapley}, A.~E., \& {Steidel}, C.~C. 2009,
  \mnras, 398, 1263

\bibitem[{{Reddy} {et~al.}(2012){Reddy}, {Dickinson}, {Elbaz}, {Morrison},
  {Giavalisco}, {Ivison}, {Papovich}, {Scott}, {Buat}, {Burgarella},
  {Charmandaris}, {Daddi}, {Magdis}, {Murphy}, {Altieri}, {Aussel},
  {Dannerbauer}, {Dasyra}, {Hwang}, {Kartaltepe}, {Leiton}, {Magnelli}, \&
  {Popesso}}]{reddy12}
{Reddy}, N., {Dickinson}, M., {Elbaz}, et~al. 2012, \apj, 744, 154

\bibitem[{{Reddy} {et~al.}(2010){Reddy}, {Erb}, {Pettini}, {Steidel}, \&
  {Shapley}}]{reddy10a}
{Reddy}, N.~A., {Erb}, D.~K., {Pettini}, M., {Steidel}, C.~C., \& {Shapley},
  A.~E. 2010, \apj, 712, 1070

\bibitem[{{Reddy} \& {Steidel}(2004)}]{reddy04}
{Reddy}, N.~A. \& {Steidel}, C.~C. 2004, \apjl, 603, L13

\bibitem[{{Reddy} \& {Steidel}(2009)}]{reddy09}
---. 2009, \apj, 692, 778

\bibitem[{{Reddy} {et~al.}(2006{\natexlab{a}}){Reddy}, {Steidel}, {Erb},
  {Shapley}, \& {Pettini}}]{reddy06b}
{Reddy}, N.~A., {Steidel}, C.~C., {Erb}, D.~K., {Shapley}, A.~E., \& {Pettini},
  M. 2006{\natexlab{a}}, \apj, 653, 1004

\bibitem[{{Reddy} {et~al.}(2006{\natexlab{b}}){Reddy}, {Steidel}, {Fadda},
  {Yan}, {Pettini}, {Shapley}, {Erb}, \& {Adelberger}}]{reddy06a}
{Reddy}, N.~A., {Steidel}, C.~C., {Fadda}, D., et~al. 2006{\natexlab{b}},
  \apj, 644, 792

\bibitem[{{Reddy} {et~al.}(2008){Reddy}, {Steidel}, {Pettini}, {Adelberger},
  {Shapley}, {Erb}, \& {Dickinson}}]{reddy08}
{Reddy}, N.~A., {Steidel}, C.~C., {Pettini}, M., et~al. 2008, \apjs, 175, 48

\bibitem[{{Renzini}(2009)}]{renzini09}
{Renzini}, A. 2009, \mnras, 398, L58

\bibitem[{{Riechers} {et~al.}(2010){Riechers}, {Carilli}, {Walter}, \&
  {Momjian}}]{riechers10}
{Riechers}, D.~A., {Carilli}, C.~L., {Walter}, F., \& {Momjian}, E. 2010,
  \apjl, 724, L153

\bibitem[{{Salpeter}(1955)}]{salpeter55}
{Salpeter}, E.~E. 1955, \apj, 121, 161

\bibitem[{{Sawicki}(2011)}]{sawicki11}
{Sawicki}, M. 2011, ArXiv e-prints

\bibitem[{{Sawicki} \& {Yee}(1998)}]{sawicki98}
{Sawicki}, M. \& {Yee}, H.~K.~C. 1998, \aj, 115, 1329

\bibitem[{{Schaerer} \& {de Barros}(2009)}]{schaerer09}
{Schaerer}, D. \& {de Barros}, S. 2009, \aap, 502, 423

\bibitem[{{Schaerer} \& {de Barros}(2010)}]{schaerer10}
---. 2010, \aap, 515, A73

\bibitem[{{Schechter}(1976)}]{schechter76}
{Schechter}, P. 1976, \apj, 203, 297

\bibitem[{{Schmidt}(1959)}]{schmidt59}
{Schmidt}, M. 1959, \apj, 129, 243

\bibitem[{{Seibert} {et~al.}(2002){Seibert}, {Heckman}, \&
  {Meurer}}]{seibert02}
{Seibert}, M., {Heckman}, T.~M., \& {Meurer}, G.~R. 2002, \aj, 124, 46

\bibitem[{{Shapley} {et~al.}(2001){Shapley}, {Steidel}, {Adelberger},
  {Dickinson}, {Giavalisco}, \& {Pettini}}]{shapley01}
{Shapley}, A.~E., {Steidel}, C.~C., {Adelberger}, K.~L., et~al. 2001, \apj, 562, 95

\bibitem[{{Shapley} {et~al.}(2005){Shapley}, {Steidel}, {Erb}, {Reddy},
  {Adelberger}, {Pettini}, {Barmby}, \& {Huang}}]{shapley05}
{Shapley}, A.~E., {Steidel}, C.~C., {Erb}, D.~K., et~al. 2005, \apj, 626, 698

\bibitem[{{Shapley} {et~al.}(2003){Shapley}, {Steidel}, {Pettini}, \&
  {Adelberger}}]{shapley03}
{Shapley}, A.~E., {Steidel}, C.~C., {Pettini}, M., \& {Adelberger}, K.~L. 2003,
  \apj, 588, 65

\bibitem[{{Stark} {et~al.}(2009){Stark}, {Ellis}, {Bunker}, {Bundy}, {Targett},
  {Benson}, \& {Lacy}}]{stark09}
{Stark}, D.~P., {Ellis}, R.~S., {Bunker}, A., et~al. 2009, \apj, 697, 1493

\bibitem[{{Steidel} {et~al.}(2003){Steidel}, {Adelberger}, {Shapley},
  {Pettini}, {Dickinson}, \& {Giavalisco}}]{steidel03}
{Steidel}, C.~C., {Adelberger}, K.~L., {Shapley}, A.~E., et~al. 2003, \apj, 592, 728

\bibitem[{{Steidel} {et~al.}(2004){Steidel}, {Shapley}, {Pettini},
  {Adelberger}, {Erb}, {Reddy}, \& {Hunt}}]{steidel04}
{Steidel}, C.~C., {Shapley}, A.~E., {Pettini}, M., et~al. 2004, \apj, 604, 534

\bibitem[{{Tacconi} {et~al.}(2010){Tacconi}, {Genzel}, {Neri}, {Cox}, {Cooper},
  {Shapiro}, {Bolatto}, {Bouch{\'e}}, {Bournaud}, {Burkert}, {Combes},
  {Comerford}, {Davis}, {Schreiber}, {Garcia-Burillo}, {Gracia-Carpio}, {Lutz},
  {Naab}, {Omont}, {Shapley}, {Sternberg}, \& {Weiner}}]{tacconi10}
{Tacconi}, L.~J., {Genzel}, R., {Neri}, R., et~al. 2010, \nat, 463, 781

\bibitem[{{Tinsley}(1980)}]{tinsley80}
{Tinsley}, B.~M. 1980, \fcp, 5, 287

\bibitem[{{van der Wel} {et~al.}(2011){van der Wel}, {Straughn}, {Rix},
  {Finkelstein}, {Koekemoer}, {Weiner}, {Wuyts}, {Bell}, {Faber}, {Trump},
  {Koo}, {Ferguson}, {Scarlata}, {Hathi}, {Dunlop}, {Newman}, {Dickinson},
  {Jahnke}, {Salmon}, {de Mello}, {Kocevski}, {Lai}, {Grogin}, {Rodney}, {Guo},
  {McGrath}, {Lee}, {Barro}, {Huang}, {Riess}, {Ashby}, \&
  {Willner}}]{vanderwel11}
{van der Wel}, A., {Straughn}, A.~N., {Rix}, H.-W., et~al. 2011, \apj, 742, 111

\bibitem[{{van Dokkum} {et~al.}(2006){van Dokkum}, {Quadri}, {Marchesini},
  {Rudnick}, {Franx}, {Gawiser}, {Herrera}, {Wuyts}, {Lira}, {Labb{\'e}},
  {Maza}, {Illingworth}, {F{\"o}rster Schreiber}, {Kriek}, {Rix}, {Taylor},
  {Toft}, {Webb}, \& {Yi}}]{vandokkum06}
{van Dokkum}, P.~G., {Quadri}, R., {Marchesini}, D., et~al. 2006, \apjl, 638, L59

\bibitem[{{Weinmann} {et~al.}(2011){Weinmann}, {Neistein}, \&
  {Dekel}}]{weinmann11}
{Weinmann}, S.~M., {Neistein}, E., \& {Dekel}, A. 2011, \mnras, 1463

\bibitem[{{Williams} {et~al.}(2010){Williams}, {Dalcanton}, {Stilp}, {Gilbert},
  {Ro{\v s}kar}, {Seth}, {Weisz}, {Dolphin}, {Gogarten}, {Skillman}, \&
  {Holtzman}}]{williams10}
{Williams}, B.~F., {Dalcanton}, J.~J., {Stilp}, A., et~al. 2010, \apj, 709, 135

\bibitem[{{Wuyts} {et~al.}(2011){Wuyts}, {Forster Schreiber}, {Lutz}, {Nordon},
  {Berta}, {Altieri}, {Andreani}, {Aussel}, {Bongiovanni}, {Cepa}, {Cimatti},
  {Daddi}, {Elbaz}, {Genzel}, {Koekemoer}, {Magnelli}, {Maiolino}, {McGrath},
  {Perez Garcia}, {Poglitsch}, {Popesso}, {Pozzi}, {Sanchez-Portal}, {Sturm},
  {Tacconi}, \& {Valtchanov}}]{wuyts11}
{Wuyts}, S., {Forster Schreiber}, N.~M., {Lutz}, D., et~al. 2011, ArXiv e-prints
\end{thebibliography}



\begin{appendices}

\section{Conversion from UV Luminosity to Star Formation Rate}
\label{sec:uvconv}

\begin{figure}[tbp]
\plotone{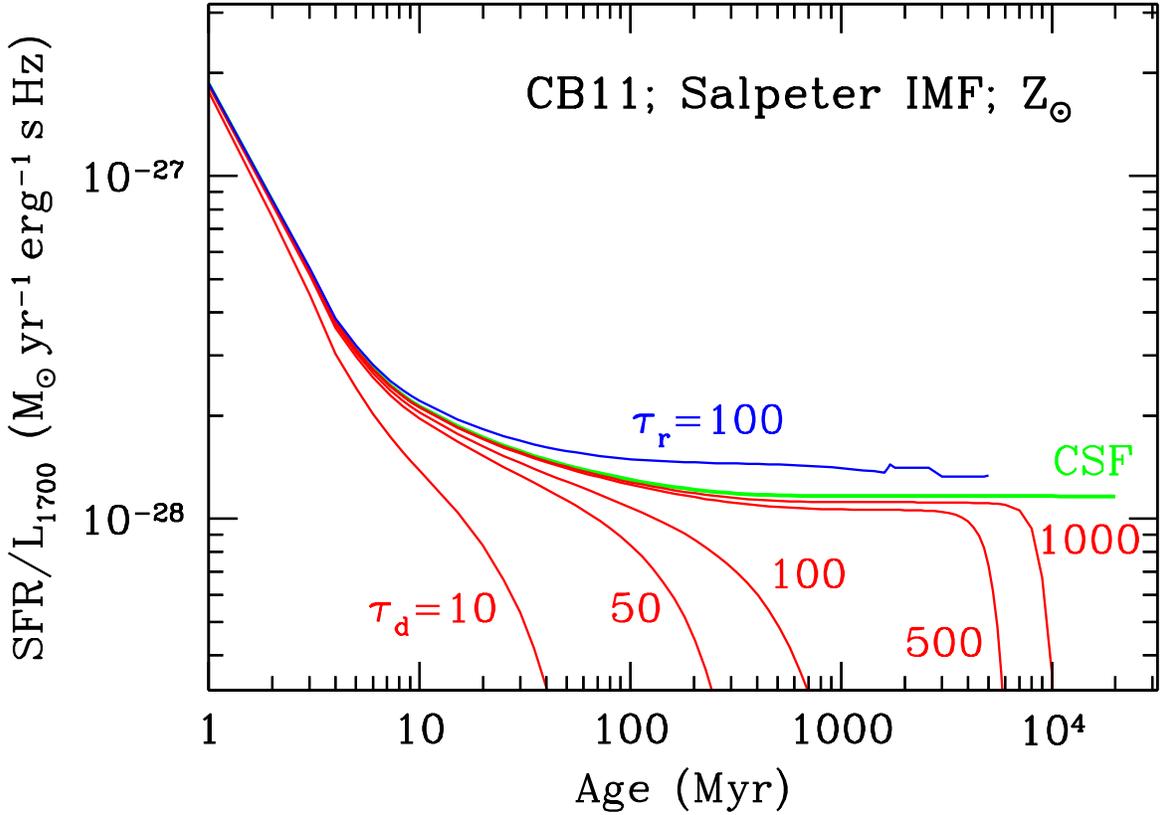}
\caption{Ratio of SFR to specific luminosity at 1700\,\AA\, as a
  function of age for different star formation histories.  The red
  curves show results for exponentially declining star formation
  histories with $\tau_{\rm d}=10$, $50$, $100$, $500$, and
  $1000$\,Myr.  The green curve shows the result for a CSF history.
  The blue curve shows the result for an exponentially rising star
  formation history with $\tau_{\rm r}=100$\,Myr.  Other rising
  histories with $\tau_{\rm r}>100$\,Myr will be bracketed by the CSF
  and $\tau_{\rm r}=100$\,Myr curves.}
\label{fig:uvconv}
\end{figure}

The conversion from UV luminosity to SFR will depend on the star
formation history and age of a galaxy.  This is illustrated in
Figure~\ref{fig:uvconv}, where we show the ratio of SFR to the
$1700$\,\AA\, luminosity as a function of age for different star
formation histories.  For ages $\la 10$\,Myr, the factor needed to
convert a UV luminosity to a SFR is larger than the \citet{madau98,
  kennicutt98} values because the ratio of O stars to B stars (both of
which contribute significantly to the rest-frame $1700$\,\AA\,
luminosity) is larger than the equilibrium ratio reached after the B
star main sequence lifetime of $\approx 100$\,Myr.  For older ages and
exponentially declining star formation histories, the UV luminosity
does not fall off as quickly with time as the SFR.  In these cases,
the ratio of O stars to B stars is {\em lower} than the case of
continuous star formation (i.e., B stars will still contribute
significantly to the UV luminosity even after the production rate of O
stars declines).  As the decay timescale $\tau_{\rm d}$ becomes larger and
$t_{\rm age}/\tau_{\rm d}<1$, the conversion factor approaches that of the CSF
history.  Similarly, exponentially {\em rising} star formation
histories have a conversion factor similar to, but slightly larger
than, those obtained for a CSF history.  In the rising histories, the
ratio of O and B stars will stabilize to a value larger than in the
CSF case since the SFR is continually increasing.  The conversion
factor for rising histories with larger $\tau_{\rm r}$ approaches the
factor for a CSF case.  For convenience, we have tabulated the form of
the simple polynomial fits that describe the relationship between the
ratio of SFR to $1700$\,\AA\, luminosity as a function of age for the
different star formation histories, assuming the latest CB11 models
with a Salpeter IMF and solar metallicity (Table~\ref{tab:uvconv}).
In this parameterization, the conversion factor between UV luminosity
and SFR is $C_{\rm UV}$, where
\begin{equation}
\frac{\rm SFR}{\rm M_{\odot}\,yr^{-1}} = C_{\rm UV}\times \frac{L_{1700}}{\rm erg\,s^{-1}\,Hz^{-1}},
\end{equation}
and $C_{\rm UV}$ is approximated with a polynomial function of
$\xi\equiv\log(t_{\rm age})$:
\begin{equation}
\log [C_{\rm UV}(\xi)] = \sum_{\rm i=0}^{\rm 5} a_{\rm i}[f(\xi)]^{\rm i}
\label{eq:uvconv}
\end{equation}
where $f(\xi)=\xi$ or $1/\xi$, and the coefficients $a_{\rm i}$ are
given in Table~\ref{tab:uvconv}.

In the present context, changes in the conversion between UV
luminosity and SFR will not significantly affect the bolometric SFRs
for two reasons.  First, as discussed in
Section~\ref{sec:sfrcomparison}, the range of ages and star formation
histories that are plausible at $z\sim 2$ rule out conversion factors
that vary by more than $\approx 20\%$ from the equilibrium value.
Secondly, the bolometric SFRs are less sensitive to changes in the
UV-SFR conversion factor given that, for most of the galaxies in our
sample, a significant fraction of the bolometric luminosity ($\simeq
80\%$) is emergent in the IR and not the UV \citep{reddy06a,
  reddy10a}.  For simplicity, we assume the \citet{kennicutt98}
relation to convert UV luminosity to SFR, and we discuss in
Section~\ref{sec:sfrcomparison} how changing the conversion factor
between these quantities affects (or does not significantly affect)
our results.

\begin{deluxetable}{lccrrrrrrr}[!b]
\tabletypesize{\footnotesize}
\tablewidth{0pc}
\tablecaption{Coefficients of Polynomial Fits to Logarithm of SFR/UV Conversion Factor, $\log[C_{\rm UV}]$}
\tablehead{
\colhead{} &
\colhead{$\tau$} &
\colhead{} &
\colhead{} &
\colhead{} &
\colhead{} &
\colhead{} &
\colhead{} &
\colhead{} &
\colhead{Age Range\tablenotemark{c}} \\
\colhead{SFH\tablenotemark{a}} &
\colhead{(Myr)} &
\colhead{$f(\xi)$\tablenotemark{b}} &
\colhead{$a_{0}$} &
\colhead{$a_{1}$} &
\colhead{$a_{2}$} &
\colhead{$a_{3}$} &
\colhead{$a_{4}$} &
\colhead{$a_{5}$} &
\colhead{(Myr)}}
\startdata
Constant & $\infty$ & $1/\xi$ & -27.822 & -0.969 & 2.511 & -2.000 & 0.704 & -0.089 & $\ge 3$\\ 
\\
Declining & 10 & $\xi$ & -26.658 & -1.662 & 0.568 & -0.689 & 0.969 & -0.395 & 3 -- 255 \\
          & 20 & $\xi$ & -26.433 & -3.065 & 3.808 & -3.592 & 1.936 & -0.425 & 3 -- 508 \\
          & 50 & $\xi$ & -25.634 & -7.271 & 11.088 & -8.624 & 3.237 & -0.471 & 3 -- 1433 \\
          & 100 & $\xi$ & -26.309 & -3.416 & 3.665 & -2.245 & 0.716 & -0.093 & 3 -- 905 \\
          & 200 & $1/\xi$ & -28.906 & 3.768 & -5.127 & 3.615 & -1.176 & 0.140 & 3 -- 1900 \\
          & 500 & $1/\xi$ & -28.008 & -0.238 & 1.404 & -1.218 & 0.449 & -0.059 & 3 -- 4000 \\
          & 1000 & $1/\xi$ & -27.889 & -0.725 & 2.162 & -1.764 & 0.630 & -0.081 & 3 -- 7000 \\
          & 2000 & $1/\xi$ & -27.847 & -0.888 & 2.409 & -1.939 & 0.687 & -0.088 & $\ge 3$ \\
          & 5000 & $1/\xi$ & -27.846 & -0.855 & 2.311 & -1.843 & 0.649 & -0.082 & $\ge 3$ \\
\\
Rising & 100 & $1/\xi$ & -27.877 & -0.049 & 0.390 & -0.138 & 0.013 & 0.0002 & $\ge 3$ \\
       & 5000 & $1/\xi$ & -27.826 & -0.921 & 2.401 & -1.902 & 0.667 & -0.085 & $\ge 3$
\enddata
\tablenotetext{a}{Star formation history, either exponentially declining,
exponentially rising, or constant.}
\tablenotetext{b}{Functional dependence of polynomial fit on 
$\xi\equiv\log(t_{\rm age})$ (see Equation~\ref{eq:uvconv}).}
\tablenotetext{c}{Range of ages over which the polynomial form has been fit.}
\label{tab:uvconv}
\end{deluxetable}

\section{The Effect of Malmquist Bias on the Mean SFR at a Given Stellar Mass}
\label{sec:malmquist}

In this section, we explore the effect of Malmquist bias from a
flux-limited sample on the mean SFRs in bins of stellar mass.  To
illustrate this bias, we simulated the effect of imposing a
spectroscopic limit on a SFR versus stellar mass trend with an
intrinsic slope of unity, and an rms dispersion of $0.38$\,dex (based
on the dispersion around the best-fit correlation shown in
Figure~\ref{fig:sfrm}).  Because our sample is selected based on
(unattenuated) UV emission, we first quantified the relationship
between bolometric SFR and bolometric luminosity using the 302
galaxies in the MIPS $24$\,$\mu$m sample.  \citet{reddy10a} give the
relationship between bolometric and UV luminosity for this sample
(with a mean redshift of $\langle z\rangle \simeq 2$) as
\begin{equation}
\log[\luv/{\rm L}_\odot] = \log[\lbol/{\rm L}_\odot] - \log[{\lbol}^{1/a} 10^{-b/a} + 1],
\end{equation}
where $a=0.69\pm0.03$ and $b=10.91\pm0.04$.  This relationship is
redshift dependent, because the ratio of bolometric to UV luminosity
at a given SFR decreases with increasing redshift \citep{reddy06a,
  reddy10a}.  The mean trend between $\luv$ and SFR implies that
galaxies with fainter UV luminosities also have on average lower
bolometric SFRs \citep{reddy10a}.

Second, to ensure a realistic weighting of points, we simulated a
population of galaxies that has a \citet{schechter76} distribution in
bolometric SFR, with a characteristic
SFR$^{\ast}=30$\,M$_{\odot}$\,yr$^{-1}$ and faint-end slope $\alpha=
-1.6$, based on the bolometric luminosity function at $z\sim 2$
\citep{reddy09}.  An $\luv$ was assigned to each galaxy based on its
bolometric SFR and the observed relationship between SFR and $\luv$,
as discussed above.  Similarly, a stellar mass was assigned to each
galaxy based on its SFR and the assumed intrinsic trend between SFR
and $M_{\ast}$.  In assigning $\luv$ and $M_{\ast}$, we randomly
perturbed these values according to the dispersions measured in the
SFR-$\luv$ ($\approx 0.5$\,dex) and SFR-$M_{\ast}$ ($\approx
0.4$\,dex) relations.  The final step in the simulation was to remove
all galaxies with $\luv\la 10^{10}$\,L$_\odot$, corresponding
approximately to the spectroscopic limit at $z\sim 2$.  The remaining
galaxies are binned according their stellar masses.  The average (or
median) SFR for each bin is then calculated.

\begin{figure}[tbp]
\plottwo{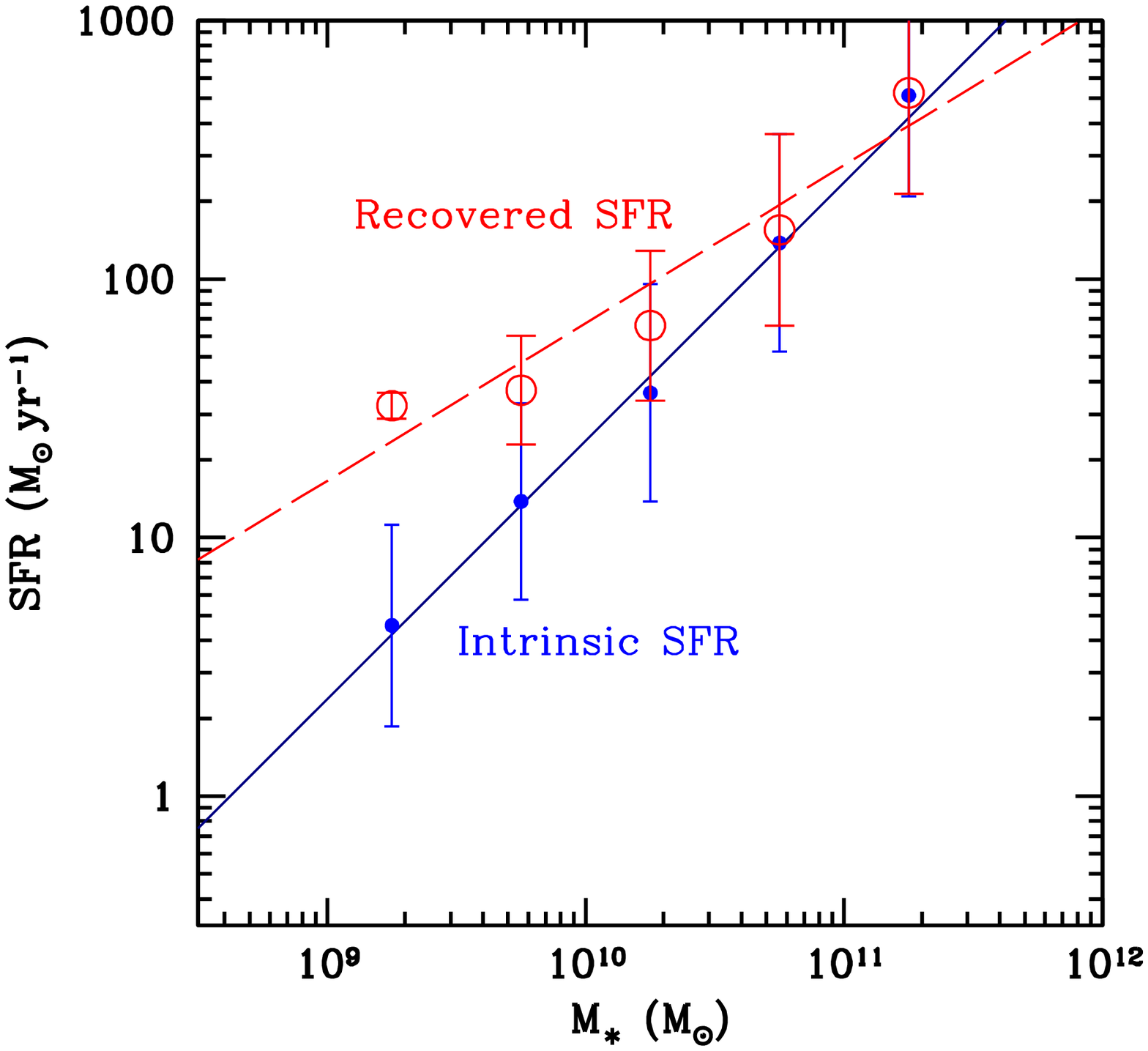}{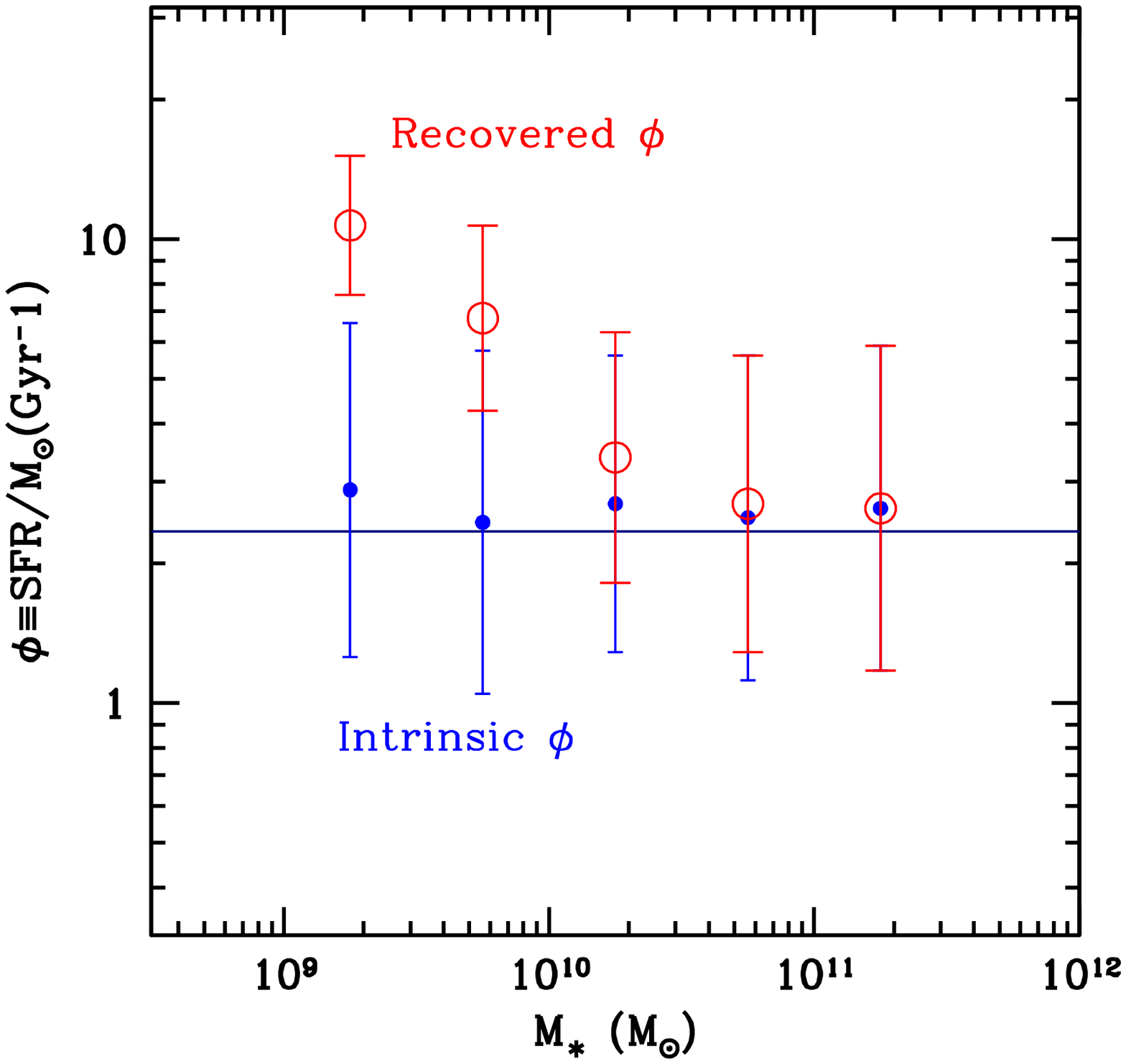}
\caption{Effects of Malmquist bias on the measurements of the SFR
  vs. stellar mass trend, and specific SFR $\phi$.  The solid line
  indicates an assumed intrinsic specific SFR of $\phi\simeq 2.4\times
  10^{-9}$\,yr$^{-1}$.  Accounting for the scatter between SFR and
  stellar mass results in recovered SFR and $\phi$ denoted by the
  solid blue symbols.  The hollow red symbols indicate the measured
  SFR and $\phi$ when we fold in the scatter between SFR and UV
  luminosity, and impose a (spectroscopic) limit of
  $\luv=10^{10}$\,L$_{\odot}$.  Errorbars reflect the dispersion in
  SFR and $\phi$ in each bin of stellar mass.}
\label{fig:simsfrm}
\end{figure}

The results of this simple simulation are shown in
Figure~\ref{fig:simsfrm}, which suggests that the UV flux limit will
result in an artificial trend of increasing specific SFR ($\phi$) at
low stellar masses, on an otherwise intrinsically flat relationship
between $\phi$ and $M_{\ast}$.  For example, in the lowest mass bin
with $9.0\le \log [M_{\ast}/{\rm M}_{\odot}]<9.5$, the recovered
specific SFR is $\approx 3-4\times$ larger than the intrinsic value.
Based on this simple simulation, and evidence that galaxies fainter
than our spectroscopic limit have specific SFRs at $M_{\ast}\simeq
10^{9}$\,M$_{\odot}$ that are lower than what we would have inferred
had we relied on the SED fitting of the spectroscopic sample only (see
the next section), we conclude that the true {\em median} specific
star formation is lower than that measured for the lowest mass
galaxies in our sample if these SFRs are also determined from SED
fitting (Section~\ref{sec:mlratio}).
\footnote{There is also a small systematic effect in the range of
  specific SFR probed at {\em large} stellar masses.  This stems from
  the fact that the most massive galaxies are also more heavily
  star-forming (e.g., Figure~\ref{fig:sfrm}) and dustier.  The
  increase in dust obscuration means that at a given (large) stellar
  mass, we are more likely to miss galaxies that are heavily
  attenuated (because they fall below our UV spectroscopic limit).
  The effect of this bias is mitigated by the fact that these
  ultraluminous galaxies make up only a small fraction of the total in
  our sample, and thus do not significantly affect the fits between
  stellar mass and absolute magnitude.}  It is evident that the
magnitude of this bias increases with the intrinsic scatter in the
SFR-M$_{\ast}$ relationship.  More generally, these results highlight
the importance of factoring in the UV luminosity limit of
high-redshift dropout samples when determining mean SFRs (or specific
SFRs) in bins of stellar mass.  We show in Section~\ref{sec:mlratio}
and Appendix~\ref{sec:massnir} how this bias can also affect our
determination of the near-IR luminosity dependence of the
mass-to-light $M/L_{1.1\mu m}$ ratio.  As such, the simulation
discussed above does not tell the full story because we cannot
directly measure the stellar masses of galaxies; the masses must be
inferred from luminosity at a given wavelength (UV or near-IR), and
therefore the position of the recovered galaxies along the stellar
mass axis in the left panel of Figure~\ref{fig:simsfrm} (and along
both the $\phi$ and $M_{\ast}$ axes in the right panel) may not be
accurate.  In Appendix~\ref{sec:sfrmbias} we discuss the combined
effects of these biases on determinations of the SFR-$M_{\ast}$
relation.

\section{UV Dependence of the Relationship Between Stellar Mass and Near-IR Magnitude}
\label{sec:massnir}

\subsection{Spectroscopic Sample}

The Malmquist bias discussed in Appendix~\ref{sec:malmquist} can
result in a slope between SFR and stellar mass that is artificially
shallow compared to the intrinsic value.  Similarly, there is a simple
explanation for why this bias manifests itself onto the best-fit
relation between stellar mass and near-IR luminosity as determined
from a UV selected sample.  Specifically, a flux-limited sample, like
ours, will tend to miss galaxies with lower SFRs at a fixed (low)
stellar mass (Section~\ref{sec:sfrm} and
Appendix~\ref{sec:malmquist}).  Therefore, the median near-IR
magnitude in a given bin of stellar mass will be biased brighter
because at a fixed stellar mass, galaxies with larger SFRs (and hence
those that satisfy the UV flux limit) will have larger near-IR
luminosities (i.e., current star formation will contribute at near-IR
wavelengths).

\begin{figure}[tbp]
\plotone{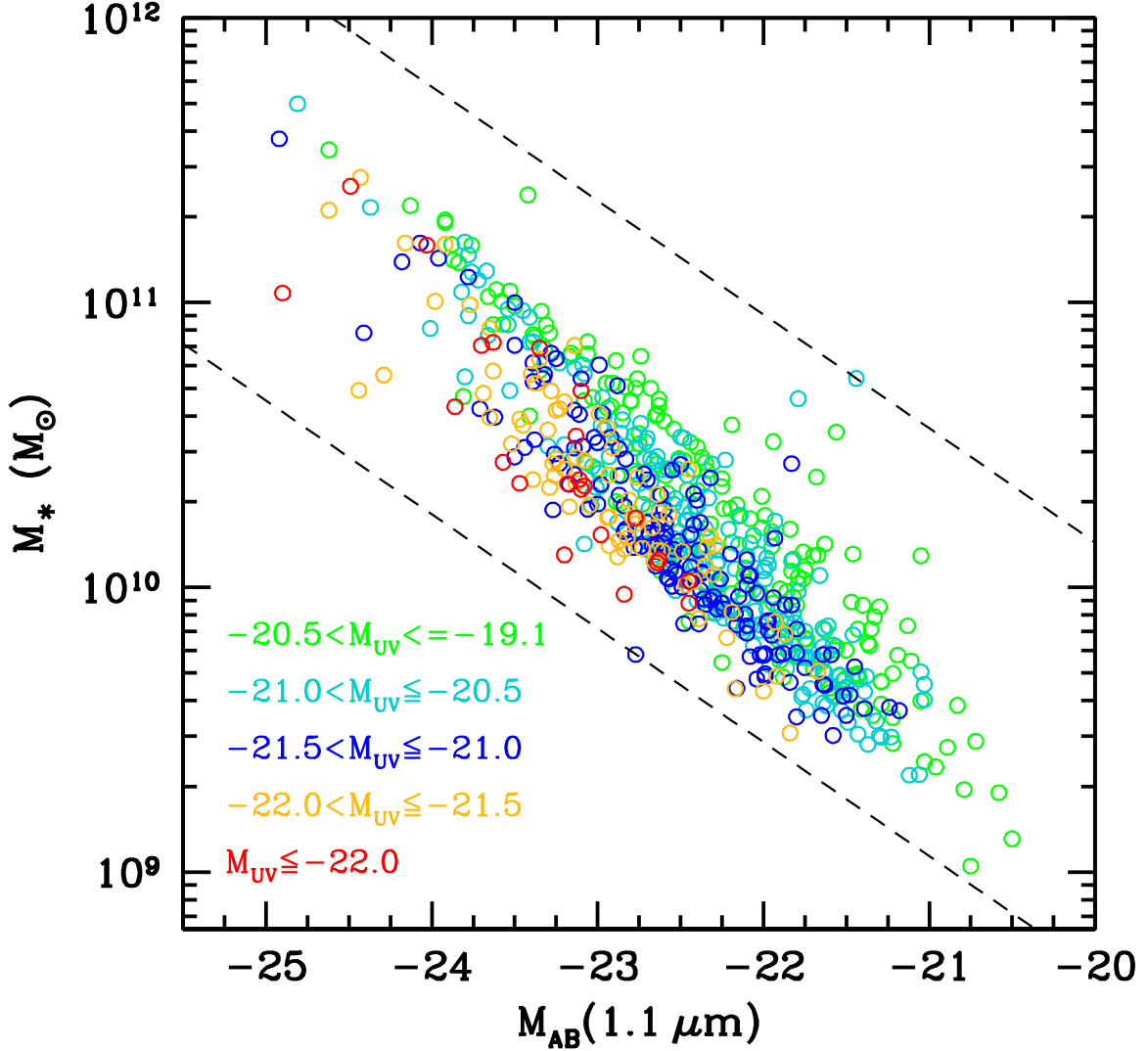}
\caption{Same as upper left panel of Figure~\ref{fig:masstraceirac},
  color coded by the absolute UV magnitude (at $1700$\,AA), for
  galaxies directly detected in the channel 1 IRAC data.  There is a
  clear trend between UV luminosity and the relation between stellar
  mass and near-IR magnitude, such that at a given stellar mass,
  UV-luminous galaxies exhibit brighter near-IR magnitudes.}
\label{fig:ssfrcode}
\end{figure}

\begin{figure}[tbp]
\plotone{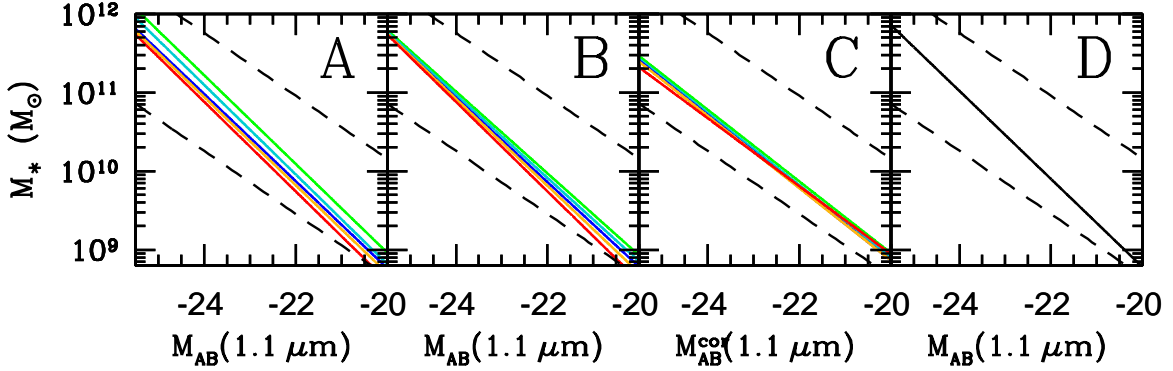}
\caption{{\em Panel A:} Best-fit linear relations between stellar mass
  and near-IR magnitude for galaxies in the UV magnitude bins
  indicated in Figure~\ref{fig:ssfrcode}.  {\em Panel B:} Best-fit
  linear relations between stellar mass and near-IR magnitude for
  galaxies brighter than the faint magnitude limit of each of the UV
  magnitude bins indicated in Figure~\ref{fig:ssfrcode}.  {\em Panel
    C:} Same as Panel B, where we have corrected the near-IR magnitude
  for the affects of dust extinction, assuming the best-fit $\ebmv$
  and either the \citet{calzetti00} or SMC attenuation curve depending
  on whether the galaxy is ``young'' (Section~\ref{sec:sfrsyoung}).
  {\em Panel D:} Best-fit linear relation between stellar mass and
  near-IR magnitude (uncorrected for dust) for all galaxies with
  $M_{1.1}<-22.0$.  In all panels, the dotted lines indicate the
  maximum and minimum mass-to-light ratio found in the sample.  }
\label{fig:nirbestfit}
\end{figure}

This effect is demonstrated clearly in Figure~\ref{fig:ssfrcode}.  In
particular, the large scatter in near-IR luminosity (or stellar mass)
at a given UV luminosity (see Section~\ref{sec:uvmass}), implies that
there will be galaxies just below our flux limit that have similar
stellar masses as galaxies that lie just above our flux limit.  The
result is that the median or average near-IR luminosity corresponding
to a bin of stellar mass will be {\em lower} than what we would have
predicted from a UV (bright) selected sample.  Conversely, UV faint
galaxies have a smaller contribution of current star formation to the
near-IR light, and hence a given near-IR magnitude will correspond to
a larger stellar mass for such galaxies.  The inclusion of UV fainter
galaxies results in a slope between $\log M_{\ast}$ and $M_{1.1}$ that
is less negative than the one derived for a sample of UV-brighter
galaxies (Figure~\ref{fig:nirbestfit}.  This has important
implications for (1) determining the luminosity dependence of the
mass-to-light ratio based on a UV/optical selected samples (e.g.,
\citealt{gonzalez11, lee11b}), and (2) translating the mean stellar
mass found for galaxies of a given UV luminosity.

To illustrate these points, the best-fit linear relation between log
stellar mass and near-IR magnitude ($\log M_{\ast}-M_{1.1}$) for
galaxies in the different UV luminosity bins denoted in
Figure~\ref{fig:ssfrcode} are shown in Panel A of
Figure~\ref{fig:nirbestfit}.  Panel B indicates the best-fit relations
when we fit for all galaxies above the faint magnitude limits of the
bins in UV luminosity.  As we add UV faint galaxies, the best-fit
slope between stellar mass and near-IR magnitude becomes shallower
(less steep).  The slope of the $\log M_{\ast}-M_{1.1}$ relation can
also be affected by dust, because galaxies with larger stellar mass
have on average larger SFRs and hence larger dust obscuration.  Hence,
the Panel C of Figure~\ref{fig:nirbestfit} shows that correcting the
near-IR magnitudes for the effects of dust attenuation results in a
slope of the $\log M_{\ast}-M_{1.1}$ relation that is very close to
$0.4$, implying a dust-corrected $M/L$ ratio at $1.1$\,$\mu$m that is
only very weakly dependent on near-IR luminosity.

Finally, Figure~\ref{fig:ssfrcode} suggests that our UV selected
galaxies sample approximately the full range of stellar masses down to
$M_{1.1}\simeq -22.0$\footnote{Note that we have not included IRAC
  non-detections in Figure~\ref{fig:ssfrcode}.  As we show in
  Section~\ref{sec:uvmass}, the majority ($\ga 70\%$) of galaxies
  brighter than $M_{\rm UV}=-20$ at $z\sim 2$ are detected in the IRAC
  bands.  For galaxies with fainter UV luminosities, the detection
  fraction is lower (e.g., Figure~\ref{fig:iracundet}), but the vast
  majority of the IRAC upper limits for such UV faint galaxies lie
  fainter than $M_{1.1}\simeq -22$.}  Therefore, Panel D of
Figure~\ref{fig:nirbestfit} indicates the best-fit correlation between
near-IR magnitude (uncorrected for dust) and stellar mass for all
galaxies in our sample with $M_{1.1}<-22$ The best-fit linear
relations and the rms dispersion of the data about these linear fits
are tabulated in Table~\ref{tab:nirbestfit}.  The important point
demonstrated in Figures~\ref{fig:ssfrcode} and \ref{fig:nirbestfit} is
that the systematic biases introduced by virtue of having a flux
limited sample, combined with the effects of dust attenuation, can
easily influence our interpretation of the presence of a luminosity
dependence of the $M/L$ ratio.

\begin{deluxetable}{lccc}
\tabletypesize{\footnotesize}
\tablewidth{0pc}
\tablecaption{Best-Fit Linear Relations Between Stellar Mass and Near-IR ($1.1$\,$\mu$m) Magnitude}
\tablehead{
\colhead{Panel\tablenotemark{a}} &
\colhead{$M_{\rm UV}$ or $M_{1.1}$ Range} &
\colhead{$\log[M_{\ast}/{\rm M}_{\odot}] = $} &
\colhead{RMS}}
\startdata
A & $-20.5<M_{\rm UV}\le -19.1$ & $-(0.56\pm0.02)M_{1.1}-(2.33\pm0.39)$ & 0.19 \\
  & $-21.0<M_{\rm UV}\le -20.5$ & $-(0.57\pm0.02)M_{1.1}-(2.58\pm0.39)$ & 0.17 \\
  & $-21.5<M_{\rm UV}\le -21.0$ & $-(0.55\pm0.02)M_{1.1}-(2.32\pm0.46)$ & 0.16 \\
  & $-22.0<M_{\rm UV}\le -21.5$ & $-(0.56\pm0.03)M_{1.1}-(2.51\pm0.67)$ & 0.16 \\
  & $M_{\rm UV}\le 22.0$ & $-(0.57\pm0.07)M_{1.1}-(2.80\pm1.64)$ & 0.19 \\
\\
B & $M_{\rm UV}\le -19.1$ & $-(0.52\pm0.01)M_{1.1}-(1.49\pm0.24)$ & 0.20 \\
  & $M_{\rm UV}\le -20.5$ & $-(0.53\pm0.01)M_{1.1}-(1.75\pm0.26)$ & 0.18 \\
  & $M_{\rm UV}\le -21.0$ & $-(0.54\pm0.02)M_{1.1}-(1.98\pm0.35)$ & 0.17 \\
  & $M_{\rm UV}\le -21.5$ & $-(0.55\pm0.03)M_{1.1}-(2.36\pm0.62)$ & 0.17 \\
\\
C & $M_{\rm UV}\le -19.1$ & $-(0.46\pm0.01)M^{\rm cor}_{1.1}-(0.15\pm0.23)$ & 0.22 \\
  & $M_{\rm UV}\le -20.5$ & $-(0.46\pm0.01)M^{\rm cor}_{1.1}-(0.26\pm0.27)$ & 0.20 \\
  & $M_{\rm UV}\le -21.0$ & $-(0.46\pm0.02)M^{\rm cor}_{1.1}-(0.37\pm0.36)$ & 0.19 \\
  & $M_{\rm UV}\le -21.5$ & $-(0.45\pm0.03)M^{\rm cor}_{1.1}-(0.32\pm0.64)$ & 0.20 \\
  & $M_{\rm UV}\le -22.0$ & $-(0.43\pm0.07)M^{\rm cor}_{1.1}-(0.31\pm1.67)$ & 0.24 \\
\\
D & $M_{1.1}<-22.0$ & $-(0.56\pm0.09)M_{1.1}-(2.42\pm1.94)$ & 0.12
\enddata
\tablenotetext{a}{Indicates panel of Figure~\ref{fig:nirbestfit} that shows the
best-fit linear relations between $M_{\ast}$ and $M_{1.1}$.}
\label{tab:nirbestfit}
\end{deluxetable}

\subsection{UV Faint Sample}

We have several options for inferring the stellar masses of UV faint
galaxies that lie below our spectroscopic limit.  One possibility is
to simple assume the same relationship between near-IR magnitude and
stellar mass as was found for the faintest UV luminosity bin in our
spectroscopic sample.  A second option is to attempt to fit for the
average SED of these UV faint galaxies.  This option is beyond the
scope of this paper and is not explored further.  A third option is to
extrapolate UV luminosity dependence of the $\log M_{\ast}-M_{1.1}$
relation to UV faint galaxies.  While the marginalized uncertainties
in the slope and intercept of the $\log M_{\ast}-M_{1.1,0.9}$
relations for the different UV luminosity bins in the spectroscopic
sample suggest very little evolution (Table~\ref{tab:nirbestfit}),
there is some indication that the relation systematically shifts in
normalization towards higher masses at fainter UV-luminosities (Panel
A of Figure~\ref{fig:nirbestfit}), as would be expected if these
UV faint galaxies have a lower contribution of current star formation
to the near-IR magnitude (see discussion above and in
Section~\ref{sec:uvmass}).  By extrapolation from the UV-bright bins,
the intercepts of the $\log M_{\ast}-M_{1.1,0.9}$ relation are
computed as $b=0.14M_{\rm UV}+0.55$.  The slope is kept fixed at
$\delta\log [M_{\ast}/{\rm M}_{\odot}]/\delta M_{1.1,0.9} = -0.56$
(similar to that found for the more UV luminous bins).  Converting the
stacked IRAC channel 1 magnitudes listed in
Table~\ref{tab:stackfaint}, we find that the median stellar masses
obtained in this way are $\log [M_{\ast}/{\rm M}_{\odot}] = 9.02$ and
$8.71$ for the two UV faint bins at $z\sim 2$.  Similarly, we find
$\log [M_{\ast}/{\rm M}_{\odot}] = 9.19$ and $9.12$ for the two faint
bins at $z\sim 3$.  The uncertainties in these stellar masses include:
(1) measurement uncertainties in the stacked IRAC fluxes, (2)
statistical uncertainties in the fit between near-IR magnitude and
stellar mass, and (3) systematic uncertainties in the relation used to
convert between near-IR magnitude and stellar mass (e.g., arising from
the use of other of the other options mentioned above).  Combining the
errors from these effects in quadrature results in stellar mass
uncertainties of $\approx 0.15-0.29$\,dex.

\section{IRAC Stacking Method}
\label{sec:iracstack}

As noted in Section~\ref{sec:sedmodeling}, we modeled the stellar
populations only for those galaxies that were directly detected at
wavelengths longward of rest-frame $4000$\,\AA\, to ensure that robust
mass determinations included information from the age-sensitive Balmer
and $4000$\,\AA\, breaks.  However, a substantial fraction of galaxies
in the {\em spectroscopic} sample are undetected at these longer
wavelengths, particularly those with fainter UV luminosities, as
demonstrated in Figure~\ref{fig:iracundet}.  For an unbiased view of
the mass distribution, we stacked the IRAC data in bins of UV
luminosity.

Our stacking analysis proceeded as follows.  Cutouts around each
galaxy in our sample were assembled from the IRAC images.  We included
in the stack any galaxy that (1) does not show evidence of AGN, (2)
has a spectroscopic redshift in the range $1.4\le z<3.7$, (3) has an
IRAC exposure time of more than 7200 sec, (4) is not blended with any
nearby neighbor, and (5) lies more than $2\farcs4$ away from any
nearby source as identified in higher resolution optical and near-IR
data.  The cutouts were then median combined, though average
combination yielded similar results.  Given the field-dependent
variation of the IRAC PSF, we constructed a ``stacked PSF'' made in
exactly the same way as the regular stack --- i.e., for each galaxy
that went into the stack, we also stacked the PSF appropriate for the
field in which that galaxy lies.  Photometry on the stacked image was
measured by fitting the stacked PSF to the stacked IRAC signal.
Table~\ref{tab:iracstack} summarizes the stacked photometry, converted
to absolute magnitudes based on the median redshifts of the objects
that went into the stacks.

\begin{deluxetable}{lcllll}
\tabletypesize{\footnotesize}
\tablewidth{0pc}
\tablecaption{Stacked IRAC Magnitudes for Spectroscopic Sample}
\tablehead{
\colhead{Redshift Interval} &
\colhead{M(1700\AA) Range} &
\colhead{3.6\,$\mu$m\tablenotemark{a}} &
\colhead{4.5\,$\mu$m\tablenotemark{a}} &
\colhead{5.8\,$\mu$m\tablenotemark{a}} &
\colhead{8.0\,$\mu$m\tablenotemark{a}}}
\startdata
$1.4\le z < 2.7$ & -22.5\,\,\,\,\,-22.0 & $-23.92\pm0.13$ (4) & ... & ... & $-24.06\pm0.16$ (2) \\
& -22.0\,\,\,\,\,-21.5 & $-22.87\pm0.21$ (28) & $-22.93\pm0.19$ (26) & $-23.19\pm0.20$ (12) & $-22.95\pm0.20$ (22) \\
& -21.5\,\,\,\,\,-21.0 & $-22.20\pm0.17$ (72) & $-22.27\pm0.17$ (48) & $-22.41\pm0.19$ (31) & $-22.25\pm0.21$ (55) \\
& -21.0\,\,\,\,\,-20.5 & $-21.95\pm0.20$ (192) & $-21.85\pm0.19$ (138) & $-22.14\pm0.20$ (78) & $-21.78\pm0.23$ (147) \\
& -20.5\,\,\,\,\,-20.0 & $-21.61\pm0.19$ (199) & $-21.79\pm0.20$ (180) & $-21.99\pm0.21$ (93) & $-21.68\pm0.22$ (189) \\
& -20.0\,\,\,\,\,-19.5 & $-21.14\pm0.21$ (113) & $-21.15\pm0.19$ (79) & $-21.35\pm0.26$ (39) & $-21.02\pm0.29$ (93) \\
& -19.5\,\,\,\,\,-19.0 & $-20.85\pm0.21$ (22) & $-20.96\pm0.26$ (22) & ... & ... \\
\\
$2.7\le z < 3.7$ & -23.0\,\,\,\,\,-22.5 & $-23.51\pm0.27$ (2) & ... & ... & ... \\
& -22.5\,\,\,\,\,-22.0 & $-22.45\pm0.13$ (13) & $-23.18\pm0.12$ (6) & $-23.06\pm0.32$ (6) & $-23.54\pm0.21$ (6) \\
& -22.0\,\,\,\,\,-21.5 & $-22.35\pm0.12$ (45) & $-22.49\pm0.13$ (34) & $-22.32\pm0.25$ (21) & $-22.57\pm0.16$ (34) \\
& -21.5\,\,\,\,\,-21.0 & $-21.83\pm0.13$ (104) & $-21.83\pm0.12$ (78) & $-21.87\pm0.18$ (52) & $-21.99\pm0.19$ (79) \\
& -21.0\,\,\,\,\,-20.5 & $-21.37\pm0.12$ (108) & $-21.49\pm0.13$ (84) & $-21.48\pm0.25$ (65) & $-21.13\pm0.30$ (81) \\
& -20.5\,\,\,\,\,-20.0 & $-20.97\pm0.16$ (68) & $-21.12\pm0.12$ (57) & ... & ... \\
\enddata
\tablenotetext{a}{Uncertainties in absolute magnitude reflect the stacked flux measurement
uncertainty combined in quadrature with the dispersion in absolute magnitude given the range
of redshifts of objects in each bin.  Parentheses indicate the number of galaxies in the stack.}
\label{tab:iracstack}
\end{deluxetable}

\section{Combined Effect of Biases on the SFR-$M_{\ast}$ Relation}
\label{sec:sfrmbias}

In this section, we discuss the combined affect of the biases
discussed in Appendices~\ref{sec:malmquist} and \ref{sec:massnir}.  To
recap, there is Malmquist bias of selecting galaxies with larger star
formation rates at a given stellar mass in a flux limited sample.
Second, there is a bias introduced by the conversion that one used to
convert near-IR luminosity to stellar mass.  To investigate jointly
these effects, we built upon the simulations discussed in
Appendix~\ref{sec:malmquist}.  For each galaxy in the simulation, we
assigned it a near-IR magnitude $M_{1.1}$ based on its UV magnitude,
using the correlation between these two quantities, taking into
account upper limits in near-IR magnitude with the EM parametric
estimator (Figure~\ref{fig:uvch1}):
\begin{equation}
M_{1.1}=0.90M_{1700}-3.41
\end{equation}
with a dispersion of 1\,dex.  This near-IR magnitude was then
converted to a ``measured'' stellar mass, $M^{\rm meas}_{\ast}$, using
three different options.  In the first option, $M^{\rm meas}_{\ast}$
is estimated for each galaxy using the $\log M_{\ast}-M_{1.1}$
relation appropriate for the UV luminosity of that galaxy (Panel A of
Figure~\ref{fig:nirbestfit}, and Table~\ref{tab:nirbestfit}), and
assuming that galaxies fainter than $M_{1700}=-19.1$ follow the
relation for galaxies with $-20.5<M_{1700}\le -19.1$ .  In the second
option, $M^{\rm meas}_{\ast}$ is estimated for each galaxy using the
$\log M_{\ast}-M_{1.1}$ relation found for all galaxies in our {\em
  spectroscopic} sample with $M_{1700}\le -19.1$ (Panel B of
Figure~\ref{fig:nirbestfit}, and Table~\ref{tab:nirbestfit}).  In the
third option, $M^{\rm meas}_{\ast}$ is estimated for each galaxy using
a $\log M_{\ast}-M_{1.1}$ relation with a fixed slope of $\delta\log
[M_{\ast}/{\rm M}_{\odot}]/\delta M_{1.1} = -0.56$ and an intercept
that evolves with $M_{1700}$ (see Appendix~\ref{sec:massnir}).

\begin{figure}[tbp]
\epsscale{0.90}
\plotone{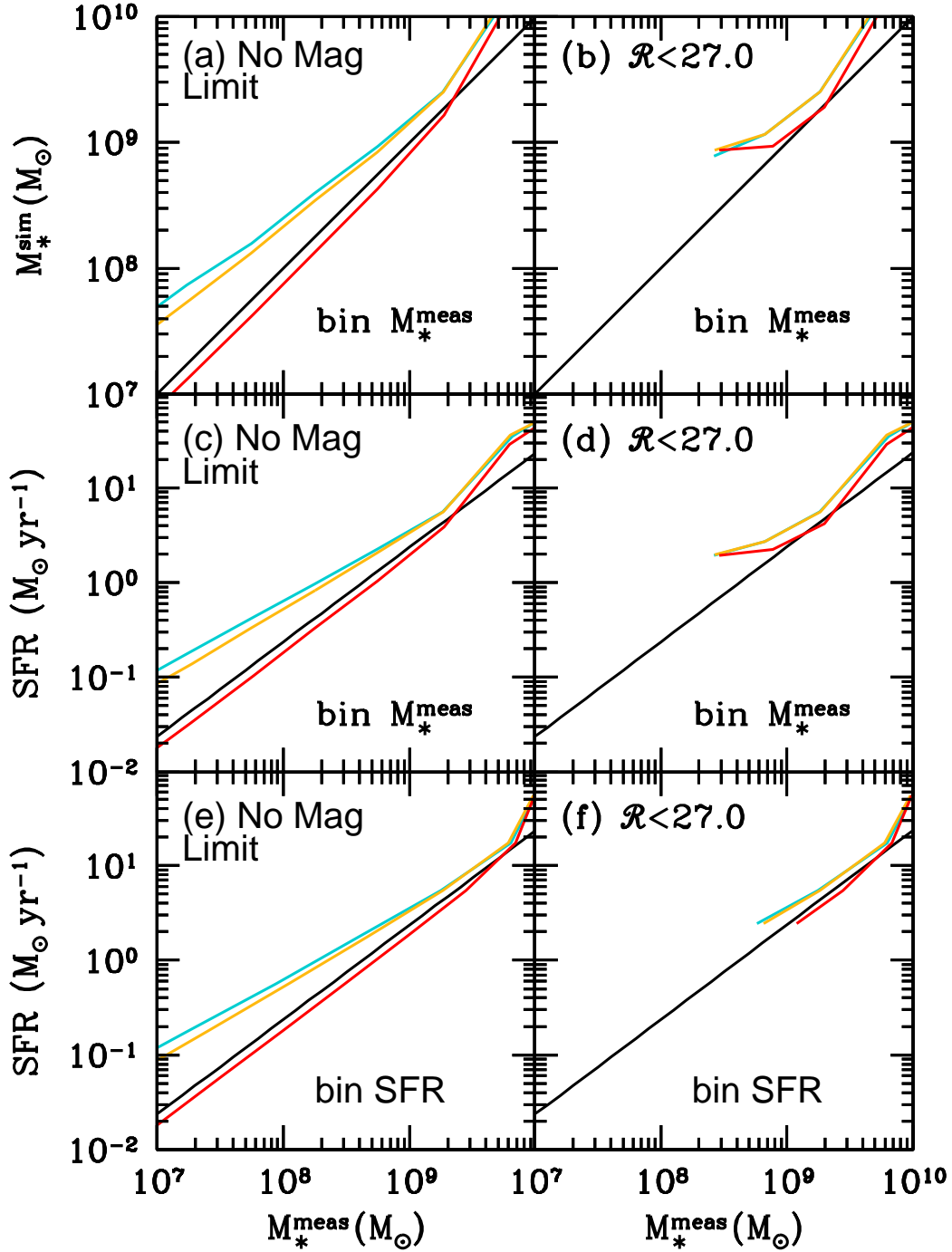}
\caption{Comparison between simulated (black lines) and measured
  stellar mass ($M^{\rm meas}_{\ast}$), binned by the latter, for
  galaxies in the full simulated sample (top row, panel a) and those
  brighter than $\rs=27.0$ (top row, panel b).  The cyan, orange, and
  red lines assume options one, two, and three described in the text
  for converting near-IR magnitude to stellar mass.  The middle row
  indicates the same for the median input SFR in bins of $M^{\rm
    meas}_{\ast}$.  The bottom row indicates the same, except binned
  in terms of SFR.}
\label{fig:simsum}
\end{figure}

The comparison between the simulated ($M^{\rm sim}_{\ast}$) and
measured stellar masses is shown in Panel (a) of
Figure~\ref{fig:simsum}.  For the first two options, there is a
tendency to underestimate $M^{\rm sim}_{\ast}$ because we attribute too
large a fraction of the near-IR light to star formation rather than
stellar mass.  With the third option, we obtain a reasonable agreement
between $M^{\rm sim}_{\ast}$ and $M^{\rm meas}_{\ast}$.  Applying a
magnitude cut results in an artificial offset between $M^{\rm
  meas}_{\ast}$ and $M^{\rm sim}_{\ast}$ because fainter galaxies will
be excluded by the magnitude limit, leaving only brighter (and more
massive simulated) galaxies in the bins of $M^{\rm meas}_{\ast}$.  The
middle row shows the recovered SFR-$M_{\ast}$ relation, where galaxies
are binned by $M^{\rm meas}_{\ast}$.  Because options 1 and 2 result
in an underestimation of stellar mass, there is tendency to predict a
shallower slope of the SFR-$M_{\ast}$ relation.  Introducing a
$\rs<27.0$ limit then causes us to overestimate the median SFR in bins
of $M^{\rm meas}_{\ast}$, due to Malmquist bias
(Appendix~\ref{sec:malmquist}).  Finally, the bottom row shows the
results when binning by SFR, instead of $M^{\rm meas}_{\ast}$.  In this
case, as expected, the Malmquist bias is less noticeable, because at a
given low SFR (or faint UV luminosity) we can probe the full range of
stellar mass, or at least quantify the average stellar mass using
stacking analyses (Figure~\ref{fig:uvch1} and
Appendix~\ref{sec:iracstack}).  The results from the simulations
underscore how the obvious Malmquist bias and the subtler bias arising
from the conversion between near-IR light and stellar mass can affect
our interpretation of the SFR-$M_{\ast}$ relation.  In practice,
estimating the SFR-$M_{\ast}$ relation in bins of SFR may yield more
accurate results that are largely immune to the effects of Malmquist
bias, though one will still have to account for systematics in the
conversion between near-IR light and stellar mass, particularly if
SED-fitting for {\em all} the objects in a sample is not an option.
In our case, for the UV faint samples, we determined median stellar
masses in bins of UV luminosity, which roughly translates into bins of
SFR, so our estimates should be largely unaffected by the Malmquist
bias discussed above.

\end{appendices}

\clearpage


\clearpage


\clearpage

\end{document}